\documentclass[preprint,12pt]{elsarticle}
\usepackage{amsfonts}
\usepackage{amsmath}
\usepackage{amssymb}
\usepackage{amsthm}
\usepackage{fullpage}
\usepackage{mathtools}
\usepackage{color}
\usepackage{tikz}
\usepackage{pgfplots}
\usepackage{comment}
\usepackage{graphicx}
\usepackage{caption}
\usepackage{subcaption}
\usepackage{hyperref}
\usepackage{mathabx}

\allowdisplaybreaks
\RequirePackage{siunitx}
\def \ymax{\bar y}
\def \xmax{\bar x}
\def \zmax{z_{\rm{max}}}
\def \d{\textup{d}}
\def \R{\mathbb R}
\def \E{\mathbb E}
\def \P{\mathbb P}

\newtheorem{theorem}{Theorem}
\newtheorem{remark}[theorem]{Remark}

\newcommand{\eps}{\varepsilon}
\newcommand{\ud}{\operatorname{d}\! }
\newcommand{\bepo}{K-BEPO}

\newcommand{\m}[1]{\begin{pmatrix}{#1}\end{pmatrix}}

\usepackage{lipsum}

\pgfplotsset{compat=1.18}

\begin{document}

\title{A control variate method for threshold crossing probabilities of plastic deformation driven by transient coloured noise}

\author{Harry L.F. \textsc{Ip}$^2$}
\author{Charlie \textsc{Mathey}$^1$}
\author{Laurent \textsc{Mertz}$^{2,3}$}
\author{Jonathan J. \textsc{Wylie}$^2$}
\address{$^1$Laboratoire Interdisciplinaire Carnot de Bourgogne, UMR 6303 CNRS,

Université Bourgogne Europe, Le Creusot, France}
\address{$^2$Department of Mathematics, City University of Hong Kong, Hong Kong, China}
\address{$^3$City University of Hong Kong Shenzhen Research Institute, Shenzhen, China}

\begin{abstract}
We propose a hybrid method combining partial differential equation (PDE) and Monte Carlo (MC) techniques to obtain efficient estimates of statistics for plastic deformation related to kinematic hardening models driven by transient coloured noise. 
Our approach employs a control variate strategy inspired by [CPAM, 75 (3), 455-492, 2022] and relies on a class of PDEs with non-standard boundary conditions, which we derive here. The solutions of those PDEs represent the statistics of models driven by transient white noise and are significantly easier to solve than the coloured noise version.	
    Our approach uses a coupling between the white-noise-driven process and the coloured-noise-driven process, yielding a variance-reduced estimator through control variate techniques.
	We apply our method to threshold-crossing probabilities, which are used as failure criteria known as ultimate and serviceability limit states under non-stationary excitation.
	Our contribution provides solid grounds for such calculations and is significantly more computationally efficient in terms of variance reduction compared to standard MC simulations.
\end{abstract}

\maketitle
\section{Introduction}
We propose a hybrid computational approach that integrates partial differential equations (PDEs) with Monte Carlo (MC) techniques to obtain efficient estimates of statistics for plastic deformation related to kinematic hardening models driven by transient \footnote{i.e. lasting for a finite period} coloured noise. Our approach uses a coupling between the white-noise-driven process and the coloured-noise-driven process, yielding a variance-reduced estimator through control variate techniques.
Materials typically respond to forces by reversible (elastic) or permanent (plastic) deformations \cite{chaboche1979modelization,lipinski1989elastoplasticity,dirrenberger2012elastoplasticity}.
In many situations, the action of cyclic, repeated, or alternating stresses or deformations can modify local properties, such as the threshold of the transition from elastic to plastic behavior, known as the yield strength of a material, and lead to degradation of performance (integrity and material strength \cite{milligan1966bauschinger,sowerby1979review,patinet2020origin}).
This raises a fundamental problem for the risk analysis of failure of mechanical structures under vibration.

The simplest structures are those that can be described by models with a single degree of freedom (1-DoF), for example, rocking blocks~\cite{housner1963behavior} or single pipes~\cite{campbell1983inelastic}.
In fact, these simple models can also be remarkably successful in describing real structures composed of large numbers of interacting components.
For example, \cite{touboul1999seismic} showed experimentally that a complicated system of pipes excited by random shaking could be accurately modeled by a 1-DoF elasto-plastic oscillator. Furthermore, \cite{feau2015experimental} showed that the response of structures as complicated as a crane bridge to earthquakes could also be well described by simple 1-DoF models.
A detailed discussion of the circumstances under which attachments to buildings or industrial facilities can realistically be modeled as a 1DoF system is given in \cite{kasinos2015performance}.

An additional difficulty in earthquake engineering results from the random nature of seismic forces \cite{rossi2015importance,qiang2019seismic,li2020isotropic}.
Two main characteristics of earthquakes, established by seismology models, play an important role in the response of the structure. 
The first is an ``envelope'' describing the amplitude of the acceleration expected over time (typically, earthquake excitations last for a short period of time, usually $10$ to $30$s).
The second is the signal's autocorrelation, often represented by its Power Spectral Density (PSD), the Fourier transform of the autocorrelation function.
Both the envelope and autocorrelation can vary widely depending on the location and magnitude of the epicenter~\cite{alamilla2001evolutionary}. Moreover, local geological structures can greatly amplify narrow ranges of frequencies~\cite{liu1969spectral, bazzurro2004nonlinear}.
As a result, it is important to be able to deal with complicated models of envelopes and autocorrelation structures in the signal.

In the context of structural engineering, the ``limit state'' methodology (used in building regulations in the US \cite{aci2008building} and the EU \cite{eurocode8}) defines failure as the point at which the structure or material reaches a state beyond which it can no longer fulfill its purpose.
Two kinds of limits are commonly used: the ultimate limit and the serviceability limit. 
The ultimate limit refers to the situation when the level of stress, instability, or damage exceeds capacity such that the structure may fail or collapse.
The serviceability limit on the other hand does not refer to the structural integrity but rather to the structure's ability to perform its function efficiently enough.

Our main objective is to compute the probability of failure for a given structural 1D model, namely the bilinear elastoplastic oscillator with kinematic hardening (KBEPO), earthquake model expressed in terms of transient coloured noise, and failure definition expressed in terms of threshold crossing events.
For non-trivial situations, explicit formulas are not available, requiring the use of computational techniques. Two main methodologies are widely used:
\begin{itemize}
\item
The Kolmogorov backward equation (KBE) is a PDE governing the evolution of the probability of failure from all possible initial states and initial times.
Such an approach requires numerical methods for PDEs, for example, finite differences which are suitable when the domain geometry is simple. These methods have the advantage of being able to accurately compute small probabilities, and the numerical efficiency is such that they are computationally possible for a 1-DoF elastoplastic model with white noise excitation (requiring three state space variables and one-time variable).
However, for higher-dimensional models or for correlated noise, this approach is likely infeasible with current computers.
It should be pointed out that recent studies have begun to address the challenges involved in analyzing probability density functions (PDF) solutions of high-dimensional systems, including aspects related to correlated noise and transient behaviors (see, e.g., \cite{er2011, er2012, er2017, er2025, luo2024}).
\item
The standard MC approach estimates the probability of failure by directly simulating the response for a large number of randomly generated signals.
This is easy to implement, but it can be extremely expensive in terms of the number of MC samples when computing small probabilities.
It is worth mentioning that a number of modifications can improve the efficiency of MC methods (e.g. \cite{au2003subset, bourinet2011assessing, katzgraber2006feedback,GOODMAN20097127,Giles_2015}).
\end{itemize}
The hybrid method described in this paper leverages the accuracy (in low dimension) and efficiency (in any dimension) of the KBE and MC approaches, respectively, to estimate the probability of failure of a KBEPO driven by coloured noise.
If one wishes to estimate the expectation $\mathbb{E}(X)$ of a random variable $X$
of finite variance, a naive way would be to use a simple MC estimator to find the average of $X(\omega)$ over $N$ samples $\omega$.
The variance of this estimator is $N^{-1} \textup{var}(X)$, and the efficiency of the method is guaranteed 
by the Strong Law of Large Numbers under
appropriate conditions. 
However, suppose there is a random variable $Y$, called the control variate, that is correlated with $X$ and whose expectation is known or cheaply computable. Then to estimate $\mathbb{E}(X)$, we write $X$ in the form $X = \lambda Y + \{ X-\lambda Y\}$ where $\lambda$ is a parameter. Hence $\mathbb{E}(X) = \lambda \mathbb{E}(Y) + \mathbb{E}(X-\lambda Y)$ and since $\mathbb{E}(Y)$ is cheaply computable, the main task becomes to estimate $\mathbb{E}(X-\lambda Y)$. We can do this by taking an average of $X(\omega) - \lambda Y (\omega)$ over
the $N$ samples $\omega$. The variance of this estimator is $N^{-1} \textup{var}(X - \lambda Y)$. 
If we can adjust $\lambda$ to be the theoretical optimal parameter 
$\lambda^\star = \textup{cov}(X,Y)/\textup{var}(Y)$ 
then the variance is minimal and reduced by a factor $1-\rho^2$ compared to the simple MC
estimator where $\rho$ = $\textup{cov}(X,Y)/\sqrt{\textup{var}(X) \textup{var}(Y)}$.  In some situations, the gain can be particularly dramatic. This is especially true if  $Y$ can be designed in a way such that $\rho$ is close to unity.
In this paper, we will estimate the probability of threshold crossing events for a KBEPO driven by coloured noise. We will achieve this by introducing a random variable that describes the equivalent elasto-plastic system driven by white noise. 
In our context, the latter will be coupled (sharing the same source of randomness) with the coloured noise original system and will play the role of the control variate. 
Then the probability of failure of the mechanical model driven by white noise is computed to high accuracy with a KBE approach, and the residual expectation can be computed using a regular MC average but with significant variance reduction.

\section{Kinematic hardening models with noise and risk of failure}
\subsection{Formulation of kinematic hardening models for engineering applications}
\begin{figure}[h]
	\centering
		\begin{subfigure}[t]{.5\textwidth}
            \includegraphics{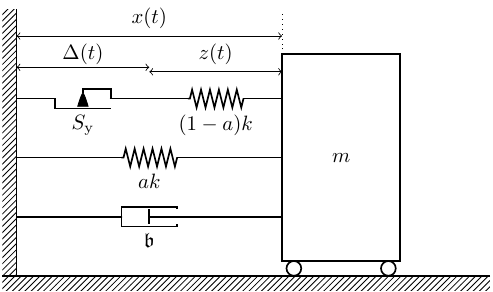}
			\caption{Rheological model.}
			\label{fig:KBEPO_rheo_simple}
		\end{subfigure}%
		\hfill%
    	\begin{subfigure}[t]{.5\textwidth}
				\hspace{-5cm}
				\hfill
            \includegraphics{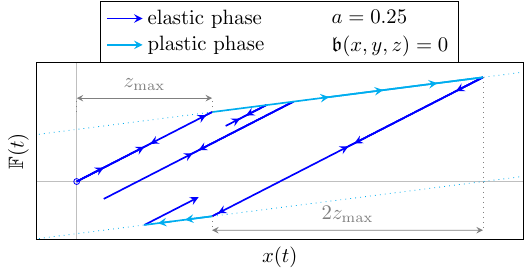}
			\caption{Archetypal evolution of $\mathbb{F}(t)$ and $x(t)$ when
$\mathfrak{b}=0$. }
			\label{fig:KBEPO_graph}
		\end{subfigure}
	\caption{Description of the kinematic hardening bilinear elastoplastic model.(a) In the rheological model, $x(t)$ is the total displacement, that is the sum of the elastic $z(t)$ and plastic $\Delta(t)$ ones. The friction component prevents the elastic line from exceeding yield strength $S_\text{y}$. 
    The damping term $\mathfrak{b}$ resists the total velocity $y$. (b) For the archetypal evolution of $\mathbb{F}(t)$ and $x(t)$ when $\mathfrak{b}=0$, 
 starting from a neutral position, the behaviour starts being elastic (first elastic phase). Once $x$ exceeds $z_\text{max}$ with a positive velocity $y$, $z$ is frozen and $y$ becomes the rate of change of the plastic displacement $\Delta$ (plastic phase). Then, when the velocity becomes negative, $\Delta$ is frozen and $y$ becomes the rate of change of the elastic displacement (second elastic phase).  
 }
	\label{fig:KBEPO}
\end{figure}

The simplest way to model elasto-plastic behavior is to consider the time-dependent one-dimensional deformation $x(\tau)$ at time $\tau \geq 0$ of an oscillator with mass $m$. See Figure \ref{fig:KBEPO_rheo_simple}. The velocity is denoted by $y = \ud x/\ud \tau$. From Newton’s law, 
\begin{equation}
\label{newton}
m \: \ud y/\ud \tau + \mathbb{F} = f,
\end{equation}
where $\mathbb{F}$ is the elasto-plastic force, and $f$ represents all the other forces, internal (e.g. damping) and external (e.g. seismic forcing), some of which are random. It is important to emphasize that at each time $\tau$, $\mathbb{F}$ cannot simply be expressed in terms of $x(\tau)$. Instead, it must be expressed using two variables: the elastic deformation, which we denote as $z$, and the plastic deformation, $\Delta$. See Figure \ref{fig:KBEPO_graph}. These two variables arise from the decomposition of the deformation as $x = z + \Delta$ \cite{Karnopp1966PlasticDI}
where
\begin{eqnarray*}
	\ud z = \ud x, \quad \ud \Delta = 0 &\quad \mbox{in the elastic phase when} \quad  |z|<z_{\textup{max}},\\
	\ud \Delta = \ud x, \quad \ud z = 0 &\quad \mbox{in the plastic phase when} \quad |z|=z_{\textup{max}}.
\end{eqnarray*}
Here, $z_{\textup{max}}>0$ is the yielding displacement.
Then, in this context, the expression of the force is $\mathbb{F} = k z + k_p \Delta$ where $k \geq 0$ is the elastic stiffness and $k_p \in [0,k)$ is the plastic stiffness. For convenience, we denote the plastic-to-elastic stiffness ratio by $a \triangleq k_p/k \in [0,1)$.
Equivalently, we can formulate $\mathbb{F}$ as $\mathbb{F} = a k x + (1-a)k z$. Here, the last term remains bounded in $[-S_\text{y},S_\text{y}]$, where $S_\text{y} \triangleq (1-a)k z_{\textup{max}}$ is the yield strength. See Figure \ref{fig:KBEPO_graph}. It is worth mentioning that we exclude the value $a = 1$ because the evolution of $x$ is the same as that of a linear oscillator, which is not affected by plastic deformation. When $a = 0$, the dynamics of $x$ is the same as that of an elastic-perfectly-plastic oscillator which only requires the tracking of the pair $(y,z)$. This is simpler than the case where $a \in (0,1)$ which requires the whole triple $(x,y,z)$.

The other forces $f$ are of the form $\mathfrak{b}(x,y) + \sigma \xi$, where  $\mathfrak{b}: \mathbb{R}^2 \to \mathbb{R}$ is a function that 
describes the deterministic forces experienced by the oscillator
and $\sigma(\cdot)$ is a positive transient time-dependent function, i.e. vanishing outside of a finite time interval that modulates a Gaussian noise $\xi$, which can be white or coloured. The noise strength is defined as $\mathfrak{s} \triangleq \max \limits_{[0,\infty)} |\sigma|$.  An important example for $\mathfrak{b}$ is when it represents a damping force of the form $\mathfrak{b}(x,y) = -\gamma y$ and $\gamma>0$ is a damping coefficient. 

Below, we derive the effective parameters and the corresponding nondimensional system.
The parameters $m, k$, and $\mathfrak{s}$ are expressed in \textup{kg}, \textup{kg} $\textup{s}^{-2}$ and \textup{kg m} $\textup{s}^{-3/2}$, respectively.
We can introduce the characteristic time and space length, $\tau_c \triangleq \sqrt{m k^{-1}}$ and $\ell \triangleq \mathfrak{s} m^{-1} \tau_c^{3/2}$.
We thus introduce the nondimensional variables 
$$
t \triangleq \tau \tau_c^{-1}, \: \:
X(t) \triangleq \ell^{-1} x(\tau_c t), \: \:
Z(t) \triangleq \ell^{-1} z(\tau_c t), \: \:
Y(t) \triangleq \ell^{-1} \tau_c y(\tau_c t)  
$$ 
and 
$$
\hat \xi (t) \triangleq  \tau_c^{1/2} \xi(\tau_c t). 
$$
Therefore, we can recast Equation \eqref{newton} into the nondimensional form 
\begin{equation}
\label{eq:nondim}
\ud Y/ \ud t + a \hat{k} X + (1-a) \hat{k} Z = \hat{\mathfrak{b}}(X,Y) + \hat{\sigma} \hat{\xi}
\end{equation}
where 
$
\hat{k} \triangleq k m^{-1} \tau_c^2,\:
\hat{\mathfrak{b}}(X,Y) \triangleq  \tau_c m^{-1} \mathfrak{b}(\ell X, \ell \tau_c^{-1} Y),
$
and the elastic bound for $Z$ is $\hat{z}_{\textup{max}} \triangleq \ell^{-1} z_{\textup{max}}$.
We define the normalization of $\sigma$ as $\hat{\sigma}(\cdot) \triangleq \sigma(\cdot) / \mathfrak{s}$.
Below, we will drop the hats for notational simplicity and proceed with the nondimensional system \eqref{eq:nondim} unless stated otherwise.

\subsection{Random forces}
In earthquake engineering, realistic modeling of seismic excitation requires accounting for both its transient duration and frequency-dependent characteristics. While white noise models offer analytical convenience, they fail to capture the spectral content and autocorrelation structure observed in real ground motion records. Coloured noise models, characterized by non-flat PSDs, provide a more accurate representation of seismic forcing. These models reflect the fact that energy is unevenly distributed across frequencies and that local geological conditions can amplify specific frequency bands \cite{liu1969spectral,bazzurro2004nonlinear}
Early work by Saragoni and Hart \cite{Rodolfo1973} introduced artificial earthquake simulations using modulated coloured noise, where the envelope function captures the transient nature of the signal and the PSD encodes its frequency content. This approach was extended by Der Kiureghian and Crempien \cite{DerKiureghian1989}, who proposed evolutionary models combining time-varying envelopes with coloured noise to simulate nonstationary ground motion. These models are particularly relevant for assessing structural response under realistic seismic scenarios. Furthermore, Bormann and Wielandt \cite{Bormann2013} emphasized the importance of distinguishing seismic signals from ambient noise, which often exhibits coloured characteristics due to environmental and instrumental factors.
Recent studies have further demonstrated the utility of coloured noise modeling in seismic applications. Maranò et al. \cite{marano2017} showed that incorporating coloured noise into spectral fitting significantly improves the estimation of earthquake source parameters, especially under low signal-to-noise conditions. Chen et al. \cite{chen2023} proposed an adaptive Kalman filter that models Global Navigation Satellite System (GNSS) displacement noise as coloured noise, enhancing the fusion of GNSS and strong motion data for coseismic deformation monitoring. Hillers et al. \cite{hillers2015} applied noise-based imaging techniques to detect aseismic transient deformation, highlighting the role of coloured ambient noise in passive seismic monitoring.
In this work, we adopt a general framework for coloured noise based on Ornstein–Uhlenbeck processes, which allows flexible modeling of PSDs while retaining analytical tractability. This choice enables coupling with white-noise-driven systems in the small-correlation limit, facilitating variance reduction via control variates. Such modeling is essential for capturing the stochastic nature of earthquake excitation and for improving the reliability of failure probability estimates in structural systems.
Our framework for coloured noise encompasses popular models of ground motion adopted in seismic engineering, namely the Kanai-Tajimi model and the Clough-Penzien model \cite{er2019,er2018,er2014}.

\subsubsection*{Comments on the modulation function $\sigma(\cdot)$ used in noise modeling}
A common example used in the earthquake literature \cite{Rodolfo1973} for the function $\sigma(\cdot)$ is
	\begin{equation}
		\sigma(t) = \alpha \, t^\beta e^{-\gamma t}, 
        \: \: t \geq  0.
		\label{eq:sigma}
	\end{equation}
The specific parameters $\alpha, \beta, \gamma$ can be chosen to obtain a best fit between modulation function and real seismic signal, as shown in \cite{DerKiureghian1989}.
The coloured noise component $\xi$ is characterized by its PSD. In practice, it is possible to extract an empirical PSD from a given data of a real seismic signal \cite{Bormann2013}. However, this goes beyond the scope of the paper. Instead, we will directly consider coloured noise arising from two general classes of PSDs.


\subsubsection*{Ornstein-Ulhenbeck processes}
From a modeling point of view, using white noise excitations can only be realistic for a very restricted class of seismic applications.
To meet the need for more advanced modeling, a natural generalization consists of considering coloured noise, which we model by a class of zero-mean stationary Gaussian processes.
The latter can be represented by sums of Ornstein-Uhlenbeck (OU) processes. For the purpose of implementing our hybrid method, we will further require the limiting distribution (as $\eps \rightarrow 0$) of the coloured noise to be that of a white noise. In particular, given $\eps > 0$, we will consider noise of the form $
\xi^\eps = \boldsymbol{r} \cdot \eta^\eps/ \eps
$, with $d \geq 1$, $\boldsymbol{r} = (r_1 , \cdots , r_d ) \in \mathbb{R}^d$ and $\eta^\eps \in \mathbb{R}^d$ is an OU process, i.e. satisfying
\begin{equation}
\label{eq:noise}
{\rm d} \eta^\eps = - \frac{A}{\eps^2} \eta^\eps {\rm d} t + \frac{K}{\eps} {\rm d} W_t
\end{equation}
where $A \in \mathbb{R}^{d \times d}$ has eigenvalues with positive real parts, 
$K \in \mathbb{R}^{d \times d'}$, and $W \in \mathbb{R}^{d'}$ is a multidimensional standard Brownian motion.
In this way, the unique stationary probability measure of $\eta^\eps$ is given by a multivariate centered Gaussian distribution with the following covariance matrix  
\begin{equation}
\label{eq:cov_matrix}
C(A,K) \triangleq \int_0^\infty e^{-As} K K^T e^{-A^T s} \textup{d} s.
\end{equation}
The parameter $\eps$ allows us to adjust the level of autocorrelation in the noise.
Using the process $\xi^\eps$ above, we can construct two types of real-valued noise depending on the parameter $\eps$ 
whose PSD are of the form $S^\eps(\omega) = S(\eps^2 \omega)$.   
We distinguish two cases where the matrices $A,K$ and $\boldsymbol{r}$ have different structures: either $d = q$ or $d = 2 q$ for $q\geq 1$.
For the first type, we consider a sum of centered Lorentzians
\begin{equation}
\label{eq:PSD1}
S(\omega) = \sum \limits_{k=1}^q \frac{r_k^2}{1+\omega^2/\lambda_k^2},
\end{equation}
where $\lambda_1, \dots, \lambda_q \in \R$ are constants. Here $S^\eps(\cdot)$ is the PSD of  $
\xi^\eps = \boldsymbol{r} \cdot \eta^\eps/ \eps
$, $\eta^\eps$ is the solution of \eqref{eq:noise} with $d=d'=q$ and $A = K = \textup{diag} (\lambda_1, \cdots, \lambda_q)$.

\noindent For the second type, we consider a sum of uncentered Lorentzian
\begin{equation}
\label{eq:PSD2}
S(\omega) = \frac{1}{2} \left ( \sum \limits_{k=1}^q \frac{r_k^2}{1+(\omega-\omega_k)^2/\lambda_k^2} + \sum \limits_{k=1}^q \frac{r_k^2}{1+(\omega+\omega_k)^2/\lambda_k^2} \right ),
\end{equation}
where $\boldsymbol{r} = (r_1, 0 , \cdots , r_q, 0 ) \in \mathbb{R}^{2 q}$ and $\lambda_1, \dots, \lambda_q, \omega_1, \dots, \omega_q \in \R$ are constant. Here $S^\eps(\cdot)$ is the PSD of 
$
\xi^\eps = \boldsymbol{r} \cdot \eta^\eps/ \eps
$, $\eta^\eps$ is solution of (\ref{eq:noise}) with $d=d'=2q$ and 
$$
A = \oplus_{k=1}^q \begin{pmatrix} \lambda_k & -\omega_k \\  \omega_k & \lambda_k  \end{pmatrix} ,
\quad 
K =  \oplus_{k=1}^q \begin{pmatrix}  \lambda_k & 0 \\  0 &  \lambda_k  \end{pmatrix},
$$
where the notation $\oplus$ is the direct sum of matrices, i.e. $A, K \in \mathbb{R}^{2q\times2q}$.
We note that in both cases when $\eps \downarrow 0$, $S^\eps$ goes to a constant, that is the PSD of white noise, precisely $
\xi^0 = \boldsymbol{r} \cdot A^{-1} K \ud W/ \ud t
= \boldsymbol{r} \cdot \ud W/ \ud t$ for the first case and $\xi^0 = \boldsymbol{r} \cdot A^{-1} K \ud W/ \ud t$ for the second case. This can also be seen intuitively by writing Equation \eqref{eq:noise} as
$\eps^2 \left ( \ud \eta^\eps/\eps \right ) = - A \eta^\eps \ud t + K \ud W_t$ and neglecting the left hand side when $\eps$ is small. In the second case, we assume that 
$$
\sum_{k=1}^q r_k^2 = 1 
\: 
\mbox{ and }
\:
\sum_{k=1}^q \frac{r_k^2}{1+\omega_k^2/\lambda_k^2} = 1. 
$$
Hence, in both cases, $\xi^0$ has the same law as $\ud W / \ud t$.

\begin{remark}[Choice of PSDs]
The coloured noise models used in this work are based on two classes of PSDs, given in Equations~(6) and~(7). The first corresponds to a sum of centered Lorentzian functions, which allows for broadband noise modeling with tunable decay rates and leads to a diagonal structure in the OU representation. The second corresponds to a sum of uncentered Lorentzians, enabling the modeling of narrowband noise with peaks at prescribed frequencies, which is particularly relevant in earthquake engineering where local site effects can amplify specific frequency bands. Both choices ensure that the coloured noise converges to white noise in the limit $\varepsilon \to 0$, a property that is essential for the control variate strategy employed in our hybrid method. This design facilitates coupling between the coloured and white noise-driven systems and enables efficient variance reduction.
\end{remark}

\subsection{Formulation of kinematic hardening models using stochastic variational inequalities} 
Inspired by \cite{MR0521262}, we adopt a mathematical formulation of the models as described in the previous section by representing them with variational inequalities.
To describe the dynamics of the kinematic hardening models shown in \eqref{newton} and driven by the noise $\xi^\eps$, we consider the $\mathbb{R}^3$ valued process $\{ (X^\eps(s),Y^\eps(s),Z^\eps(s)), \: s \geq t \}$ which solves the variational inequality: for a.e. $s>t$,
\begin{equation}
\label{eq:svi}
\begin{cases}
& \dot X^\eps(s) = Y^\eps(s),\\
& \dot Y^\eps(s) = \mathfrak{b}(X^\eps(s),Y^\eps(s)) - a k X^\eps(s) - (1-a) k Z^\eps(s) + \sigma(s)\xi^\eps(s),\\  
& |Z^\eps(s)| \leq z_{\textup{max}}, \: \forall \varphi \in [-z_{\textup{max}},z_{\textup{max}}] , \: (\dot Z^\eps(s) - Y^\eps(s))(\varphi - Z^\eps(s)) \geq 0,
\end{cases}
\end{equation}
with the initial state $(X^\eps(t),Y^\eps(t),Z^\eps(t)) = (x,y,z) \in \mathbb{R}^2 \times [-z_{\textup{max}},z_{\textup{max}}]$.
The problem is well-posed \cite{MR3308895} in both cases $\eps>0$ and $\eps = 0$.

\subsection{Risk of failure}
The risk of failure problem consists of finding the probability that, either over a certain interval of time $[0,T]$ or at a given time $T$, the state variable $(x,y,z) \in \R^2 \times [-z_{\textup{max}},z_{\textup{max}}]$ reaches a given domain of failure, say $D_\text{f} \subset \R^2 \times [-z_{\textup{max}},z_{\textup{max}}]$. In this paper, both types of failures will be considered:
\begin{subequations}
	\begin{itemize}
		\item the ultimate limit state (ULS) probability can be expressed as the probability that the total deformation reaches a threshold $X_{\rm f}$ at any time over a finite time period $[0,T]$.
	      Given that the system is in the state $(x_0,y_0,z_0) \in \mathbb{R}^2\times[-z_{\max},z_{\max}]$ at $t=0$, this probability is given by:
	      \begin{equation}
		      \label{eq:P1}
		      P_1^\eps = \P \left(\max_{t \in [0,T]}{{|X^\eps(t)|}} \geq X_{\rm f} \:\middle|\: \big(X^\eps(0), Y^\eps(0), Z^\eps(0)\big) = (x_0,y_0,z_0)\right).
	      \end{equation}
	      For this quantity,
        $D_{\rm f} = \left\{(x,y,z), |x| \geq X_{\rm f} \right\}$.
	    Recall that the total displacement $X^\eps(t)$ is the sum of the elastic and plastic displacements, $Z^\eps(t)$ and $X^\eps(t)-Z^\eps(t)$, respectively.
		\item the serviceability limit state (SLS) probability can be expressed as the probability that the plastic deformation exceeds a threshold $\Delta_{\rm f}$ at the final time $T$.
	    Given that the system is in the state $(x_0,y_0,z_0) \in \mathbb{R}^2\times[-z_{\max},z_{\max}]$ at $t=0$, this probability is given by:
	      \begin{equation}
		      \label{eq:P2}
		      P_2^\eps = \P \left({|X^\eps(T)-Z^\eps(T)|} \geq \Delta_\text{f} \:\middle|\: \big(X^\eps(0), Y^\eps(0), Z^\eps(0)\big) = (x_0,y_0,z_0)\right).
	      \end{equation}
	      For this quantity, $D_{\rm f} = \left\{(x,y,z), |x-z| \geq \Delta_{\rm f} \right\}$.

	\end{itemize}
\end{subequations}

\begin{remark}
The KBEs related to the risk of failure for the white noise $(\eps=0)$ driven kinematic hardening models are non-standard boundary value problems. For this type of non-standard PDEs where the value of the solution on the boundary depends on the values of the solution inside the domain, only partial existence and uniqueness results are available, mainly for the case where $a=0$ {\normalfont{\cite{BT08}}}. 
This is because standard PDE theory techniques do not apply due to the challenging non-standard boundary conditions and the degeneracy of the differential operators involved. Therefore, we adopt the same philosophy as \cite{MWS18} who used numerical techniques. By performing careful convergence tests, their results are highly suggestive of the existence of solutions. We also have performed convergence tests (on sequences of grids with increasing resolution) and all of our results are suggestive that the continuous problem has a solution.
\end{remark}
\section{Hybrid method combining PDEs and MC for the coloured noise system statistics}
It is possible to relate the threshold-crossing probabilities $P_1^0$ and $P_2^0$ of the white noise-driven system to the solution of parabolic PDEs with non-standard boundary conditions in the same spirit as \cite{MWS18}. In the first part, we provide a formal presentation of these PDEs. In the second part, we present our hybrid approach which combines the solution of the PDE with our control variate Monte Carlo estimator. 

\subsection{PDEs for \texorpdfstring{$P_1^0$ and $P_2^0$}{P10 and P20} of the white noise driven system}
\label{sec:pde}
Fix $T>0$. For $P_1^0$, we define
$$D_1 \triangleq (-X_f,X_f) \times (-\infty,\infty) \times (-z_{\textup{max}},z_{\textup{max}}) 
\: \:
\mbox{ and }
\: \:
D^{\pm}_1 \triangleq (-X_f,X_f) \times (-\infty,\infty) \times \{ \pm z_{\textup{max}} \}.
$$
For $P_2^0$, we define
$$
D_2 \triangleq (-\infty,\infty) \times (-\infty,\infty) \times (-z_{\textup{max}},z_{\textup{max}}) 
\: \:
\mbox{ and }
\: \:
D^{\pm}_2 \triangleq (-\infty,\infty) \times (-\infty,\infty) \times \{ \pm z_{\textup{max}} \}.
$$ 
We introduce the notation $\mathcal{C}_{t,x,z}^1 \mathcal{C}_y^2$ for the set of functions continuously differentiable with respect to (w.r.t.) the variables $x,z,t$, and twice continuously differentiable w.r.t. the variable $y$. Below, we use the notation $\mathbf{B}(x,y,z) = \mathfrak{b}(x,y) - k a x - k(1-a) z$.
Using Ito's lemma, we can derive the generator of $(X^0, Y^0, Z^0)$, the solution of \eqref{eq:svi} driven by the modulated transient white noise $\sigma(t)\xi^0(t)$, which in law is the same as $\sigma(t) \ud W (t) / \ud t$. Since $\sigma(.)$ is time-dependent, the generator is also time-dependent. It is defined for any function $\varphi \in \mathcal{C}_{t,x,z}^1 \mathcal{C}_y^2$ as follows
\begin{align*}
& \lim \limits_{h \to 0} \frac{\mathbb{E}\varphi(X^0(t+h),Y^0(t+h),Z^0(t+h)) - \varphi(x,y,z)}{h}\\ 
& =
\left \{ 
\begin{array}{ll}
L(t) \varphi(x,y,z), \quad & |z| < z_{\textup{max}}, \: x,y \in (-\infty,\infty)\\
L_\pm(t) \varphi(x,y, \pm z_{\textup{max}}), \quad & z = \pm z_{\textup{max}}, \: x,y \in (-\infty,\infty),
\end{array}
\right .
\end{align*}
where for any test function $\varphi$
$$
L(t) \varphi = \frac{\sigma^2(t)}{2}\varphi_{yy} + \mathbf{B}(x,y,z) \varphi_y + y \varphi_x + y \varphi_z,
$$
$$
L^+(t) \varphi = 
\frac{\sigma^2(t)}{2} \varphi_{yy} + \mathbf{B}(x,y,z_{\textup{max}}) \varphi_y 
+ y \varphi_x + \min(y,0) \varphi_z,
$$
and
$$
L^-(t) \varphi = \frac{\sigma^2(t)}{2} \varphi_{yy} +  \mathbf{B}(x,y,-z_{\textup{max}}) \varphi_y 
+ y \varphi_x + \max(y,0) \varphi_z.
$$
Here the notation $\varphi_x$ stands for the partial derivative of $\varphi$ w.r.t. $x$, etc.

\subsubsection{PDE problem for \texorpdfstring{$P_1^0$}{P10}}
Consider a function $v \in \mathcal{C}_{t,x,z}^1 \mathcal{C}_y^2$ that satisfies the following PDE,
\begin{equation}
\label{eq:pdeBEPO1}
\begin{cases}
    & v_t+L(t)v=0 \; {\rm on} \; D_1\times[0,T),\\
    & v_t+L^\pm(t)v=0 \; { \rm on} \; D_1^\pm \times[0,T),\\
    & v(\pm X_f,y,z,t) = 1 \; {\rm on} \; (-\infty,\infty) \times (-z_{\rm max},z_{ \rm max}) \times [0,T),\\
    & v(x,y,z,T) = 0 \; {\rm on} \; D_1.
\end{cases}
\end{equation}
Then $v(x,y,z,t) = \mathbb{P}[\tau_t \leq T | (X^0_t,Y^0_t,Z^0_t)=(x,y,z)]$ where $\tau_t = \inf \{ s \geq t, \: |X^0_s| = X_{\rm f} \}$.
Hence, $v(x,y,z,0) = P_1^0$.

\subsubsection{PDE problem for \texorpdfstring{$P_2^0$}{P20}}
Consider a function $w \in \mathcal{C}_{t,x,z}^1 \mathcal{C}_y^2$.  Assume $w$ satisfies 
\begin{equation}
\label{eq:pdeBEPO2}
\begin{cases}
    & w_t + L(t)w = 0 \: \mbox{ in } \: D_2 \times [0,T), \\
    & w_t + L^\pm(t) w = 0 \: \mbox{ in } \: D^{\pm}_2 \times [0,T), \\
    & w(T) = f \: \mbox{ in } \: D_2 \cup D_2^- \cup D_2^+.
\end{cases}
\end{equation}
Then $w(x,y,z,t) = \mathbb{E}[ f(X^0(T),Y^0(T),Z^0(T)) | (X^0(0),Y^0(0),Z^0(0)) = (x,y,z) ]$. As a consequence, if $f(x,y,z) = \mathbf{1}_{\{ |x-z| \geq \Delta_{\rm f}\}}$ then $w(x,y,z,0) = P_2^0$.

\begin{remark}
Equations~\eqref{eq:pdeBEPO1} and~\eqref{eq:pdeBEPO2} are derived via Itô’s lemma, applied to the stochastic variational inequality formulation presented in Equation~\eqref{eq:svi}. Specifically, the functions $v$ and $w$ correspond to the solutions of backward Kolmogorov equations associated with two distinct failure criteria: $v$ relates to the ultimate limit state (ULS) probability $P_1^0$, while $w$ pertains to the serviceability limit state (SLS) probability $P_2^0$. This connection can be established using an approach similar to that employed in Section~3 of~\cite{MWS18}.
\end{remark}

\subsection{Control variate MC estimator for \texorpdfstring{$P_1^\eps$}{P1eps} and \texorpdfstring{$P_2^\eps$}{P2eps} of the coloured noise driven system}
\label{sec:approx_diffusion}
We now turn to the case $\epsilon > 0$, corresponding to the system driven by coloured noise. In this regime, we estimate the threshold crossing probabilities $P^\epsilon_1$ and $P^\epsilon_2$ using a control variate Monte Carlo approach.
In this section, we present our MC estimator for the computation of the threshold crossing probabilities $P_1^\eps$ and $P_2^\eps$. We explain how it fits into the formalism and the convergence results of \cite{MR4373175}.

The probabilities $P_1^\eps$ and $P_2^\eps$ are similar to quantities of the form
$$
I^\eps \triangleq \E (F\{ (X^\eps(t),Y^\eps(t),Z^\eps(t)), 0 \leq t \leq T \})
$$
where $F$ is a real-valued function on the space of continuous functions on $[0,T]$. For $P_1^\eps$ and $P_2^\eps$, the underlying functions are respectively $$
F_1(x,y,z) = \mathbf{1}_{ \left \{ \max \limits_{t \in [0,T]}|x(s)| \geq X_{ \rm f} \right \}}
\: \mbox{ and } \:
F_2(x,y,z) = \mathbf{1}_{\{ |x(T)-z(T)| \geq \Delta_{\rm f} \}}.
$$
Other possible examples (not treated in this work) include functions of the form 
$$
F(x,y,z) = f(x(T),y(T),z(T)) + \int_0^T g(x(s),y(s),z(s)) \d s
$$
where $f,g$ are real valued functions on $\R^3$.
Our estimator is built using a coupling approach as follows. Let $W$ be a given Wiener process. We use the notation $U^0(W) \triangleq(X^0,Y^0,Z^0)$ and $U^\epsilon(W)\triangleq(X^\epsilon,Y^\epsilon,Z^\epsilon)$, where $\epsilon > 0$, to denote two processes satisfying \eqref{eq:svi} with random forces given by $\xi^0=W$ and $\xi^\epsilon$ driven by $W$ (See Equation \eqref{eq:noise}, respectively. Of particular note, $U^0(W)$ and $U^\epsilon(W)$ are then driven by the same underlying Wiener process $W$ (same randomness) through $\xi^0$ and $\xi^\epsilon$. This coupling ensures that $U^0(W)$ and $U^\epsilon(W)$ are correlated, which forms the basis of our control variate approach. For conciseness, when taking expectations, we drop the explicit dependence of $U^0(W), U^\epsilon(W)$ on any Wiener process $W$ and simply write $\mathbb{E}(F(U^0))$ or $\mathbb{E}(F(U^\epsilon))$, where $F$ is a function.
For $\E (F (U^\eps))$, the simple MC estimator by control variate is defined as
\begin{equation}
\label{eq:simple_cv_estimator}
\hat J_N^\eps
\triangleq 
\E (F (U^0)) + \frac{1}{N} \sum_{k=1}^N F(U^\eps(W^k)) - F(U^0(W^k))  
\end{equation}
where $W^k, k=1 \dots N$ are $N$ independent and identically distributed Wiener processes. 
The asymptotic distribution of the estimator is normal with asymptotic variance $\sigma_{J}^\eps = \textup{var} \left ( F(U^\eps) - F(U^0) \right )$.

We also consider the practical optimal MC estimator by control variate defined as
\begin{equation}
\label{eq:optimal_cv_estimator}
\hat K_N^\eps
\triangleq 
\hat \rho_N^\epsilon \E (F (U^0)) + \frac{1}{N} \sum_{k=1}^N F(U^\eps(W^k)) - \hat \lambda_N^\epsilon F(U^0(W^k)),
\end{equation}
where 
$$
\hat \lambda_N^\epsilon 
\triangleq \frac{\sum \limits_{k=1}^N (F(U^\eps(W^k) - \hat E_{MC}^{N,\eps})(F(U^0(W^k) - \hat I_N^0)}{\sum \limits_{k=1}^N (F(U^0(W^k) - \hat I_N^0)^2}
$$
with
$$
\hat I_N^0 \triangleq \frac{1}{N} \sum \limits_{k=1}^N F(U^0(W^k) 
\: 
\mbox{ and }
\: 
\hat I_N^\eps \triangleq \frac{1}{N} \sum \limits_{k=1}^N F(U^\eps(W^k). 
$$
This estimator contrasts with the theoretical optimal control variate estimator as we replace the optimal coefficient $\hat \lambda_N^\epsilon$ with the empirical one $\lambda^\epsilon \triangleq \mathrm{cov}(F(U^\eps),F(U^0)) / \mathrm{var}(F(U^0))$.
Here, the asymptotic distribution is normal with asymptotic variance $\sigma_{K}^\eps = \textup{var} \left ( F(U^\eps) - \lambda^\epsilon F(U^0) \right )$.

In comparison, the standard Monte Carlo estimator
\begin{equation}
\label{eq:standard_estimator}
\hat I_N^\eps
\triangleq 
\frac{1}{N} \sum_{k=1}^N F(U^\eps(W^k))
\end{equation}
is asymptotically normal but with asymptotic variance $\sigma_{I}^\eps = \textup{var} \left ( F(U^\eps) \right )$.

\begin{remark}
For all the estimators discussed above, asymptotic normality is ensured by the Central Limit Theorem. The conditions required for its application—such as independence and finite variance—are well satisfied in our framework, guaranteeing that the estimators exhibit predictable behavior in the large-sample regime.
\end{remark}

Let us remark that our Equation \eqref{eq:svi} is contained in the class of variational inequalities that appear in Equation (2.31) of \cite{MR4373175}. 
When $\sigma(\cdot) \equiv \sigma_0$ is constant, Proposition 5.2 of \cite{MR4373175} implies the convergence in probability of the continuous process $U^\eps-U^0$ to $0$. This property is also valid when $\sigma(\cdot)$ is continuous on $[0,T]$ which is the case of our present paper. Allowing for a few minor modifications, the proof is the same as in the case in which $\sigma(\cdot)$ is constant. As a consequence, when $F(\cdot)$ is bounded and continuous, $\sigma_{J}^\eps$ vanishes as $\eps \to 0$. On the other hand, $\sigma_{I}^\eps \to \textup{var} \left ( F(U^0) \right )$, thus making the variance reduction possible when $\eps$ is small enough. The estimator $\hat K_N^\eps$ inherits this property, as it is optimized so that its variance is always smaller than those of the estimators $\hat I_N^\eps$ and $\hat J_N^\eps$. In fact, it is also possible to characterize the rate of variance reduction when $F(\cdot)$ has a particular structure such as $
F(x,y,z) = f(x(T),y(T),z(T))$
where $f$ is smooth with bounded derivative. Indeed as shown in Lemma 6.1 of \cite{MR4373175}, both variances 
$\sigma_{J}^\eps$ and $\sigma_{K}^\eps$ are of the order $O(\eps)$ when $\eps \to 0$.
Although $F_1$ and $F_2$ do not exactly fall into the class of continuous functions on the path space, our estimator for $P_1^\eps$ and $P_2^\eps$ appears to be highly efficient in this context.    
We will illustrate this using the numerical results shown below. 

\begin{remark}
Equations (12), (13), and (14) define estimators for quantities of the form $\mathbb{E}[F(U^\epsilon)]$, where $F$ is a real-valued function defined on the space $C([0,T]; \mathbb{R}^3)$ of continuous paths over the fixed time interval $[0,T]$. The process $U^\epsilon = (X^\epsilon, Y^\epsilon, Z^\epsilon)$ evolves in time, but the estimators aggregate entire sample paths to produce scalar values. Since $T$ is fixed, the results of these equations do not depend on time.
\end{remark}

\section{Numerical simulations}
\subsection{Truncated PDEs for finite difference discretization}
\label{sec:fdm_pde}
We recall that the PDEs associated with $P_1^0$ and $P^0_2$ correspond to Equations (10) and (11), respectively, as introduced in Section 3.1.

For the computation of $P_1^0$,
we need to truncate $D_1$ and $D^{\pm}_1$
and add artificial boundary conditions at $y= \pm \ymax$ with $\ymax>0$.
Let 
$\mathcal{R} \triangleq [-X_f,X_f]
\times
(-\ymax,\ymax)
\times
[-\zmax,\zmax]$ and define
\begin{align*}
& D_{1,\mathcal{R}} \triangleq 
D_1 \cap \mathcal{R} 
=
(-X_f,X_f) 
\times
(-\ymax,\ymax)
\times
(-\zmax,\zmax),\\ 
& D_{1,\mathcal{R}}^\pm 
\triangleq 
D^\pm_1 \cap \mathcal{R} =(-X_f,X_f) 
\times
(-\ymax,\ymax)
\times
\{ \pm \zmax \},\\
& 
\Gamma^{X_f}_1
\triangleq 
\{ \pm X_f \}
\times 
[-\ymax,\ymax]
\times
[-\zmax,\zmax]
, \: \mbox{ and }
\Gamma^{\ymax}_1 \triangleq 
(-X_f,X_f)
\times \{ - \ymax,\ymax \}
\times 
[-\zmax,\zmax].
\end{align*}
The value of $\ymax$ is chosen sufficiently large such that the probability of the underlying process outside $D_{1,\mathcal{R}}$ is sufficiently small that it does not affect the solution at the desired accuracy.
We impose artificial reflecting boundary conditions to each of the four equations in \eqref{eq:pdeBEPO1}. Thus we consider the problem of finding $v$ satisfying
\begin{equation}
\label{eq:truncatedpde1}
\begin{cases}
v_t + L(t) v = 0, 
& \: \mbox{on} \: D_{1,\mathcal{R}} \times [0,T),\\
v_t + L^{\pm}(t) v = 0, 
& \: \mbox{on} \: D_{1,\mathcal{R}}^{\pm} \times [0,T),\\
v_y = 0, 
& \: \mbox{on} \: \Gamma^{\ymax}_1 \times [0,T),\\
v = 1,
& \: \mbox{on} \: \Gamma^{X_f}_1 \times [0,T),\\
v(T) = 0,
& \: \mbox{on} \: D_{1,\mathcal{R}},
\end{cases}
\end{equation}
where $L(t)$ and $L^\pm(t)$ are defined in Section \ref{sec:pde},  and
$$
L^{\pm \xmax}(t) \varphi = \frac{\sigma^2(t)}{2}\varphi_{yy} + \mathbf{B}(\pm \xmax,y,z) \varphi_y \pm \min(\pm y,0) \varphi_x + y \varphi_z,
$$
$$
L^{\pm \xmax,+}(t) \varphi = \frac{\sigma^2(t)}{2}\varphi_{yy} + \mathbf{B}(\pm \xmax,y,\zmax) \varphi_y \pm \min(\pm y,0) \varphi_x + \min(y,0) \varphi_z,
$$
$$
L^{\pm \xmax,-}(t) \varphi = \frac{\sigma^2(t)}{2}\varphi_{yy} + \mathbf{B}(\pm \xmax,y,-\zmax) \varphi_y \pm \min(\pm y,0) \varphi_x + \max(y,0) \varphi_z.
$$
Here we recall the elementary identity
$-\min(-y,0) = \max(y,0)$.
The truncation of the PDE associated with $P_2^0$ follows similarly but with an additional artificial reflecting boundary condition at $x=\pm \xmax$, for sufficiently large $\xmax$.
Let 
$\mathcal{S} \triangleq (-\xmax,\xmax) 
\times
(-\ymax,\ymax)
\times
[-\zmax,\zmax]$ and similarly define 
\begin{align*}
& D_{2,\mathcal{S}} \triangleq 
D_2 \cap \mathcal{S}, 
\: \:
D_{2,\mathcal{S}}^\pm 
\triangleq 
D^\pm_2 \cap \mathcal{S},
\: \:
\mathrm{and} \: \:
\Gamma^{\ymax}_2 \triangleq 
[-\xmax,\xmax]
\times \{ \pm \ymax \}
\times 
[-\zmax,\zmax].
\end{align*}
For the purpose of the artificial reflecting boundary at $\pm \xmax$, we define two additional domains
\begin{align*}
& 
\Gamma^{\pm \xmax}_2
\triangleq 
\{ \pm \xmax\}
\times 
(-\ymax,\ymax)
\times
(-\zmax,\zmax), 
\: \: \mathrm{and} \: \:
\Gamma^{\pm \xmax, \pm}_2
\triangleq 
\{ \pm \xmax\}
\times 
(-\ymax,\ymax)
\times
\{ \pm \zmax \}.
\end{align*}
The solution of the PDE now corresponds to solving for $w$ satisfying the set of equations as above but with the last two equations replaced by 
\begin{equation}
\label{eq:truncatedpde2}
\begin{cases}
w_t + L^{\pm \xmax}(t) w = 0, 
& \: \mbox{on} \: \Gamma^{ \pm \xmax}_2 \times [0,T),\\
w_t + L^{\pm \xmax, \pm}(t) w = 0, 
& \: \mbox{on} \: \Gamma^{\pm \xmax, \pm}_2 \times [0,T),\\
w(T) = f, 
& \: \mbox{on} \: D_{2,\mathcal{S}}.
\end{cases}
\end{equation}
along with the appropriate changes made corresponding to the three equations \eqref{eq:pdeBEPO2}.
Finite difference implementation details are given in \ref{sec:fd_details}.

\subsection{Numerical examples}
\subsubsection{Numerical Solution of \texorpdfstring{$P_1^0$}{P10} and \texorpdfstring{$P_2^0$}{P20} (white noise driven probabilities)}
\label{sec:num_soln_PDE}
To obtain solutions of the two PDEs in Section \ref{sec:pde}, we numerically solve the linear systems from Section \ref{sec:fdm_pde} backward in time between the time steps $n=N_T$ and $n=0$.
In particular, at each time step, we solve the linear systems by the method of Generalized Minimal Residuals with tolerance $=10^{-6}$, and precondition each system with Incomplete LU Factorization. 
For the discretized PDE corresponding to $P_1^0$, convergence of the solutions has been obtained with the choice of artificial boundary $\ymax = 2.5$, space step sizes of $\delta x = \delta z = 2^{-9}$, $\delta y = 2^{-6}$, and $N_T = 2^{12}$ time steps. Similarly, convergence of solutions can be shown with the same choice of parameters but with $\xmax = 2.5$ for the discretized PDE corresponding to $P_2^0$.

\noindent Figure \ref{fig:pnsn} presents the results for the probabilities $P_1^0$ and $P_2^0$ for a range of threshold values $X_f$ and $\Delta_f$ by the numerical KBE method (circles, triangles, and squares) for three choices of elasto-plastic stiffness ratios $a$: $a=0$ (elasto-perfectly-plastic oscillator), $a=0.5$, and $a=1$ (linear elastic oscillator) (respectively). Note that the linear elastic oscillator ($a=1$) does not consist of plastic deformation; hence, the probability $P_2^0$ is excluded for $a=1$.
For comparison, approximations of the same probabilities $P_1^0$ and $P_2^0$ (dotted, dash-dotted, and dashed lines, for $a=0$, $a=0.5$, and $a=1$  respectively) have been computed with the intensive Monte Carlo method with time step $\delta t = 10^{-4}$ and $10^8$ samples.
All results were obtained using the parameters $\alpha = 2.84$, $\beta = 2$, and $\gamma = 1.25$ for the modulation function \eqref{eq:sigma} (these parameters are commonly used e.g. \cite{MATHEY2018}); elastic stiffness $k=1$; and damping function to be $\mathfrak{b}(x,y) \triangleq -y$.
These results highlight an excellent agreement between the two methods across the range of elasto-plastic regimes considered.

\begin{figure}[h!]
	\centering
	\begin{subfigure}{.5\textwidth}
		\centering
%
%
\definecolor{mycolor1}{rgb}{0.00000,0.44700,0.74100}%
\definecolor{mycolor2}{rgb}{0.85000,0.32500,0.09800}%
\begin{tikzpicture}

\begin{axis}[%
width=0.8\linewidth,
height=0.64\linewidth,
scale only axis,
xmin=0,
xmax=2,
ymode = log,
ymax=1,
xtick distance = 0.5,
xlabel = $X_f$,
ylabel = $P_1^0$,
ylabel style={yshift=-10pt},
yminorticks=true,
axis background/.style={fill=white},
legend style={at={(0.05,0.05)}, anchor=south west, legend cell align=left, align=left, draw=white!15!black, font=\scriptsize}
]
\addplot [color=red, line width=1pt, dotted]
  table[row sep=crcr]{%
0	1\\
0.03125	1\\
0.0625	0.99999407\\
0.09375	0.99977706\\
0.125	0.99832272\\
0.15625	0.99382527\\
0.1875	0.98459907\\
0.21875	0.96960261\\
0.25	0.948644\\
0.28125	0.92207087\\
0.3125	0.8905175\\
0.34375	0.8548557\\
0.375	0.8160224\\
0.40625	0.77500434\\
0.4375	0.73253534\\
0.46875	0.68936559\\
0.5	0.64609653\\
0.53125	0.60329229\\
0.5625	0.5613427\\
0.59375	0.52062162\\
0.625	0.48136236\\
0.65625	0.44376352\\
0.6875	0.40796687\\
0.71875	0.37405895\\
0.75	0.34211001\\
0.78125	0.31210204\\
0.8125	0.28404453\\
0.84375	0.25793642\\
0.875	0.23366891\\
0.90625	0.21120685\\
0.9375	0.19044387\\
0.96875	0.17136841\\
1	0.15385193\\
1.03125	0.13797745\\
1.0625	0.12377662\\
1.09375	0.11107278\\
1.125	0.0997163\\
1.15625	0.08953096\\
1.1875	0.08039153\\
1.21875	0.07218786\\
1.25	0.06482563\\
1.28125	0.0582019\\
1.3125	0.05225571\\
1.34375	0.04690822\\
1.375	0.0420906\\
1.40625	0.03776935\\
1.4375	0.03389092\\
1.46875	0.03040432\\
1.5	0.02727192\\
1.53125	0.02445633\\
1.5625	0.02192181\\
1.59375	0.01965469\\
1.625	0.01761032\\
1.65625	0.01577622\\
1.6875	0.01413032\\
1.71875	0.01264496\\
1.75	0.01131608\\
1.78125	0.01011753\\
1.8125	0.0090434\\
1.84375	0.00808735\\
1.875	0.00722289\\
1.90625	0.00644766\\
1.9375	0.00575476\\
1.96875	0.00513327\\
2	0.00458093\\
};
\addlegendentry{MC $a = 0$}

\addplot[only marks, mark=o, mark options={}, mark size=2.5pt, color=red, fill=red] table[row sep=crcr]{%
x	y\\
0	1\\
0.25	0.953481771121291\\
0.5	0.654924988229376\\
0.75	0.349795191544659\\
1	0.159528982198142\\
1.25	0.0673967125260622\\
1.5	0.02802912354143\\
1.75	0.0113266217181278\\
2	0.00458357011934289\\
};
\addlegendentry{PDE $a = 0$}

\addplot [color=blue, line width=1pt, dashdotted]
  table[row sep=crcr]{%
0.03125	1\\
0.0625	0.99999405\\
0.09375	0.99977867\\
0.125	0.99832111\\
0.15625	0.99382592\\
0.1875	0.98461838\\
0.21875	0.96966167\\
0.25	0.94874179\\
0.28125	0.92214957\\
0.3125	0.89057881\\
0.34375	0.85495683\\
0.375	0.8162172\\
0.40625	0.77521884\\
0.4375	0.73275791\\
0.46875	0.68964821\\
0.5	0.64645905\\
0.53125	0.60365398\\
0.5625	0.56172857\\
0.59375	0.52100234\\
0.625	0.48173685\\
0.65625	0.44413161\\
0.6875	0.4083198\\
0.71875	0.37442771\\
0.75	0.342457\\
0.78125	0.31243532\\
0.8125	0.28439039\\
0.84375	0.2582368\\
0.875	0.23395366\\
0.90625	0.21148724\\
0.9375	0.19072426\\
0.96875	0.17161664\\
1	0.15408411\\
1.03125	0.13811246\\
1.0625	0.12367758\\
1.09375	0.11065047\\
1.125	0.09888595\\
1.15625	0.08827867\\
1.1875	0.07872855\\
1.21875	0.0701352\\
1.25	0.06239356\\
1.28125	0.05545036\\
1.3125	0.04921251\\
1.34375	0.04362817\\
1.375	0.03863215\\
1.40625	0.03415211\\
1.4375	0.03016833\\
1.46875	0.0265956\\
1.5	0.02343108\\
1.53125	0.02061866\\
1.5625	0.01811397\\
1.59375	0.01589404\\
1.625	0.01392689\\
1.65625	0.01218238\\
1.6875	0.01064019\\
1.71875	0.00929137\\
1.75	0.00809909\\
1.78125	0.00704684\\
1.8125	0.00612326\\
1.84375	0.005316\\
1.875	0.00461243\\
1.90625	0.00399004\\
1.9375	0.00344559\\
1.96875	0.00297149\\
2	0.00256145\\
};
\addlegendentry{MC $a = 0.5$}

\addplot[only marks, mark=triangle, mark options={}, mark size=3pt, color=blue, fill=mycolor2] table[row sep=crcr]{%
x	y\\
0	1\\
0.25	0.950420390110104\\
0.5	0.650690817431888\\
0.75	0.346497300622497\\
1	0.15709930517865\\
1.25	0.0637154501664273\\
1.5	0.0236282036477021\\
1.75	0.00813950678430997\\
2	0.0025598492115159\\
};
\addlegendentry{PDE $a = 0.5$}

\addplot [color=black, line width=1pt, loosely dashed]
  table[row sep=crcr]{%
0	1\\
0.03125	1\\
0.0625	0.9999937\\
0.09375	0.99977732\\
0.125	0.99831618\\
0.15625	0.9938093\\
0.1875	0.98457764\\
0.21875	0.96961165\\
0.25	0.9486312\\
0.28125	0.92205665\\
0.3125	0.89048562\\
0.34375	0.85483116\\
0.375	0.81606991\\
0.40625	0.77505651\\
0.4375	0.73260428\\
0.46875	0.6894573\\
0.5	0.64620926\\
0.53125	0.60339674\\
0.5625	0.56150323\\
0.59375	0.52077854\\
0.625	0.48151489\\
0.65625	0.44391581\\
0.6875	0.40812994\\
0.71875	0.37420221\\
0.75	0.34225384\\
0.78125	0.31222457\\
0.8125	0.28416228\\
0.84375	0.25803098\\
0.875	0.23376781\\
0.90625	0.2112736\\
0.9375	0.19052492\\
0.96875	0.17143423\\
1	0.15391202\\
1.03125	0.13786747\\
1.0625	0.12321303\\
1.09375	0.10987711\\
1.125	0.09774542\\
1.15625	0.0867747\\
1.1875	0.07684891\\
1.21875	0.06791007\\
1.25	0.05986433\\
1.28125	0.05265605\\
1.3125	0.04619935\\
1.34375	0.04045264\\
1.375	0.03532729\\
1.40625	0.03078026\\
1.4375	0.02676271\\
1.46875	0.02320042\\
1.5	0.02007509\\
1.53125	0.01733014\\
1.5625	0.01492091\\
1.59375	0.01281965\\
1.625	0.01098165\\
1.65625	0.00939071\\
1.6875	0.00800916\\
1.71875	0.00682072\\
1.75	0.00578825\\
1.78125	0.00489672\\
1.8125	0.00413537\\
1.84375	0.00348037\\
1.875	0.00292219\\
1.90625	0.00244774\\
1.9375	0.00204433\\
1.96875	0.0017044\\
2	0.00141512\\
};
\addlegendentry{MC $a = 1$}

\addplot[only marks, mark=square, mark options={}, mark size=2pt, color=black, fill=black] table[row sep=crcr]{%
x	y\\
0	1\\
0.25	0.949389080891402\\
0.5	0.648091006282931\\
0.75	0.343941345271802\\
1	0.155032567256653\\
1.25	0.0604993959484692\\
1.5	0.0203769356290443\\
1.75	0.00589030703950436\\
2	0.00141913578372257\\
};
\addlegendentry{PDE $a = 1$}
\end{axis}

\end{tikzpicture}%
		\caption{}
		\label{fig:p1}
	\end{subfigure}%
	\hfill
	\begin{subfigure}{.5\textwidth}
		\centering
%
%
\definecolor{mycolor1}{rgb}{0.00000,0.44700,0.74100}%
\definecolor{mycolor2}{rgb}{0.85000,0.32500,0.09800}%
\begin{tikzpicture}

\begin{axis}[%
width=0.8\linewidth,
height=0.64\linewidth,
scale only axis,
xmin=0,
xmax=1,
ymode=log,
ymax=0.15,
xtick distance = 0.25,
xlabel = $\Delta_f$,
ylabel = $P_2^0$,
ylabel style={yshift=-10pt},
yminorticks=true,
axis background/.style={fill=white},
legend style={at={(0.05,0.05)}, anchor=south west, legend cell align=left, align=left, draw=white!15!black, font=\scriptsize}
]
\addplot [color=red, line width=1pt, dotted]
  table[row sep=crcr]{%
0.015625	0.14574056\\
0.03125	0.13802949\\
0.046875	0.13072661\\
0.0625	0.12383617\\
0.078125	0.11731401\\
0.09375	0.11114744\\
0.109375	0.10530295\\
0.125	0.09975793\\
0.140625	0.09452108\\
0.15625	0.08956874\\
0.171875	0.08486919\\
0.1875	0.08042764\\
0.203125	0.07621227\\
0.21875	0.07222047\\
0.234375	0.0684365\\
0.25	0.06485358\\
0.265625	0.06144576\\
0.28125	0.05822709\\
0.296875	0.05516623\\
0.3125	0.05226411\\
0.328125	0.04951653\\
0.34375	0.04691111\\
0.359375	0.04444098\\
0.375	0.0421173\\
0.390625	0.03989681\\
0.40625	0.03779231\\
0.421875	0.03579922\\
0.4375	0.03390062\\
0.453125	0.03211042\\
0.46875	0.0304099\\
0.484375	0.02880407\\
0.5	0.02727582\\
0.515625	0.02583196\\
0.53125	0.02445738\\
0.546875	0.02315325\\
0.5625	0.02192584\\
0.578125	0.02076413\\
0.59375	0.01965632\\
0.609375	0.01860785\\
0.625	0.01761254\\
0.640625	0.01666704\\
0.65625	0.01577056\\
0.671875	0.01492142\\
0.6875	0.01411509\\
0.703125	0.01335577\\
0.71875	0.01263661\\
0.734375	0.01195335\\
0.75	0.01130579\\
0.765625	0.0106932\\
0.78125	0.01011293\\
0.796875	0.00956373\\
0.8125	0.00904685\\
0.828125	0.00855355\\
0.84375	0.00808577\\
0.859375	0.00764778\\
0.875	0.00723029\\
0.890625	0.00683384\\
0.90625	0.00645833\\
0.921875	0.00610091\\
0.9375	0.0057681\\
0.953125	0.00544897\\
0.96875	0.00514583\\
0.984375	0.00485878\\
1	0.00459194\\
};
\addlegendentry{MC $a = 0$}

\addplot[only marks, mark=o, mark options={}, mark size=3pt, color=red, fill=red] table[row sep=crcr]{%
x	y\\
0.125	0\\
0.25	0.0640431750746675\\
0.375	0.041251528236265\\
0.5	0.0274645553667461\\
0.625	0.0177085710651574\\
0.75	0.0115236397787987\\
0.875	0.00754523210084392\\
1	0.00418573645927055\\
};
\addlegendentry{PDE $a = 0$}

\addplot [color=blue, line width=1pt, dashdotted]
  table[row sep=crcr]{%
0.015625	0.1458883\\
0.03125	0.13809217\\
0.046875	0.13068275\\
0.0625	0.12364766\\
0.078125	0.11695967\\
0.09375	0.11061624\\
0.109375	0.10457951\\
0.125	0.09885027\\
0.140625	0.0934031\\
0.15625	0.08824318\\
0.171875	0.08333652\\
0.1875	0.0786941\\
0.203125	0.07428289\\
0.21875	0.07010207\\
0.234375	0.06612759\\
0.25	0.06236179\\
0.265625	0.05880594\\
0.28125	0.05541975\\
0.296875	0.05221502\\
0.3125	0.04918259\\
0.328125	0.04632343\\
0.34375	0.04359969\\
0.359375	0.04103404\\
0.375	0.03860472\\
0.390625	0.03630179\\
0.40625	0.03412615\\
0.421875	0.03208013\\
0.4375	0.03014314\\
0.453125	0.02830776\\
0.46875	0.0265709\\
0.484375	0.02494519\\
0.5	0.02340745\\
0.515625	0.02196414\\
0.53125	0.02059602\\
0.546875	0.01930855\\
0.5625	0.0180917\\
0.578125	0.01694949\\
0.59375	0.01587163\\
0.609375	0.01485345\\
0.625	0.01390433\\
0.640625	0.01300569\\
0.65625	0.01216002\\
0.671875	0.01136413\\
0.6875	0.01061753\\
0.703125	0.00992265\\
0.71875	0.00926866\\
0.734375	0.00865346\\
0.75	0.008077\\
0.765625	0.00753486\\
0.78125	0.00702401\\
0.796875	0.0065476\\
0.8125	0.00609974\\
0.828125	0.00568069\\
0.84375	0.00529174\\
0.859375	0.00492854\\
0.875	0.00458749\\
0.890625	0.00426345\\
0.90625	0.00396348\\
0.921875	0.00368083\\
0.9375	0.00341811\\
0.953125	0.00317298\\
0.96875	0.00294327\\
0.984375	0.00272956\\
1	0.00253202\\
};
\addlegendentry{MC $a = 0.5$}

\addplot[only marks, mark=triangle, mark options={}, mark size=3.5pt, color=blue, fill=mycolor2] table[row sep=crcr]{%
x	y\\
0.125   0.0984450101848239\\
0.25	0.0615055026447198\\
0.375	0.0376358938817175\\
0.5	0.0225155929670405\\
0.625	0.0130678207398722\\
0.75	0.00726353738896694\\
0.875	0.00456469026278403\\
1	0.0025657701987118\\
};
\addlegendentry{PDE $a = 0.5$}
\end{axis}

\end{tikzpicture}%
		\caption{}
		\label{fig:p2}
	\end{subfigure}%
\caption{Probabilities $P_1^0$ (left) and $P_2^0$ (right) as functions of the displacement threshold $X_f$ and plastic deformation threshold $\Delta_f$, respectively, taken from the numerical solution of the PDEs and from the Monte Carlo Simulations, with elasto-plastic stiffness ratios $a=0$, $a=0.5$, and $a=1$.} 
				
			\label{fig:pnsn}
\end{figure}

\subsubsection{Estimators and Variance Reduction}
In this section, for the coloured-noise-driven system with $a=0.5$, we estimate the small probabilities $P^\epsilon_1$ corresponding to $X_f=2$ with the three types of estimators: the standard MC estimator $\hat I^\epsilon_N$ in Equation \eqref{eq:standard_estimator}, the simple control variate estimator $\hat J^\epsilon_N$ in Equation \eqref{eq:simple_cv_estimator}, and the optimal control variate estimator $\hat K^\epsilon_N$ in Equation \eqref{eq:optimal_cv_estimator} using the values $P_1^0$ obtained from the KBE approach in Section \ref{sec:num_soln_PDE}. We consider two types of coloured noise $\xi^\epsilon$: the first with PSD of the form Equation \eqref{eq:PSD1} at $\lambda=1$ (PSD1), and the second with PSD of the form Equation \eqref{eq:PSD2} with $\lambda=\omega=1$ (PSD2), both with $q=1$. We will consider a range of values $\epsilon > 0$, which we recall, indicates the level of autocorrelation between the coloured and white noises. At each value $\epsilon$, we run two separate MC simulations at time step $\delta t=10^{-4}$ with $N=10^5$ and $N=10^6$ samples to compute the estimated value $P^\epsilon_1$ and the variance $\sigma^2$, respectively, of each of the three estimators: $\hat I_N^\eps$ (standard MC), $\hat J_N^\eps$ (simple control variate) and $\hat K_N^\eps$ (practical optimal control variate). For clarity, the empirical optimal coefficient $\lambda_\epsilon$ for the practical optimal control variate is estimated using the same set of MC trajectories; the samples sizes used to estimate the coefficient for the estimator and variance are $N=10^5$ and $N=10^6$, respectively.
The Euler-Maruyama time discretizations of the SVI \ref{eq:svi} and coloured noise dynamics \ref{eq:noise} with PSDs of the forms \ref{eq:PSD1} and \ref{eq:PSD2} are given in \ref{sec:time_discretization}.

For $P_1^\epsilon$, Figures \ref{fig:P1_estimates_PSD1}, \ref{fig:P1_estimates_PSD2} compares the estimated values for PSD1 and PSD2, respectively. Note that the estimated values of $P^\epsilon_1$ tend to $P_1^0$ from the KBE as $\eps$ tends to $0$ as expected from approximation diffusion theory in Section \ref{sec:approx_diffusion}. 
Figures \ref{fig:P1_variances_PSD1}, \ref{fig:P1_variances_PSD2} presents the corresponding empirical variances of the estimators $\hat I_N^\eps, \hat J_N^\eps$ and $\hat K_N^\eps$. Observe that indeed the optimal control variate estimator attains lower variance than both the standard MC and simple control variate estimators. In addition, it provides a robust estimate over the whole range of $\eps$ values. 
For PSD1 and PSD2, the numerical results shown in Figures \ref{fig:P1_bounds_PSD1} and \ref{fig:P1_bounds_PSD2} empirically reveal an estimate for the variance bound in the form $O(\eps^{1+\delta})$ with $\delta \approx 1$. This is finer than that predicted by the theory in $O(\eps)$ when $F(x,y,z) = f(x(T),y(T),z(T))$ with $f$ being smooth with bounded derivatives.

As shown in Figure \ref{fig:P2_PSD1_PSD2}, in the estimate of $P_2^\eps$, we observe a behavior similar to what we observe in Figures \ref{fig:P1_estimates_PSD1}, \ref{fig:P1_estimates_PSD2}, \ref{fig:P1_variances_PSD1}, and \ref{fig:P1_variances_PSD2}. However, comparing \ref{fig:P1_bounds_PSD1} and \ref{fig:P1_bounds_PSD2} with \ref{fig:P2_bounds_PSD1} and \ref{fig:P2_bounds_PSD2}, we empirically see that the $\sigma_{\hat K_N^\eps}^2/\eps^2$ remains bounded for \eqref{eq:P1} but becomes unbounded as $\eps \to 0^+$ for \eqref{eq:P2}. Moreover, for the limiting cases of $a=0$ and $a=1$, similar patterns can be observed in the estimation of $P_1^\epsilon$ using the three estimators (see Figures \ref{fig:a0_PSD1_PSD2} and \ref{fig:a1_PSD1_PSD2}, respectively).

We apply our method to test the dependence of $P_1^\epsilon$ on the elasto-plastic stiffness ratio $a$ for systems driven by coloured noise of the form PSD1 and PSD2 with $\epsilon=0.5$. We estimate the probability $P_1^\epsilon$ using the control variate $P_1^0$ obtained from the KBE method (with same discretizations as in \ref{sec:num_soln_PDE}) and the optimal control variate estimator $\hat K_N^\epsilon$ (with $N=10^6$ samples) for five choices of $a$: $a=0$, $a=0.25$, $a=0.5$, $a=0.75$, and $a=1$. Note that when $\epsilon=0.5$, the coloured noises have the same variance, and our optimal control variate estimators yield strictly lower variance than the standard MC estimator.

As shown in Figure \ref{fig:compare_P1_a}, in both coloured noise cases, the ULS probability $P_1^\epsilon$ decreases as the elasto-plastic stiffness ratio increases, consistent with standard theory. Of particular note, however, is that at all ratios $a$, the system driven by coloured noise of form PSD1 observes lower probability than that of the systems driven by coloured noise of form PSD2, indicating the non-trivial influence of the noise structure on the ULS probability.

\begin{figure}[h!]
    \centering
    \begin{subfigure}{0.45\textwidth}
        \centering
%
%
\definecolor{mycolor1}{rgb}{0.00000,0.44700,0.74100}%
\definecolor{mycolor2}{rgb}{0.85000,0.32500,0.09800}%
\definecolor{mycolor3}{rgb}{0.92900,0.69400,0.12500}%
\begin{tikzpicture}[scale=0.95]

\begin{axis}[%
width=0.8\linewidth,
height=0.64\linewidth,
scale only axis,
xmin=0,
xmax=1.2,
xtick distance = 0.3,
xlabel = $\epsilon$,
y label style={at={(axis description cs:0.1,.5)},anchor=south},
axis background/.style={fill=white},
legend style={legend cell align=left, align=left, draw=white!15!black, minimum height=0.5cm, font={\tiny\arraycolsep=2pt}},
]
\addplot [color=mycolor1, line width=1pt]
  table[row sep=crcr]{%
0.012	0.00258\\
0.024	0.00263\\
0.036	0.00225\\
0.048	0.00244\\
0.06	0.00246\\
0.072	0.00243\\
0.084	0.00247\\
0.096	0.0028\\
0.108	0.00252\\
0.12	0.00235\\
0.132	0.00266\\
0.144	0.0024\\
0.156	0.00232\\
0.168	0.00222\\
0.18	0.00241\\
0.192	0.00215\\
0.204	0.00226\\
0.216	0.00232\\
0.228	0.00233\\
0.24	0.00226\\
0.252	0.00232\\
0.264	0.0021\\
0.276	0.00255\\
0.288	0.00252\\
0.3	0.00239\\
0.312	0.00236\\
0.324	0.00232\\
0.336	0.00205\\
0.348	0.00208\\
0.36	0.00213\\
0.372	0.00193\\
0.384	0.00204\\
0.396	0.00245\\
0.408	0.0021\\
0.42	0.00172\\
0.432	0.00212\\
0.444	0.00178\\
0.456	0.00169\\
0.468	0.00167\\
0.48	0.00169\\
0.492	0.00127\\
0.504	0.00155\\
0.516	0.00138\\
0.528	0.00138\\
0.54	0.00149\\
0.552	0.00138\\
0.564	0.0012\\
0.576	0.00138\\
0.588	0.0014\\
0.6	0.00125\\
0.612	0.00112\\
0.624	0.00121\\
0.636	0.00099\\
0.648	0.00072\\
0.66	0.00089\\
0.672	0.0007\\
0.684	0.00066\\
0.696	0.00083\\
0.708	0.00072\\
0.72	0.00066\\
0.732	0.00081\\
0.744	0.00048\\
0.756	0.00053\\
0.768	0.00038\\
0.78	0.00045\\
0.792	0.00033\\
0.804	0.00033\\
0.816	0.00037\\
0.828	0.00032\\
0.84	0.00034\\
0.852	0.00031\\
0.864	0.00035\\
0.876	0.00038\\
0.888	0.00028\\
0.9	0.00031\\
0.912	0.00027\\
0.924	0.00022\\
0.936	0.00021\\
0.948	9e-05\\
0.96	0.00017\\
0.972	6e-05\\
0.984	3e-05\\
0.996	0.00013\\
1.008	5e-05\\
1.02	6e-05\\
1.032	8e-05\\
1.044	7e-05\\
1.056	0.0001\\
1.068	2e-05\\
1.08	4e-05\\
1.092	5e-05\\
1.104	4e-05\\
1.116	4e-05\\
1.128	3e-05\\
1.14	3e-05\\
1.152	3e-05\\
1.164	3e-05\\
1.176	4e-05\\
1.188	5e-05\\
1.2	1e-05\\
};
\addlegendentry{$\hat I^\epsilon_N$}

\addplot [color=mycolor2, line width=1pt]
  table[row sep=crcr]{%
0.012	0.0025598492115159\\
0.024	0.0025598492115159\\
0.036	0.0025498492115159\\
0.048	0.0025698492115159\\
0.06	0.0025398492115159\\
0.072	0.0025398492115159\\
0.084	0.0025698492115159\\
0.096	0.0025298492115159\\
0.108	0.0025598492115159\\
0.12	0.0025198492115159\\
0.132	0.0025598492115159\\
0.144	0.0025398492115159\\
0.156	0.0025198492115159\\
0.168	0.0024798492115159\\
0.18	0.0025098492115159\\
0.192	0.0024498492115159\\
0.204	0.0025198492115159\\
0.216	0.0025498492115159\\
0.228	0.0024798492115159\\
0.24	0.0026398492115159\\
0.252	0.0023898492115159\\
0.264	0.0023798492115159\\
0.276	0.0023898492115159\\
0.288	0.0025198492115159\\
0.3	0.0023698492115159\\
0.312	0.0023398492115159\\
0.324	0.0022498492115159\\
0.336	0.0023498492115159\\
0.348	0.0022098492115159\\
0.36	0.0022098492115159\\
0.372	0.0020398492115159\\
0.384	0.0021398492115159\\
0.396	0.0021798492115159\\
0.408	0.0021398492115159\\
0.42	0.0018898492115159\\
0.432	0.0020198492115159\\
0.444	0.0019198492115159\\
0.456	0.0018898492115159\\
0.468	0.0019298492115159\\
0.48	0.0017898492115159\\
0.492	0.0015698492115159\\
0.504	0.0018098492115159\\
0.516	0.0016098492115159\\
0.528	0.0015298492115159\\
0.54	0.0018698492115159\\
0.552	0.0014498492115159\\
0.564	0.0014798492115159\\
0.576	0.0012198492115159\\
0.588	0.0013998492115159\\
0.6	0.0012298492115159\\
0.612	0.0010998492115159\\
0.624	0.0011398492115159\\
0.636	0.0012898492115159\\
0.648	0.000849849211515902\\
0.66	0.000969849211515902\\
0.672	0.000809849211515902\\
0.684	0.000759849211515902\\
0.696	0.000559849211515902\\
0.708	0.000759849211515902\\
0.72	0.000829849211515902\\
0.732	0.000709849211515902\\
0.744	0.000619849211515902\\
0.756	0.000729849211515902\\
0.768	0.000639849211515902\\
0.78	0.000549849211515902\\
0.792	0.000629849211515902\\
0.804	0.000589849211515902\\
0.816	0.000599849211515902\\
0.828	0.000469849211515902\\
0.84	0.000719849211515902\\
0.852	0.000379849211515902\\
0.864	0.000629849211515902\\
0.876	0.000219849211515902\\
0.888	0.000279849211515902\\
0.9	0.000289849211515902\\
0.912	0.000249849211515902\\
0.924	0.000149849211515902\\
0.936	0.000509849211515902\\
0.948	0.000219849211515902\\
0.96	0.000249849211515902\\
0.972	0.000169849211515902\\
0.984	0.000129849211515902\\
0.996	-0.000140150788484098\\
1.008	8.98492115159019e-05\\
1.02	0.000229849211515902\\
1.032	-2.01507884840979e-05\\
1.044	0.000209849211515902\\
1.056	0.000299849211515902\\
1.068	0.000279849211515902\\
1.08	0.000139849211515902\\
1.092	0.000349849211515902\\
1.104	0.000299849211515902\\
1.116	0.000269849211515902\\
1.128	0.000179849211515902\\
1.14	0.000409849211515902\\
1.152	9.9849211515902e-05\\
1.164	0.000309849211515902\\
1.176	-0.000120150788484098\\
1.188	4.98492115159018e-05\\
1.2	-1.01507884840979e-05\\
};
\addlegendentry{$\hat J^\epsilon_N$}

\addplot [color=mycolor3, line width=1.0pt]
  table[row sep=crcr]{%
0.012	0.00255984921469554\\
0.024	0.00256011677901671\\
0.036	0.00254851054255262\\
0.048	0.00256984700670861\\
0.06	0.00253920506212513\\
0.072	0.00253895194532571\\
0.084	0.00256944091449619\\
0.096	0.00253463447479426\\
0.108	0.00255873946800876\\
0.12	0.00251557980601436\\
0.132	0.00256135962806607\\
0.144	0.00253637516333271\\
0.156	0.00251221456652304\\
0.168	0.00246740654717012\\
0.18	0.00250659868736302\\
0.192	0.00242726509520793\\
0.204	0.0025085284798906\\
0.216	0.00253501491127857\\
0.228	0.00246926411546943\\
0.24	0.00262757073544365\\
0.252	0.00238254929456389\\
0.264	0.00234544526556445\\
0.276	0.00240635434493255\\
0.288	0.00251986160658357\\
0.3	0.0023722728574339\\
0.312	0.00234297700397951\\
0.324	0.00226293189778653\\
0.336	0.00230468992917655\\
0.348	0.00218578930886521\\
0.36	0.00219502949781824\\
0.372	0.00201114029238551\\
0.384	0.0021142567502484\\
0.396	0.0022410192259045\\
0.408	0.00212829318838142\\
0.42	0.00183083533845247\\
0.432	0.00204924110377997\\
0.444	0.00186838130035505\\
0.456	0.00181783139618196\\
0.468	0.00183602056474703\\
0.48	0.00175208507135993\\
0.492	0.00142386571179835\\
0.504	0.00170586066452382\\
0.516	0.00149635560008937\\
0.528	0.00145520925788777\\
0.54	0.00170420540688695\\
0.552	0.0014139308188748\\
0.564	0.00132883141771205\\
0.576	0.00130996520387487\\
0.588	0.00139992639548682\\
0.6	0.00124180313958576\\
0.612	0.00111258357168859\\
0.624	0.00118173690156729\\
0.636	0.00109740970236277\\
0.648	0.000753652661425354\\
0.66	0.000916396084048391\\
0.672	0.000725542313750756\\
0.684	0.00068312669036328\\
0.696	0.000770873443388492\\
0.708	0.000729007513955454\\
0.72	0.000698355077663279\\
0.732	0.000785165712819312\\
0.744	0.000504262742890134\\
0.756	0.000568090169273289\\
0.768	0.000418400944263482\\
0.78	0.000467044354245436\\
0.792	0.000364473515434234\\
0.804	0.000360487513887004\\
0.816	0.000398588761157378\\
0.828	0.000336780418649884\\
0.84	0.000381777422788606\\
0.852	0.000316165102047124\\
0.864	0.000381886287635018\\
0.876	0.00036057815412041\\
0.888	0.000279984688568094\\
0.9	0.000308048616699291\\
0.912	0.000268126114431036\\
0.924	0.000214934191891883\\
0.936	0.000233871827911514\\
0.948	9.26665693077378e-05\\
0.96	0.000174827976008378\\
0.972	6.17912452904949e-05\\
0.984	3.12176683640446e-05\\
0.996	0.000121419690769015\\
1.008	5.07906584591625e-05\\
1.02	6.35515852511151e-05\\
1.032	7.69879575989389e-05\\
1.044	7.4045192355008e-05\\
1.056	0.000106770428252384\\
1.068	2.22594918756849e-05\\
1.08	4.16235578187261e-05\\
1.092	5.66335676122807e-05\\
1.104	4.33866334077369e-05\\
1.116	4.1968302151353e-05\\
1.128	3.1865325460085e-05\\
1.14	3.34808174759789e-05\\
1.152	3.05603365071337e-05\\
1.164	3.24519283440178e-05\\
1.176	3.76448636141809e-05\\
1.188	4.999705491242e-05\\
1.2	9.92189618099048e-06\\
};
\addlegendentry{$\hat K^\epsilon_N$}

\addplot [color=black, dashed, line width=1.0pt]
  table[row sep=crcr]{%
0	0.0025598492115159\\
0.133333333333333	0.0025598492115159\\
0.266666666666667	0.0025598492115159\\
0.4	0.0025598492115159\\
0.533333333333333	0.0025598492115159\\
0.666666666666667	0.0025598492115159\\
0.8	0.0025598492115159\\
0.933333333333333	0.0025598492115159\\
1.06666666666667	0.0025598492115159\\
1.2	0.0025598492115159\\
};
\addlegendentry{$P^0_1$}

\end{axis}

\node[] at (2.85,0.65) {(a)};

\end{tikzpicture}%
        \captionlistentry{}
        \label{fig:P1_estimates_PSD1}
    \end{subfigure}%
    \setcounter{subfigure}{0}
    \renewcommand{\thesubfigure}{\alph{subfigure}'}
    \begin{subfigure}{.45\textwidth}
        \centering
%
%
\definecolor{mycolor1}{rgb}{0.00000,0.44700,0.74100}%
\definecolor{mycolor2}{rgb}{0.85000,0.32500,0.09800}%
\definecolor{mycolor3}{rgb}{0.92900,0.69400,0.12500}%
\begin{tikzpicture}[scale=0.95]

\begin{axis}[%
width=0.8\linewidth,
height=0.64\linewidth,
scale only axis,
xmin=0,
xmax=1.8,
xtick distance = 0.45,
xlabel = $\epsilon$,
y label style={at={(axis description cs:0.1,.5)},anchor=south},
axis background/.style={fill=white},
legend style={legend cell align=left, align=left, draw=white!15!black, minimum height=0.5cm, font={\tiny\arraycolsep=2pt}},
]
\addplot [color=mycolor1, line width=1pt]
  table[row sep=crcr]{%
0.018	0.0026\\
0.036	0.0024\\
0.054	0.00214\\
0.072	0.00288\\
0.09	0.00245\\
0.108	0.00242\\
0.126	0.00233\\
0.144	0.00258\\
0.162	0.00253\\
0.18	0.00236\\
0.198	0.00238\\
0.216	0.00291\\
0.234	0.00257\\
0.252	0.00254\\
0.27	0.00274\\
0.288	0.00268\\
0.306	0.00294\\
0.324	0.00277\\
0.342	0.0026\\
0.36	0.0026\\
0.378	0.0027\\
0.396	0.00248\\
0.414	0.00307\\
0.432	0.00275\\
0.45	0.00272\\
0.468	0.00315\\
0.486	0.00292\\
0.504	0.00291\\
0.522	0.00335\\
0.54	0.00324\\
0.558	0.00295\\
0.576	0.00308\\
0.594	0.00322\\
0.612	0.00337\\
0.63	0.00337\\
0.648	0.00342\\
0.666	0.00408\\
0.684	0.00367\\
0.702	0.00378\\
0.72	0.00383\\
0.738	0.00392\\
0.756	0.00409\\
0.774	0.00381\\
0.792	0.00367\\
0.81	0.00343\\
0.828	0.00365\\
0.846	0.0034\\
0.864	0.00364\\
0.882	0.00342\\
0.9	0.00316\\
0.918	0.0034\\
0.936	0.00303\\
0.954	0.00283\\
0.972	0.00341\\
0.99	0.00247\\
1.008	0.00269\\
1.026	0.00232\\
1.044	0.00233\\
1.062	0.00196\\
1.08	0.00174\\
1.098	0.00199\\
1.116	0.00192\\
1.134	0.00169\\
1.152	0.00157\\
1.17	0.00152\\
1.188	0.00132\\
1.206	0.00128\\
1.224	0.00131\\
1.242	0.00117\\
1.26	0.00118\\
1.278	0.00088\\
1.296	0.00085\\
1.314	0.00089\\
1.332	0.0007\\
1.35	0.00056\\
1.368	0.0007\\
1.386	0.00054\\
1.404	0.0004\\
1.422	0.0006\\
1.44	0.00025\\
1.458	0.00039\\
1.476	0.00033\\
1.494	0.00022\\
1.512	0.00022\\
1.53	0.00024\\
1.548	0.0002\\
1.566	0.00017\\
1.584	0.00014\\
1.602	0.00014\\
1.62	0.00017\\
1.638	8e-05\\
1.656	0.0001\\
1.674	0.00017\\
1.692	0.0001\\
1.71	0.00013\\
1.728	5e-05\\
1.746	5e-05\\
1.764	6e-05\\
1.782	5e-05\\
1.8	5e-05\\
};
\addlegendentry{$\hat I^\epsilon_N$}

\addplot [color=mycolor2, line width=1pt]
  table[row sep=crcr]{%
0.018	0.0025598492115159\\
0.036	0.0025398492115159\\
0.054	0.0025698492115159\\
0.072	0.0025598492115159\\
0.09	0.0025798492115159\\
0.108	0.0025498492115159\\
0.126	0.0025698492115159\\
0.144	0.0025098492115159\\
0.162	0.0025398492115159\\
0.18	0.0025898492115159\\
0.198	0.0025598492115159\\
0.216	0.0025998492115159\\
0.234	0.0025498492115159\\
0.252	0.0025798492115159\\
0.27	0.0027298492115159\\
0.288	0.0025998492115159\\
0.306	0.0026998492115159\\
0.324	0.0026998492115159\\
0.342	0.0027398492115159\\
0.36	0.0026398492115159\\
0.378	0.0027598492115159\\
0.396	0.0027998492115159\\
0.414	0.0028798492115159\\
0.432	0.0028598492115159\\
0.45	0.0026998492115159\\
0.468	0.0031098492115159\\
0.486	0.0030598492115159\\
0.504	0.0033398492115159\\
0.522	0.0030298492115159\\
0.54	0.0033698492115159\\
0.558	0.0030798492115159\\
0.576	0.0033198492115159\\
0.594	0.0031498492115159\\
0.612	0.0033798492115159\\
0.63	0.0035998492115159\\
0.648	0.0035998492115159\\
0.666	0.0037698492115159\\
0.684	0.0036498492115159\\
0.702	0.0038198492115159\\
0.72	0.0038198492115159\\
0.738	0.0038398492115159\\
0.756	0.0038498492115159\\
0.774	0.0037398492115159\\
0.792	0.0038098492115159\\
0.81	0.0034698492115159\\
0.828	0.0037098492115159\\
0.846	0.0037198492115159\\
0.864	0.0034498492115159\\
0.882	0.0035298492115159\\
0.9	0.0031398492115159\\
0.918	0.0033598492115159\\
0.936	0.0031698492115159\\
0.954	0.0032598492115159\\
0.972	0.0030898492115159\\
0.99	0.0025998492115159\\
1.008	0.0028198492115159\\
1.026	0.0025598492115159\\
1.044	0.0022598492115159\\
1.062	0.0019698492115159\\
1.08	0.0019698492115159\\
1.098	0.0021698492115159\\
1.116	0.0016098492115159\\
1.134	0.0016698492115159\\
1.152	0.0016098492115159\\
1.17	0.0015098492115159\\
1.188	0.0012398492115159\\
1.206	0.0010398492115159\\
1.224	0.0012398492115159\\
1.242	0.0013098492115159\\
1.26	0.0012198492115159\\
1.278	0.000939849211515902\\
1.296	0.0011698492115159\\
1.314	0.000699849211515902\\
1.332	0.000809849211515902\\
1.35	0.000539849211515902\\
1.368	0.000659849211515902\\
1.386	0.000679849211515902\\
1.404	0.000829849211515902\\
1.422	0.000279849211515902\\
1.44	0.000379849211515902\\
1.458	0.000519849211515902\\
1.476	0.000569849211515902\\
1.494	0.000149849211515902\\
1.512	0.000229849211515902\\
1.53	0.000469849211515902\\
1.548	0.000379849211515902\\
1.566	-0.000140150788484098\\
1.584	0.000119849211515902\\
1.602	0.000179849211515902\\
1.62	0.000159849211515902\\
1.638	-1.50788484098302e-07\\
1.656	-0.000140150788484098\\
1.674	9.9849211515902e-05\\
1.692	0.000239849211515902\\
1.71	0.000169849211515902\\
1.728	0.000109849211515902\\
1.746	0.000369849211515902\\
1.764	-0.000130150788484098\\
1.782	0.000159849211515902\\
1.8	2.98492115159018e-05\\
};
\addlegendentry{$\hat J^\epsilon_N$}

\addplot [color=mycolor3, line width=1pt]
  table[row sep=crcr]{%
0.018	0.0025598492364756\\
0.036	0.00253811302281909\\
0.054	0.0025677853366138\\
0.072	0.00256209024918187\\
0.09	0.00257931004823803\\
0.108	0.00254824263952728\\
0.126	0.00256569600193256\\
0.144	0.00251198531347903\\
0.162	0.00253953962381751\\
0.18	0.0025819270068858\\
0.198	0.0025522721472536\\
0.216	0.00261396068177528\\
0.234	0.00255086698869173\\
0.252	0.00257762895067853\\
0.27	0.00273040532799435\\
0.288	0.00260502749567095\\
0.306	0.00272569004278064\\
0.324	0.00270494047206304\\
0.342	0.00272534068245253\\
0.36	0.00263509005464002\\
0.378	0.00275359721689685\\
0.396	0.0027625506395653\\
0.414	0.00289863331935606\\
0.432	0.00284318549893571\\
0.45	0.00270337580626717\\
0.468	0.0031157548625384\\
0.486	0.00303602723857891\\
0.504	0.00327678757637538\\
0.522	0.0030968998768353\\
0.54	0.00334403134744603\\
0.558	0.00305299778153346\\
0.576	0.00327199483399904\\
0.594	0.00316700673824283\\
0.612	0.00337747900495918\\
0.63	0.00354324514643532\\
0.648	0.0035481518980582\\
0.666	0.00385043265569072\\
0.684	0.00365613570371395\\
0.702	0.00380775047904534\\
0.72	0.00382310915281416\\
0.738	0.00386704425927944\\
0.756	0.00393961230456608\\
0.774	0.00376694336427582\\
0.792	0.00374942704886581\\
0.81	0.0034526914634025\\
0.828	0.00368338051228094\\
0.846	0.00356920369483704\\
0.864	0.00354639776529475\\
0.882	0.00346996326201453\\
0.9	0.00315168511544539\\
0.918	0.00338202192601765\\
0.936	0.00308460744945332\\
0.954	0.00301077933211013\\
0.972	0.00328735140144645\\
0.99	0.00251681629397318\\
1.008	0.00273625200093485\\
1.026	0.00239094083726429\\
1.044	0.00230769926710861\\
1.062	0.00196238147847587\\
1.08	0.00179200334808382\\
1.098	0.00203281655204743\\
1.116	0.00185557285691731\\
1.134	0.00168596223881481\\
1.152	0.00157611999514668\\
1.17	0.00151823351398476\\
1.188	0.0013085387132153\\
1.206	0.00125020573418548\\
1.224	0.00130126675525044\\
1.242	0.00118258028372186\\
1.26	0.00118518438844948\\
1.278	0.000885467120120498\\
1.296	0.000884072452710727\\
1.314	0.00087352917809775\\
1.332	0.000708912179068722\\
1.35	0.000559228255147957\\
1.368	0.000697086436010542\\
1.386	0.00054513787729994\\
1.404	0.000422071944461305\\
1.422	0.000585700164997066\\
1.44	0.000254252713443895\\
1.458	0.000397448431345713\\
1.476	0.000339246567790218\\
1.494	0.000218410858488451\\
1.512	0.000220113996937979\\
1.53	0.000242910984687521\\
1.548	0.000202236329095634\\
1.566	0.000166801659289968\\
1.584	0.000139689605014843\\
1.602	0.000141104125352628\\
1.62	0.000169882933082391\\
1.638	7.9397620303443e-05\\
1.656	9.5723722644746e-05\\
1.674	0.000168942213490558\\
1.692	9.99859812668098e-05\\
1.71	0.000130311868899066\\
1.728	5.02369965372692e-05\\
1.746	5.14150108040609e-05\\
1.764	5.86247303125366e-05\\
1.782	4.9994494077031e-05\\
1.8	4.98443984492452e-05\\
};
\addlegendentry{$\hat K^\epsilon_N$}

\addplot [color=black, dashed, line width=1.0pt]
  table[row sep=crcr]{%
0	0.0025598492115159\\
0.2	0.0025598492115159\\
0.4	0.0025598492115159\\
0.6	0.0025598492115159\\
0.8	0.0025598492115159\\
1	0.0025598492115159\\
1.2	0.0025598492115159\\
1.4	0.0025598492115159\\
1.6	0.0025598492115159\\
1.8	0.0025598492115159\\
};
\addlegendentry{$P^0_1$}

\end{axis}

\node[] at (2.85,0.65) {(a')};

\end{tikzpicture}%
        \captionlistentry{}
        \label{fig:P1_estimates_PSD2}
    \end{subfigure}%

    \centering
    \renewcommand{\thesubfigure}{\alph{subfigure}}
    \begin{subfigure}{.45\textwidth}
        \centering
%
%
\definecolor{mycolor1}{rgb}{0.00000,0.44700,0.74100}%
\definecolor{mycolor2}{rgb}{0.85000,0.32500,0.09800}%
\definecolor{mycolor3}{rgb}{0.92900,0.69400,0.12500}%
\begin{tikzpicture}[scale=0.95]

\begin{axis}[%
width=0.8\linewidth,
height=0.64\linewidth,
scale only axis,
xmin=0,
xmax=1.2,
xtick distance = 0.3,
xlabel = $\epsilon$,
y label style={at={(axis description cs:0.1,.5)},anchor=south},
axis background/.style={fill=white},
legend style={legend cell align=left, align=left, draw=white!15!black, minimum height=0.6cm, font={\tiny\arraycolsep=2pt}, at={(0.98,0.47)},anchor=east}
]
\addplot [color=mycolor1, line width=1pt]
  table[row sep=crcr]{%
0.012	0.002528573775\\
0.024	0.002515639516\\
0.036	0.002582297079\\
0.048	0.002540512791\\
0.06	0.0024837999\\
0.072	0.002608161775\\
0.084	0.002567374524\\
0.096	0.002538522975\\
0.108	0.0025534464\\
0.12	0.002542502599\\
0.132	0.002506684831\\
0.144	0.0024738496\\
0.156	0.002511659676\\
0.168	0.002520614271\\
0.18	0.002551456636\\
0.192	0.002494744999\\
0.204	0.002429070775\\
0.216	0.002496734991\\
0.228	0.002502704919\\
0.24	0.002452953319\\
0.252	0.002440017084\\
0.264	0.0024738496\\
0.276	0.002427080511\\
0.288	0.002456933631\\
0.3	0.002383292679\\
0.312	0.002322580416\\
0.324	0.002292719196\\
0.336	0.002311631511\\
0.348	0.002303668519\\
0.36	0.002199142384\\
0.372	0.002263851639\\
0.384	0.002169273724\\
0.396	0.002123471616\\
0.408	0.0020756736\\
0.42	0.001971099375\\
0.432	0.001974087516\\
0.444	0.0018764656\\
0.456	0.001878458076\\
0.468	0.001833625431\\
0.48	0.00179676\\
0.492	0.001723020924\\
0.504	0.001668207759\\
0.516	0.001580494111\\
0.528	0.001580494111\\
0.54	0.001491767964\\
0.552	0.001429949376\\
0.564	0.001463850844\\
0.576	0.001393053975\\
0.588	0.001357153119\\
0.6	0.001247439999\\
0.612	0.001154663664\\
0.624	0.001091805351\\
0.636	0.001074842224\\
0.648	0.001014967744\\
0.66	0.000961074556\\
0.672	0.000932129511\\
0.684	0.000831307776\\
0.696	0.000791372736\\
0.708	0.000745443484\\
0.72	0.000676541671\\
0.732	0.0006595644\\
0.744	0.000596643591\\
0.756	0.000588653079\\
0.768	0.000572671671\\
0.78	0.000525723324\\
0.792	0.000486762831\\
0.804	0.000420822759\\
0.816	0.000406834351\\
0.828	0.000382853311\\
0.84	0.000320896959\\
0.852	0.000336886431\\
0.864	0.000307905136\\
0.876	0.000305906364\\
0.888	0.000272925471\\
0.9	0.000246938991\\
0.912	0.000197960796\\
0.924	0.0002199516\\
0.936	0.000180967239\\
0.948	0.000170970759\\
0.96	0.000163973104\\
0.972	0.000143979264\\
0.984	0.000107988336\\
0.996	0.000130982839\\
1.008	9.3991164e-05\\
1.02	8.5992604e-05\\
1.032	9.6990591e-05\\
1.044	7.0994959e-05\\
1.056	6.8995239e-05\\
1.068	6.7995376e-05\\
1.08	5.5996864e-05\\
1.092	6.1996156e-05\\
1.104	4.8997599e-05\\
1.116	4.3998064e-05\\
1.128	4.1998236e-05\\
1.14	2.8999159e-05\\
1.152	2.1999516e-05\\
1.164	2.3999424e-05\\
1.176	2.5999324e-05\\
1.188	2.2999471e-05\\
1.2	2.6999271e-05\\
};
\addlegendentry{$\displaystyle \sigma^2_{\displaystyle \hat I^\epsilon_N}$}

\addplot [color=mycolor2, line width=1pt]
  table[row sep=crcr]{%
0.012	2e-06\\
0.024	9.99999e-07\\
0.036	8.999999e-06\\
0.048	1.3999984e-05\\
0.06	2.3999996e-05\\
0.072	2.2999999e-05\\
0.084	4.6999999e-05\\
0.096	6.2999999e-05\\
0.108	7.6999775e-05\\
0.12	8.3999964e-05\\
0.132	0.000103999676\\
0.144	0.000107999676\\
0.156	0.000150999471\\
0.168	0.000161999324\\
0.18	0.000212999639\\
0.192	0.0001939975\\
0.204	0.000255999424\\
0.216	0.000270995775\\
0.228	0.000290996751\\
0.24	0.000321992944\\
0.252	0.000346988551\\
0.264	0.000400991351\\
0.276	0.000423970416\\
0.288	0.000479970416\\
0.3	0.000509976896\\
0.312	0.000546957151\\
0.324	0.000562949375\\
0.336	0.000658926559\\
0.348	0.000665941436\\
0.36	0.000705919344\\
0.372	0.000748879591\\
0.384	0.000844840799\\
0.396	0.000829825276\\
0.408	0.000900754975\\
0.42	0.001033677376\\
0.432	0.000970694191\\
0.444	0.001029618076\\
0.456	0.001108565719\\
0.468	0.001145487344\\
0.48	0.001230396271\\
0.492	0.001190319375\\
0.504	0.001217379056\\
0.516	0.001340029775\\
0.528	0.001410033711\\
0.54	0.001439899599\\
0.552	0.001487743359\\
0.564	0.001521774551\\
0.576	0.0016025359\\
0.588	0.001610371824\\
0.6	0.001674330736\\
0.612	0.001737098359\\
0.624	0.0017359551\\
0.636	0.001871713856\\
0.648	0.001882643775\\
0.66	0.001837671324\\
0.672	0.001962167511\\
0.684	0.001985968919\\
0.696	0.002014923484\\
0.708	0.002059654759\\
0.72	0.002104518044\\
0.732	0.002071499359\\
0.744	0.002115386199\\
0.756	0.002128189696\\
0.768	0.0021840796\\
0.78	0.002216793399\\
0.792	0.002221739904\\
0.804	0.002193846556\\
0.816	0.002298330079\\
0.828	0.002316234511\\
0.84	0.002335062716\\
0.852	0.002329089344\\
0.864	0.002353869775\\
0.876	0.002413714599\\
0.888	0.002452420956\\
0.9	0.002382737564\\
0.912	0.002415538431\\
0.924	0.002371696191\\
0.936	0.002469206351\\
0.948	0.0024323356\\
0.96	0.002390599024\\
0.972	0.002537889216\\
0.984	0.002506923775\\
0.996	0.002451167775\\
1.008	0.002524844639\\
1.02	0.002478963151\\
1.032	0.002458075644\\
1.044	0.002455109671\\
1.056	0.002481889216\\
1.068	0.002520824775\\
1.08	0.002528644559\\
1.092	0.002502804879\\
1.104	0.0024321919\\
1.116	0.002529629424\\
1.128	0.002537629424\\
1.14	0.002511679804\\
1.152	0.002536594039\\
1.164	0.002554502599\\
1.176	0.002584348759\\
1.188	0.002613177456\\
1.2	0.002513679804\\
};
\addlegendentry{$\displaystyle \sigma^2_{\displaystyle \hat J^\epsilon_N}$}

\addplot [color=mycolor3, line width=1pt]
  table[row sep=crcr]{%
0.012	1.99960452029167e-06\\
0.024	9.99603647184698e-07\\
0.036	8.99379255515827e-06\\
0.048	1.39682226820463e-05\\
0.06	2.39511971486076e-05\\
0.072	2.29448328148492e-05\\
0.084	4.67938259342724e-05\\
0.096	6.25968370183905e-05\\
0.108	7.61772889462408e-05\\
0.12	8.33998573717352e-05\\
0.132	0.000102527976415245\\
0.144	0.000106409098837584\\
0.156	0.000148017145041393\\
0.168	0.000158534337921078\\
0.18	0.000207768991215033\\
0.192	0.000188160228065199\\
0.204	0.000248015791062828\\
0.216	0.00025999870948034\\
0.228	0.000279187366976179\\
0.24	0.000305780926766239\\
0.252	0.000326801857698129\\
0.264	0.000377264990796455\\
0.276	0.000389894255280979\\
0.288	0.000439643773534719\\
0.3	0.00046685060071372\\
0.312	0.000490904359885753\\
0.324	0.000501443027480216\\
0.336	0.000575413193702498\\
0.348	0.000585156822669685\\
0.36	0.000607472650920559\\
0.372	0.000634164386002517\\
0.384	0.000694587142328567\\
0.396	0.000677021084904485\\
0.408	0.000711753247719685\\
0.42	0.000781644746794811\\
0.432	0.000741543327878522\\
0.444	0.000758149611455022\\
0.456	0.000801410481768157\\
0.468	0.000806446802004751\\
0.48	0.000839789626186447\\
0.492	0.000792666150948876\\
0.504	0.000808840606980275\\
0.516	0.000814266910438468\\
0.528	0.0008525717790924\\
0.54	0.000831435335842776\\
0.552	0.000821897902232767\\
0.564	0.000850890942311399\\
0.576	0.000844019055089902\\
0.588	0.000820648741283952\\
0.6	0.000809274336855799\\
0.612	0.000780197892410735\\
0.624	0.000743502977877734\\
0.636	0.000766452084442789\\
0.648	0.00073871292662486\\
0.66	0.000701515080518234\\
0.672	0.00069320025034148\\
0.684	0.000637167639135341\\
0.696	0.000620890622648595\\
0.708	0.000592339285368563\\
0.72	0.000555421814004908\\
0.732	0.000536969419462387\\
0.744	0.000501654222153488\\
0.756	0.000490998569644887\\
0.768	0.000486831302810395\\
0.78	0.000450524770302804\\
0.792	0.000422377193290252\\
0.804	0.000373705662304936\\
0.816	0.000363032329198953\\
0.828	0.000344571220516459\\
0.84	0.000294003195142944\\
0.852	0.000306724673035815\\
0.864	0.000281522464664193\\
0.876	0.000282765680424776\\
0.888	0.000253785166193844\\
0.9	0.00023125595144898\\
0.912	0.000188398231516896\\
0.924	0.000206725143589599\\
0.936	0.000172649150119521\\
0.948	0.000163094805614312\\
0.96	0.000157413491764323\\
0.972	0.000139539985004448\\
0.984	0.000105257098378878\\
0.996	0.000126247004849996\\
1.008	9.21504413476352e-05\\
1.02	8.39612663394677e-05\\
1.032	9.43431169814311e-05\\
1.044	6.98323764091792e-05\\
1.056	6.7535551736214e-05\\
1.068	6.71343013103365e-05\\
1.08	5.50682473421629e-05\\
1.092	6.09399232644357e-05\\
1.104	4.85016335290565e-05\\
1.116	4.34376586113638e-05\\
1.128	4.1600835050612e-05\\
1.14	2.87133267015598e-05\\
1.152	2.1899689720626e-05\\
1.164	2.38740408601062e-05\\
1.176	2.58464130422516e-05\\
1.188	2.2862982060196e-05\\
1.2	2.67733085685722e-05\\
};
\addlegendentry{$\displaystyle \sigma^2_{\displaystyle \hat K^\epsilon_N}$}

\end{axis}

\node[] at (2.85,0.65) {(b)};

\end{tikzpicture}%
        \captionlistentry{}
        \label{fig:P1_variances_PSD1}
    \end{subfigure}%
    \setcounter{subfigure}{1}
    \renewcommand{\thesubfigure}{\alph{subfigure}'}
    \begin{subfigure}{.45\textwidth}
        \centering
%
%
\definecolor{mycolor1}{rgb}{0.00000,0.44700,0.74100}%
\definecolor{mycolor2}{rgb}{0.85000,0.32500,0.09800}%
\definecolor{mycolor3}{rgb}{0.92900,0.69400,0.12500}%
\begin{tikzpicture}[scale=0.95]

\begin{axis}[%
width=0.8\linewidth,
height=0.64\linewidth,
scale only axis,
xmin=0,
xmax=1.8,
xtick distance = 0.45,
xlabel = $\epsilon$,
y label style={at={(axis description cs:0.1,.5)},anchor=south},
axis background/.style={fill=white},
legend style={legend cell align=left, align=left, draw=white!15!black, minimum height=0.1cm, font={\tiny}, at={(0.98,0.47)},anchor=east}
]
\addplot [color=mycolor1, line width=1pt]
  table[row sep=crcr]{%
0.018	0.0025633951\\
0.036	0.002557425904\\
0.054	0.002542502599\\
0.072	0.0025235991\\
0.09	0.002600203551\\
0.108	0.0025136496\\
0.126	0.002546482191\\
0.144	0.002506684831\\
0.162	0.002532553479\\
0.18	0.0025733436\\
0.198	0.002537528064\\
0.216	0.002625072576\\
0.234	0.002559415644\\
0.252	0.002637009264\\
0.27	0.002639993391\\
0.288	0.002625072576\\
0.306	0.002686742364\\
0.324	0.00269271\\
0.342	0.002725530711\\
0.36	0.002719563471\\
0.378	0.002840883199\\
0.396	0.002721552559\\
0.414	0.002723541639\\
0.432	0.002924397511\\
0.45	0.0029015319\\
0.468	0.002913461916\\
0.486	0.003078464256\\
0.504	0.002979071856\\
0.522	0.003170881239\\
0.54	0.003170881239\\
0.558	0.003237450496\\
0.576	0.003241424496\\
0.594	0.003401351431\\
0.612	0.0034182351\\
0.63	0.003498672879\\
0.648	0.003624764956\\
0.666	0.003642633664\\
0.684	0.003638662896\\
0.702	0.003722042304\\
0.72	0.003789529584\\
0.738	0.003771666204\\
0.756	0.003737922496\\
0.774	0.003848077231\\
0.792	0.003665464959\\
0.81	0.003662487024\\
0.828	0.003651567775\\
0.846	0.003553283644\\
0.864	0.003437104399\\
0.882	0.003431145751\\
0.9	0.003372548544\\
0.918	0.003233476464\\
0.936	0.003171874876\\
0.954	0.003015849375\\
0.972	0.002956208775\\
0.99	0.002815030671\\
1.008	0.002618109375\\
1.026	0.002486784951\\
1.044	0.002392249596\\
1.062	0.002207107056\\
1.08	0.002080652775\\
1.098	0.002013927676\\
1.116	0.001875469359\\
1.134	0.0017867959\\
1.152	0.001665217776\\
1.17	0.001568531959\\
1.188	0.001446900399\\
1.206	0.0013182576\\
1.224	0.001294320384\\
1.242	0.001111761231\\
1.26	0.001032930844\\
1.278	0.000946103191\\
1.296	0.000854268975\\
1.314	0.000776396271\\
1.332	0.0007494375\\
1.35	0.000738453879\\
1.368	0.000600638799\\
1.386	0.000584657775\\
1.404	0.0005397084\\
1.422	0.000493755964\\
1.44	0.000415826944\\
1.458	0.000384851775\\
1.476	0.000306905751\\
1.494	0.000307905136\\
1.512	0.000298910599\\
1.53	0.000250936999\\
1.548	0.000212954631\\
1.566	0.000218952039\\
1.584	0.000182966511\\
1.602	0.000162973431\\
1.62	0.000141979836\\
1.638	0.000127983616\\
1.656	0.000120985359\\
1.674	0.0001099879\\
1.692	9.0991719e-05\\
1.71	6.3995904e-05\\
1.728	5.4996975e-05\\
1.746	5.0997399e-05\\
1.764	5.3997084e-05\\
1.782	4.7997696e-05\\
1.8	4.99975e-05\\
};
\addlegendentry{$\displaystyle \sigma^2_{\displaystyle \hat I^\epsilon_N}$}

\addplot [color=mycolor2, line width=1pt]
  table[row sep=crcr]{%
0.018	3.999984e-06\\
0.036	7.999996e-06\\
0.054	1.7999996e-05\\
0.072	3.2999999e-05\\
0.09	4.2999951e-05\\
0.108	6.2999975e-05\\
0.126	9.6999711e-05\\
0.144	0.000131999984\\
0.162	0.000133999936\\
0.18	0.000180999375\\
0.198	0.000207998556\\
0.216	0.000252998319\\
0.234	0.000265998844\\
0.252	0.000338992431\\
0.27	0.0003519984\\
0.288	0.000427991164\\
0.306	0.000487992604\\
0.324	0.000515992256\\
0.342	0.000578983871\\
0.36	0.000653975664\\
0.378	0.000786964279\\
0.396	0.000786960399\\
0.414	0.000838954631\\
0.432	0.001020923271\\
0.45	0.001052882351\\
0.468	0.001149878896\\
0.486	0.001327723324\\
0.504	0.001346805519\\
0.522	0.0015555775\\
0.54	0.001606662439\\
0.558	0.001704477271\\
0.576	0.001799487344\\
0.594	0.001975182784\\
0.612	0.002078220311\\
0.63	0.002271086064\\
0.648	0.002412718576\\
0.666	0.002525865775\\
0.684	0.0026027456\\
0.702	0.002789609959\\
0.72	0.002931567191\\
0.738	0.002960442496\\
0.756	0.003102691264\\
0.774	0.003271434999\\
0.792	0.003259848671\\
0.81	0.003345778975\\
0.828	0.003415989975\\
0.846	0.003360926704\\
0.864	0.003446133239\\
0.882	0.003526380631\\
0.9	0.003604332511\\
0.918	0.0035915511\\
0.936	0.0036216156\\
0.954	0.003567771516\\
0.972	0.003655811644\\
0.99	0.003598950271\\
1.008	0.00355399\\
1.026	0.003408998151\\
1.044	0.003526987679\\
1.062	0.003460887775\\
1.08	0.0034557791\\
1.098	0.003343761856\\
1.116	0.003413493056\\
1.134	0.003261449436\\
1.152	0.003308209679\\
1.17	0.003364926704\\
1.188	0.003223814079\\
1.206	0.003180341056\\
1.224	0.003246268144\\
1.242	0.003092770951\\
1.26	0.003084637631\\
1.278	0.003108065631\\
1.296	0.002936194375\\
1.314	0.002885968919\\
1.332	0.003030367164\\
1.35	0.002958658416\\
1.368	0.002883107271\\
1.386	0.002853091471\\
1.404	0.002790971951\\
1.422	0.002796850631\\
1.44	0.002839230144\\
1.458	0.0027294204\\
1.476	0.002704031559\\
1.494	0.002692155599\\
1.512	0.002684946496\\
1.53	0.002696691584\\
1.548	0.002615742151\\
1.566	0.002712373616\\
1.584	0.002641482199\\
1.602	0.002646268764\\
1.62	0.002674923775\\
1.638	0.0026141919\\
1.656	0.002682814831\\
1.674	0.002663739996\\
1.692	0.002664674775\\
1.71	0.002608714951\\
1.728	0.002674213975\\
1.746	0.002558854559\\
1.764	0.002541928704\\
1.782	0.002681198336\\
1.8	0.002592664711\\
};
\addlegendentry{$\displaystyle \sigma^2_{\displaystyle \hat J^\epsilon_N}$}

\addplot [color=mycolor3, line width=1pt]
  table[row sep=crcr]{%
0.018	3.99378399379956e-06\\
0.036	7.99646205522343e-06\\
0.054	1.7974772143561e-05\\
0.072	3.28855592337408e-05\\
0.09	4.28747579871369e-05\\
0.108	6.25413350894912e-05\\
0.126	9.6365827913213e-05\\
0.144	0.000130362841454242\\
0.162	0.000132016829659814\\
0.18	0.000178608150517331\\
0.198	0.000205101752388071\\
0.216	0.000248641785271221\\
0.234	0.000260663066898241\\
0.252	0.000332745676211476\\
0.27	0.000342626615925411\\
0.288	0.000416943131989003\\
0.306	0.000472426234640101\\
0.324	0.000498375632191956\\
0.342	0.000559275679389707\\
0.36	0.000629720032951514\\
0.378	0.000753151822381056\\
0.396	0.000752591907649127\\
0.414	0.000799815159329073\\
0.432	0.000968475267696207\\
0.45	0.00100341726865767\\
0.468	0.0010869663002857\\
0.486	0.0012643746073819\\
0.504	0.00126562981855423\\
0.522	0.00147370207400084\\
0.54	0.0015045811979615\\
0.558	0.00160804484948295\\
0.576	0.00168257587298178\\
0.594	0.00185941500649287\\
0.612	0.00193639899662683\\
0.63	0.00209993079693961\\
0.648	0.00224688967488385\\
0.666	0.00231752864462578\\
0.684	0.00238308181708832\\
0.702	0.00253298788285494\\
0.72	0.00263972424552626\\
0.738	0.00266817669177552\\
0.756	0.00273122529382476\\
0.774	0.00287655257500998\\
0.792	0.00279703013777908\\
0.81	0.00285325696690891\\
0.828	0.00286531798973609\\
0.846	0.00282253988093443\\
0.864	0.00281369578584848\\
0.882	0.00281567080403649\\
0.9	0.00284309227535428\\
0.918	0.00275818488212435\\
0.936	0.0027380983128549\\
0.954	0.00262670620799091\\
0.972	0.00262639408805167\\
0.99	0.00249943702079264\\
1.008	0.00236945401743368\\
1.026	0.00223144579850401\\
1.044	0.00220531852292512\\
1.062	0.00204417937551569\\
1.08	0.00194559733512471\\
1.098	0.0018770490688646\\
1.116	0.00176957299084972\\
1.134	0.00167746840158841\\
1.152	0.00158449694920204\\
1.17	0.00150641099578212\\
1.188	0.00139066143826934\\
1.206	0.00126575383680397\\
1.224	0.0012533722103206\\
1.242	0.00107500292369361\\
1.26	0.00100730642268983\\
1.278	0.000923388291344021\\
1.296	0.000834943831668828\\
1.314	0.000760303925803847\\
1.332	0.000736655454477609\\
1.35	0.0007271529465667\\
1.368	0.000592734667979435\\
1.386	0.000576599423409739\\
1.404	0.000531475305963142\\
1.422	0.000488897561912528\\
1.44	0.000413046200737303\\
1.458	0.000381844637015395\\
1.476	0.000305172143144921\\
1.494	0.000306503506702863\\
1.512	0.000296562341704817\\
1.53	0.000249901980534826\\
1.548	0.000212014935870634\\
1.566	0.000218153320153328\\
1.584	0.000182521186562227\\
1.602	0.000162531224044222\\
1.62	0.000141544883955126\\
1.638	0.000127780515111261\\
1.656	0.000120836398093098\\
1.674	0.000109734120008445\\
1.692	9.09290497583134e-05\\
1.71	6.39581811121886e-05\\
1.728	5.48899055321758e-05\\
1.746	5.0972849867477e-05\\
1.764	5.39410433023969e-05\\
1.782	4.79742993535411e-05\\
1.8	4.99667603301758e-05\\
};
\addlegendentry{$\displaystyle \sigma^2_{\displaystyle \hat K^\epsilon_N}$}

\end{axis}

\node[] at (2.85,0.65) {(b')};

\end{tikzpicture}%
        \captionlistentry{}
        \label{fig:P1_variances_PSD2}
    \end{subfigure}%

    \centering
    \renewcommand{\thesubfigure}{\alph{subfigure}}
    \begin{subfigure}{.45\textwidth}
        \centering
%
%
\definecolor{mycolor1}{rgb}{0.00000,0.44700,0.74100}%
\definecolor{mycolor2}{rgb}{0.85000,0.32500,0.09800}%
\definecolor{mycolor3}{rgb}{0.92900,0.69400,0.12500}%
\begin{tikzpicture}[scale=0.95]

\begin{axis}[%
width=0.8\linewidth,
height=0.64\linewidth,
scale only axis,
xmin=0,
xmax=1.2,
xtick distance = 0.3,
xlabel = $\epsilon$,
y label style={at={(axis description cs:0.1,.5)},anchor=south},
axis background/.style={fill=white},
legend style={legend cell align=left, align=left, draw=white!15!black, minimum height=0.6cm, font={\tiny\arraycolsep=2pt}}
]
\addplot [color=black, line width=1pt]
  table[row sep=crcr]{%
0.036	0.000249827570976619\\
0.048	0.000291004639209297\\
0.06	0.000399186619143461\\
0.072	0.000318678233539572\\
0.084	0.000557069356360386\\
0.096	0.000652050385608234\\
0.108	0.000705345268020748\\
0.12	0.000694998811431127\\
0.132	0.000776727094054885\\
0.144	0.000738952075261003\\
0.156	0.000948827852829445\\
0.168	0.000943656773339752\\
0.18	0.00115427217341685\\
0.192	0.000980001187839578\\
0.204	0.00121576368168053\\
0.216	0.00120369772907565\\
0.228	0.00122450599550956\\
0.24	0.00127408719485933\\
0.252	0.00129683276864337\\
0.264	0.00142903405604718\\
0.276	0.00141266034522094\\
0.288	0.00152654088032888\\
0.3	0.00155616866904573\\
0.312	0.00157341140989023\\
0.324	0.0015476636650624\\
0.336	0.00171253926697172\\
0.348	0.00168148512261404\\
0.36	0.00168742403033489\\
0.372	0.00170474297312504\\
0.384	0.00180882068314731\\
0.396	0.00170964920430425\\
0.408	0.00174449325421491\\
0.42	0.00186105892094002\\
0.432	0.00171653548120028\\
0.444	0.00170754416994374\\
0.456	0.00175747912668455\\
0.468	0.00172317692736058\\
0.48	0.00174956172122176\\
0.492	0.00161111006290422\\
0.504	0.00160484247416721\\
0.516	0.00157803664813656\\
0.528	0.00161471927858409\\
0.54	0.00153969506637551\\
0.552	0.00148894547505936\\
0.564	0.00150867188353085\\
0.576	0.00146531085953108\\
0.588	0.00139566112463257\\
0.6	0.00134879056142633\\
0.612	0.00127483315753388\\
0.624	0.00119151118249637\\
0.636	0.00120511334031885\\
0.648	0.00113998908429762\\
0.66	0.00106290163714884\\
0.672	0.00103154799157958\\
0.684	0.000931531636162779\\
0.696	0.000892084227943384\\
0.708	0.00083663740871266\\
0.72	0.000771419186117927\\
0.732	0.000733564780686322\\
0.744	0.000674266427625656\\
0.756	0.000649469007466782\\
0.768	0.000633894925534369\\
0.78	0.000577595859362569\\
0.792	0.000533304536982642\\
0.804	0.000464808037692707\\
0.816	0.000444892560292835\\
0.828	0.000416148817048864\\
0.84	0.0003500038037416\\
0.852	0.00036000548478382\\
0.864	0.000325836185953927\\
0.876	0.000322791872631023\\
0.888	0.000285794106074149\\
0.9	0.000256951057165533\\
0.912	0.00020657700824221\\
0.924	0.000223728510378354\\
0.936	0.000184454220213163\\
0.948	0.000172040934192312\\
0.96	0.000163972387254503\\
0.972	0.00014355965535437\\
0.984	0.000106968595913494\\
0.996	0.000126754020933731\\
1.008	9.1419088638527e-05\\
1.02	8.23149669994781e-05\\
1.032	9.14177490130147e-05\\
1.044	6.68892494340796e-05\\
1.056	6.39541209623239e-05\\
1.068	6.2859832687581e-05\\
1.08	5.09891179094101e-05\\
1.092	5.58057905351976e-05\\
1.104	4.39326390661744e-05\\
1.116	3.89226331643045e-05\\
1.128	3.68801729172093e-05\\
1.14	2.51871286855788e-05\\
1.152	1.90101473269323e-05\\
1.164	2.05103443815346e-05\\
1.176	2.1978242382867e-05\\
1.188	1.92449343941044e-05\\
1.2	2.23110904738101e-05\\
};
\addlegendentry{$\displaystyle \sigma^2_{\displaystyle \hat K^\epsilon_N}/\epsilon$}

\addplot [color=black,dashed, line width=0.5pt]
  table[row sep=crcr]{%
0.036	0.00693965474935052\\
0.048	0.00606259665019369\\
0.06	0.00665311031905767\\
0.072	0.00442608657693849\\
0.084	0.00663177805190936\\
0.096	0.00679219151675244\\
0.108	0.00653097470389582\\
0.12	0.00579165676192606\\
0.132	0.00588429616708246\\
0.144	0.00513161163375697\\
0.156	0.00608222982582977\\
0.168	0.00561700460321281\\
0.18	0.00641262318564916\\
0.192	0.00510417285333114\\
0.204	0.00595962589059082\\
0.216	0.00557267467164651\\
0.228	0.00537064033118226\\
0.24	0.0053086966452472\\
0.252	0.00514616178033083\\
0.264	0.00541300778805749\\
0.276	0.00511833458413384\\
0.288	0.00530048916780862\\
0.3	0.00518722889681911\\
0.312	0.00504298528810972\\
0.324	0.0047767397069827\\
0.336	0.00509684305646345\\
0.348	0.00483185380061505\\
0.36	0.00468728897315246\\
0.372	0.00458264240087378\\
0.384	0.00471047052902945\\
0.396	0.00431729597046529\\
0.408	0.00427571876033067\\
0.42	0.00443109266890482\\
0.432	0.00397346176203769\\
0.444	0.003845820202576\\
0.456	0.00385412089185209\\
0.468	0.0036820019815397\\
0.48	0.00364492025254534\\
0.492	0.00327461394899232\\
0.504	0.00318421125826828\\
0.516	0.0030582105584042\\
0.528	0.00305818045186381\\
0.54	0.00285128715995465\\
0.552	0.00269736499104957\\
0.564	0.00267495014810434\\
0.576	0.00254394246446368\\
0.588	0.00237357334121185\\
0.6	0.00224798426904389\\
0.612	0.00208306071492464\\
0.624	0.00190947304887239\\
0.636	0.00189483229609882\\
0.648	0.00175924241403954\\
0.66	0.00161045702598309\\
0.672	0.00153504165413628\\
0.684	0.00136188835696313\\
0.696	0.00128173021256233\\
0.708	0.00118169125524387\\
0.72	0.0010714155362749\\
0.732	0.00100213767853323\\
0.744	0.00090627208014201\\
0.756	0.000859085988712674\\
0.768	0.000825384017622876\\
0.78	0.000740507512003293\\
0.792	0.000673364314372022\\
0.804	0.000578119449866551\\
0.816	0.000545211470947101\\
0.828	0.000502595189672541\\
0.84	0.000416671194930477\\
0.852	0.000422541648807301\\
0.864	0.000377125215224453\\
0.876	0.000368483872866465\\
0.888	0.00032184020954296\\
0.9	0.00028550117462837\\
0.912	0.000226509877458564\\
0.924	0.000242130422487396\\
0.936	0.000197066474586713\\
0.948	0.000181477778683874\\
0.96	0.000170804570056774\\
0.972	0.000147695118677336\\
0.984	0.000108707922676315\\
0.996	0.000127263073226638\\
1.008	9.0693540315999e-05\\
1.02	8.0700948038704e-05\\
1.032	8.85830901288902e-05\\
1.044	6.40701622931797e-05\\
1.056	6.05626145476552e-05\\
1.068	5.88575212430534e-05\\
1.08	4.72121462124167e-05\\
1.092	5.1104203786811e-05\\
1.104	3.9794057125158e-05\\
1.116	3.48769114375488e-05\\
1.128	3.26951887563912e-05\\
1.14	2.20939725312095e-05\\
1.152	1.65018639990732e-05\\
1.164	1.76205707745142e-05\\
1.176	1.86889816180842e-05\\
1.188	1.61994397256771e-05\\
1.2	1.85925753948418e-05\\
};
\addlegendentry{$\displaystyle \sigma^2_{\displaystyle \hat K^\epsilon_N}/\epsilon^2$}

\end{axis}

\node[] at (2.85,0.65) {(c)};

\end{tikzpicture}%
        \captionlistentry{}
        \label{fig:P1_bounds_PSD1}
    \end{subfigure}%
    \setcounter{subfigure}{2}
    \renewcommand{\thesubfigure}{\alph{subfigure}'}
    \begin{subfigure}{.45\textwidth}
        \centering
%
%
\definecolor{mycolor1}{rgb}{0.00000,0.44700,0.74100}%
\definecolor{mycolor2}{rgb}{0.85000,0.32500,0.09800}%
\definecolor{mycolor3}{rgb}{0.92900,0.69400,0.12500}%
\begin{tikzpicture}[scale=0.95]

\begin{axis}[%
width=0.8\linewidth,
height=0.64\linewidth,
scale only axis,
xmin=0,
xmax=1.8,
xtick distance = 0.45,
xlabel = $\epsilon$,
y label style={at={(axis description cs:0.1,.5)},anchor=south},
axis background/.style={fill=white},
legend style={legend cell align=left, align=left, draw=white!15!black, minimum height=0.6cm, font={\tiny\arraycolsep=2pt}}
]
\addplot [color=black, line width=1pt]
  table[row sep=crcr]{%
0.036	0.000222123945978429\\
0.054	0.000332866150806684\\
0.072	0.0004567438782464\\
0.09	0.000476386199857076\\
0.108	0.000579086436013808\\
0.126	0.000764808158041373\\
0.144	0.000905297510098901\\
0.162	0.00081491870160379\\
0.18	0.000992267502874062\\
0.198	0.00103586743630339\\
0.216	0.00115111937625565\\
0.234	0.00111394473033436\\
0.252	0.00132041935004554\\
0.27	0.00126898746639041\\
0.288	0.00144771920829515\\
0.306	0.00154387658379118\\
0.324	0.00153819639565418\\
0.342	0.00163530900406347\\
0.36	0.00174922231375421\\
0.378	0.00199246513857422\\
0.396	0.00190048461527557\\
0.414	0.00193192067470791\\
0.432	0.00224184089744492\\
0.45	0.00222981615257259\\
0.468	0.00232257756471302\\
0.486	0.00260159384234958\\
0.504	0.00251117027490918\\
0.522	0.00282318404981004\\
0.54	0.00278626147770647\\
0.558	0.00288180080552499\\
0.576	0.00292113866837115\\
0.594	0.00313032829375904\\
0.612	0.00316405064808305\\
0.63	0.00333322348720572\\
0.648	0.00346742233778372\\
0.666	0.00347977273967835\\
0.684	0.00348403774428116\\
0.702	0.00360824484737171\\
0.72	0.00366628367434203\\
0.738	0.00361541557151155\\
0.756	0.00361273187013857\\
0.774	0.00371647619510334\\
0.792	0.00353160370931702\\
0.81	0.00352253946531964\\
0.828	0.00346052897311122\\
0.846	0.00333633555666008\\
0.864	0.00325659234473204\\
0.882	0.00319237052611847\\
0.9	0.00315899141706031\\
0.918	0.00300455869512457\\
0.936	0.00292531871031506\\
0.954	0.00275336080502191\\
0.972	0.00270205153091736\\
0.99	0.00252468385938651\\
1.008	0.00235064882681913\\
1.026	0.00217489843908773\\
1.044	0.00211237406410452\\
1.062	0.00192483933664378\\
1.08	0.00180147901400436\\
1.098	0.00170951645616084\\
1.116	0.00158563888068971\\
1.134	0.00147924903138308\\
1.152	0.00137543137951566\\
1.17	0.00128753076562574\\
1.188	0.00117059043625365\\
1.206	0.0010495471283615\\
1.224	0.0010239969038567\\
1.242	0.000865541806516597\\
1.26	0.000799449541817322\\
1.278	0.000722526049564961\\
1.296	0.000644246783695083\\
1.314	0.00057861790396031\\
1.332	0.000553044635493701\\
1.35	0.00053863181227163\\
1.368	0.000433285576008359\\
1.386	0.000416016900007027\\
1.404	0.000378543665215913\\
1.422	0.000343809818503887\\
1.44	0.000286837639400905\\
1.458	0.000261896184509873\\
1.476	0.000206756194542629\\
1.494	0.000205156296320524\\
1.512	0.000196139114884138\\
1.53	0.00016333462780054\\
1.548	0.000136960552888006\\
1.566	0.000139306079280541\\
1.584	0.000115228021819588\\
1.602	0.000101455196032598\\
1.62	8.7373385157485e-05\\
1.638	7.80100824855073e-05\\
1.656	7.29688394281993e-05\\
1.674	6.55520430157978e-05\\
1.692	5.37405731432112e-05\\
1.71	3.74024450948472e-05\\
1.728	3.17649916274165e-05\\
1.746	2.91940720890475e-05\\
1.764	3.05788227337851e-05\\
1.782	2.69216045755001e-05\\
1.8	2.77593112945421e-05\\
};
\addlegendentry{$\displaystyle \sigma^2_{\displaystyle \hat K^\epsilon_N}/\epsilon$}

\addplot [color=black, dashed, line width=0.5pt]
  table[row sep=crcr]{%
0.036	0.0061701096105119\\
0.054	0.00616418797790156\\
0.072	0.00634366497564444\\
0.09	0.00529317999841196\\
0.108	0.00536191144457229\\
0.126	0.00606990601620138\\
0.144	0.0062867882645757\\
0.162	0.00503036235557895\\
0.18	0.00551259723818923\\
0.198	0.00523165371870398\\
0.216	0.00532925637155394\\
0.234	0.00476044756553146\\
0.252	0.00523975932557753\\
0.27	0.00469995357922375\\
0.288	0.00502680280658038\\
0.306	0.00504534831304307\\
0.324	0.00474751973967341\\
0.342	0.00478160527503939\\
0.36	0.00485895087153946\\
0.378	0.0052710717951699\\
0.396	0.00479920357392821\\
0.414	0.00466647505968093\\
0.432	0.00518944652186325\\
0.45	0.00495514700571687\\
0.468	0.00496277257417313\\
0.486	0.00535307374969049\\
0.504	0.00498248070418487\\
0.522	0.00540839856285449\\
0.54	0.00515974347723421\\
0.558	0.00516451757262543\\
0.576	0.00507142129925547\\
0.594	0.00526991295245629\\
0.612	0.00517001739882851\\
0.63	0.00529083093207258\\
0.648	0.00535096039781438\\
0.666	0.00522488399351104\\
0.684	0.00509362243315959\\
0.702	0.00513994992503093\\
0.72	0.00509206065880838\\
0.738	0.00489893708876903\\
0.756	0.00477874585997166\\
0.774	0.00480164883088286\\
0.792	0.004459095592572\\
0.81	0.00434881415471561\\
0.828	0.00417938281776717\\
0.846	0.00394365905042563\\
0.864	0.00376920410269912\\
0.882	0.00361946771668761\\
0.9	0.00350999046340034\\
0.918	0.00327293975503766\\
0.936	0.00312534050247335\\
0.954	0.00288612243712989\\
0.972	0.00277988840629358\\
0.99	0.00255018571655203\\
1.008	0.00233199288374914\\
1.026	0.00211978405369174\\
1.044	0.00202334680469781\\
1.062	0.00181246641868529\\
1.08	0.00166803612407811\\
1.098	0.00155693666317017\\
1.116	0.00142082336979365\\
1.134	0.00130445240862705\\
1.152	0.00119395085027401\\
1.17	0.00110045364583397\\
1.188	0.000985345485061995\\
1.206	0.000870271250714347\\
1.224	0.000836598777660702\\
1.242	0.000696893564023025\\
1.26	0.000634483763347081\\
1.278	0.000565356846294962\\
1.296	0.000497103999764725\\
1.314	0.000440348480943919\\
1.332	0.000415198675295571\\
1.35	0.000398986527608615\\
1.368	0.000316729222228333\\
1.386	0.000300156493511563\\
1.404	0.000269617995168029\\
1.422	0.000241779056613142\\
1.44	0.000199192805139517\\
1.458	0.000179627012695386\\
1.476	0.000140078722589857\\
1.494	0.000137320144792854\\
1.512	0.000129721636828134\\
1.53	0.000106754658693163\\
1.548	8.84758093591772e-05\\
1.566	8.89566278930661e-05\\
1.584	7.27449632699418e-05\\
1.602	6.33303346021211e-05\\
1.62	5.39341883688179e-05\\
1.638	4.76252029826052e-05\\
1.656	4.40633088334537e-05\\
1.674	3.91589265327346e-05\\
1.692	3.17615680515433e-05\\
1.71	2.18727749092673e-05\\
1.728	1.83825183029031e-05\\
1.746	1.67205452972781e-05\\
1.764	1.73349335225539e-05\\
1.782	1.51075222084737e-05\\
1.8	1.5421839608079e-05\\
};
\addlegendentry{$\displaystyle \sigma^2_{\displaystyle \hat K^\epsilon_N}/\epsilon^2$}

\end{axis}

\node[] at (2.85,0.65) {(c')};

\end{tikzpicture}%
        \captionlistentry{}
        \label{fig:P1_bounds_PSD2}
    \end{subfigure}%
    
    \caption{Monte Carlo estimation of $P^\epsilon_1$ (for $a=0.5$ and $X_f=2$) using $\hat I_N^\eps$ (standard MC), $\hat J_N^\eps$ (simple control variate) and $K_N^\eps$ (practical optimal control variate).
    For $\hat J_N^\eps$ and $K_N^\eps$, the expectation of the control variate $P_1^0$ is obtained by solving the PDE \eqref{eq:pdeBEPO1}. This is done numerically by finite differences shown in Section \ref{sec:fdm_pde}. The driving coloured noise is related to PSD1 for the left column (a)-(b)-(c), and PSD2 for the right column (a')-(b')-(c'). The time discretization of the KBEPO dynamics together with the driving noise (coloured or white) is given in \eqref{eq:OU_EM_BEPO}-\eqref{eq:colorednoise_EM_BEPO}-\eqref{eq:whitenoise_EM_BEPO} in the appendix.}
    \label{fig:P1_PSD1_PSD2}
\end{figure}

\begin{figure}[h!]
    \centering
    \begin{subfigure}{0.45\textwidth}
        \centering
%
%
\definecolor{mycolor1}{rgb}{0.00000,0.44700,0.74100}%
\definecolor{mycolor2}{rgb}{0.85000,0.32500,0.09800}%
\definecolor{mycolor3}{rgb}{0.92900,0.69400,0.12500}%
\begin{tikzpicture}[scale=0.95]

\begin{axis}[%
width=0.8\linewidth,
height=0.64\linewidth,
scale only axis,
xmin=0,
xmax=1.2,
xtick distance = 0.3,
xlabel = $\epsilon$,
y label style={at={(axis description cs:0.1,.5)},anchor=south},
axis background/.style={fill=white},
legend style={legend cell align=left, align=left, draw=white!15!black, minimum height=0.5cm, font={\tiny\arraycolsep=2pt}},
]
\addplot [color=mycolor1, line width=1pt]
  table[row sep=crcr]{%
0.012	0.00255\\
0.024	0.00236\\
0.036	0.00222\\
0.048	0.00269\\
0.06	0.00225\\
0.072	0.00246\\
0.084	0.0025\\
0.096	0.00237\\
0.108	0.00233\\
0.12	0.00251\\
0.132	0.0025\\
0.144	0.00236\\
0.156	0.00245\\
0.168	0.00231\\
0.18	0.00241\\
0.192	0.00221\\
0.204	0.00269\\
0.216	0.00256\\
0.228	0.00244\\
0.24	0.00242\\
0.252	0.00231\\
0.264	0.00238\\
0.276	0.00211\\
0.288	0.00223\\
0.3	0.00231\\
0.312	0.00237\\
0.324	0.0021\\
0.336	0.0025\\
0.348	0.00251\\
0.36	0.00196\\
0.372	0.00239\\
0.384	0.00213\\
0.396	0.00198\\
0.408	0.00222\\
0.42	0.00191\\
0.432	0.00185\\
0.444	0.0018\\
0.456	0.00183\\
0.468	0.00156\\
0.48	0.00159\\
0.492	0.00184\\
0.504	0.00147\\
0.516	0.00141\\
0.528	0.00169\\
0.54	0.00153\\
0.552	0.00148\\
0.564	0.00149\\
0.576	0.00126\\
0.588	0.00126\\
0.6	0.00127\\
0.612	0.00092\\
0.624	0.00112\\
0.636	0.00121\\
0.648	0.00093\\
0.66	0.00074\\
0.672	0.00115\\
0.684	0.00085\\
0.696	0.00072\\
0.708	0.00062\\
0.72	0.00081\\
0.732	0.00082\\
0.744	0.00067\\
0.756	0.00052\\
0.768	0.00044\\
0.78	0.00047\\
0.792	0.00036\\
0.804	0.00047\\
0.816	0.00037\\
0.828	0.00033\\
0.84	0.00037\\
0.852	0.00027\\
0.864	0.00036\\
0.876	0.00022\\
0.888	0.00014\\
0.9	0.00018\\
0.912	0.00019\\
0.924	0.00013\\
0.936	0.00023\\
0.948	0.00011\\
0.96	0.00014\\
0.972	0.00014\\
0.984	0.00014\\
0.996	0.00011\\
1.008	0.0001\\
1.02	7e-05\\
1.032	6e-05\\
1.044	8e-05\\
1.056	0.0001\\
1.068	7e-05\\
1.08	6e-05\\
1.092	4e-05\\
1.104	6e-05\\
1.116	2e-05\\
1.128	3e-05\\
1.14	0\\
1.152	3e-05\\
1.164	2e-05\\
1.176	1e-05\\
1.188	2e-05\\
1.2	0\\
};
\addlegendentry{$\hat I^\epsilon_N$}

\addplot [color=mycolor2, line width=1pt]
  table[row sep=crcr]{%
0.012	0.0025657701987118\\
0.024	0.0025657701987118\\
0.036	0.0025557701987118\\
0.048	0.0025657701987118\\
0.06	0.0025457701987118\\
0.072	0.0025757701987118\\
0.084	0.0025757701987118\\
0.096	0.0025857701987118\\
0.108	0.0025457701987118\\
0.12	0.0025257701987118\\
0.132	0.0025557701987118\\
0.144	0.0025457701987118\\
0.156	0.0025657701987118\\
0.168	0.0025157701987118\\
0.18	0.0025357701987118\\
0.192	0.0026057701987118\\
0.204	0.0024957701987118\\
0.216	0.0024457701987118\\
0.228	0.0024057701987118\\
0.24	0.0024957701987118\\
0.252	0.0024157701987118\\
0.264	0.0025357701987118\\
0.276	0.0024457701987118\\
0.288	0.0024957701987118\\
0.3	0.0023257701987118\\
0.312	0.0022757701987118\\
0.324	0.0022857701987118\\
0.336	0.0022657701987118\\
0.348	0.0023457701987118\\
0.36	0.0022057701987118\\
0.372	0.0023857701987118\\
0.384	0.0021057701987118\\
0.396	0.0021757701987118\\
0.408	0.0021857701987118\\
0.42	0.0021757701987118\\
0.432	0.0019557701987118\\
0.444	0.0019757701987118\\
0.456	0.0017457701987118\\
0.468	0.0019257701987118\\
0.48	0.0017457701987118\\
0.492	0.0017057701987118\\
0.504	0.0017957701987118\\
0.516	0.0014357701987118\\
0.528	0.0017257701987118\\
0.54	0.0015357701987118\\
0.552	0.0014057701987118\\
0.564	0.0014457701987118\\
0.576	0.0011757701987118\\
0.588	0.0014157701987118\\
0.6	0.0013857701987118\\
0.612	0.0008857701987118\\
0.624	0.0009257701987118\\
0.636	0.0012157701987118\\
0.648	0.0007957701987118\\
0.66	0.0006357701987118\\
0.672	0.0011157701987118\\
0.684	0.0009957701987118\\
0.696	0.0007557701987118\\
0.708	0.0007257701987118\\
0.72	0.0011457701987118\\
0.732	0.0007957701987118\\
0.744	0.0008257701987118\\
0.756	0.0006057701987118\\
0.768	0.0005757701987118\\
0.78	0.0004757701987118\\
0.792	0.0006757701987118\\
0.804	0.0006257701987118\\
0.816	0.0006157701987118\\
0.828	0.0003457701987118\\
0.84	0.0002557701987118\\
0.852	0.0004257701987118\\
0.864	0.0002757701987118\\
0.876	0.0003657701987118\\
0.888	0.0002457701987118\\
0.9	0.0003357701987118\\
0.912	0.0002257701987118\\
0.924	0.0001957701987118\\
0.936	0.0002857701987118\\
0.948	0.0003157701987118\\
0.96	0.0001857701987118\\
0.972	0.0001957701987118\\
0.984	0.0001757701987118\\
0.996	0.0004157701987118\\
1.008	9.57701987117998e-05\\
1.02	0.0003257701987118\\
1.032	0.0002457701987118\\
1.044	0.0002757701987118\\
1.056	4.57701987117997e-05\\
1.068	-5.42298012882001e-05\\
1.08	0.0001757701987118\\
1.092	6.57701987117998e-05\\
1.104	0.0004257701987118\\
1.116	6.57701987117998e-05\\
1.128	0.0003557701987118\\
1.14	-3.42298012882001e-05\\
1.152	-0.0002442298012882\\
1.164	-5.42298012882001e-05\\
1.176	-0.0001942298012882\\
1.188	-4.22980128819998e-06\\
1.2	0.0001257701987118\\
};
\addlegendentry{$\hat J^\epsilon_N$}

\addplot [color=mycolor3, line width=1.0pt]
  table[row sep=crcr]{%
0.012	0.00256577019716981\\
0.024	0.00256576649826841\\
0.036	0.00255424756650389\\
0.048	0.00256577091336019\\
0.06	0.00254315297080614\\
0.072	0.00257576840329934\\
0.084	0.00257546420710822\\
0.096	0.00258392095740143\\
0.108	0.00254116866166617\\
0.12	0.00252546081951233\\
0.132	0.00255443418340175\\
0.144	0.0025402944381194\\
0.156	0.0025638748063545\\
0.168	0.00250528935723871\\
0.18	0.00252957265503136\\
0.192	0.00259658740775115\\
0.204	0.00250493292648724\\
0.216	0.00245430448968469\\
0.228	0.00240880063602366\\
0.24	0.00248936914634013\\
0.252	0.00240716525104491\\
0.264	0.00252281522499432\\
0.276	0.00240957793368765\\
0.288	0.00246798504569612\\
0.3	0.00232366592316244\\
0.312	0.00229030594548367\\
0.324	0.00225139413373172\\
0.336	0.00230345336901806\\
0.348	0.0023728802797369\\
0.36	0.00215170117521637\\
0.372	0.00238636346341092\\
0.384	0.00211232462792416\\
0.396	0.00212616454439526\\
0.408	0.00219380894191708\\
0.42	0.00210753500819984\\
0.432	0.00192005985942605\\
0.444	0.0019161592783363\\
0.456	0.00177788906376163\\
0.468	0.00178769155166489\\
0.48	0.00168369785281687\\
0.492	0.00175899377842171\\
0.504	0.00165608352795034\\
0.516	0.00142237289993033\\
0.528	0.00171162603153558\\
0.54	0.00153304183688576\\
0.552	0.00144402475385646\\
0.564	0.00147052686549734\\
0.576	0.00122568767662589\\
0.588	0.00133301644309759\\
0.6	0.00132006368487313\\
0.612	0.00090986786512794\\
0.624	0.00104894555073637\\
0.636	0.00121223017130217\\
0.648	0.000891242950164256\\
0.66	0.000715026630362963\\
0.672	0.00113710383803797\\
0.684	0.000889727395539173\\
0.696	0.000728620486189843\\
0.708	0.000642778241703191\\
0.72	0.000900266514382459\\
0.732	0.000813735704457768\\
0.744	0.000704882215005412\\
0.756	0.0005352103694135\\
0.768	0.000461223226264797\\
0.78	0.000470855993209371\\
0.792	0.000397862258755276\\
0.804	0.00049325116662177\\
0.816	0.00039857718629898\\
0.828	0.000331544832920628\\
0.84	0.000355939040254579\\
0.852	0.000285507544028403\\
0.864	0.000349832259621907\\
0.876	0.000231440283257148\\
0.888	0.000144726359252908\\
0.9	0.000189690458387223\\
0.912	0.000191835841813797\\
0.924	0.00013289256502508\\
0.936	0.000233328403423537\\
0.948	0.000118716856885594\\
0.96	0.00014199652726046\\
0.972	0.000142442431943915\\
0.984	0.000141554149415162\\
0.996	0.000122170055461752\\
1.008	9.98354162934366e-05\\
1.02	7.4421104823901e-05\\
1.032	6.39008164354812e-05\\
1.044	8.66081696332343e-05\\
1.056	9.81376900490022e-05\\
1.068	6.67672728383693e-05\\
1.08	6.28351729919584e-05\\
1.092	4.02023981809993e-05\\
1.104	6.49764748370567e-05\\
1.116	2.03632552425897e-05\\
1.128	3.43627865695689e-05\\
1.14	0\\
1.152	2.80716055330927e-05\\
1.164	1.97195713599834e-05\\
1.176	9.26271943118083e-06\\
1.188	1.99066915920746e-05\\
1.2	0\\
};
\addlegendentry{$\hat K^\epsilon_N$}

\addplot [color=black, dashed, line width=1.0pt]
  table[row sep=crcr]{%
0	0.0025657701987118\\
0.133333333333333	0.0025657701987118\\
0.266666666666667	0.0025657701987118\\
0.4	0.0025657701987118\\
0.533333333333333	0.0025657701987118\\
0.666666666666667	0.0025657701987118\\
0.8	0.0025657701987118\\
0.933333333333333	0.0025657701987118\\
1.06666666666667	0.0025657701987118\\
1.2	0.0025657701987118\\
};
\addlegendentry{$P^0_2$}

\end{axis}

\node[] at (2.85,0.65) {(a)};

\end{tikzpicture}%
        \captionlistentry{}
        \label{fig:P2_estimates_PSD1}
    \end{subfigure}%
    \setcounter{subfigure}{0}
    \renewcommand{\thesubfigure}{\alph{subfigure}'}
    \begin{subfigure}{.45\textwidth}
        \centering
%
%
\definecolor{mycolor1}{rgb}{0.00000,0.44700,0.74100}%
\definecolor{mycolor2}{rgb}{0.85000,0.32500,0.09800}%
\definecolor{mycolor3}{rgb}{0.92900,0.69400,0.12500}%
\begin{tikzpicture}[scale=0.95]

\begin{axis}[%
width=0.8\linewidth,
height=0.64\linewidth,
scale only axis,
xmin=0,
xmax=1.8,
xtick distance = 0.45,
xlabel = $\epsilon$,
y label style={at={(axis description cs:0.1,.5)},anchor=south},
axis background/.style={fill=white},
legend style={legend cell align=left, align=left, draw=white!15!black, minimum height=0.5cm, font={\tiny\arraycolsep=2pt}},
]
\addplot [color=mycolor1, line width=1pt]
  table[row sep=crcr]{%
0.018	0.00246\\
0.036	0.00295\\
0.054	0.00261\\
0.072	0.00231\\
0.09	0.00247\\
0.108	0.00233\\
0.126	0.00223\\
0.144	0.0026\\
0.162	0.0024\\
0.18	0.00254\\
0.198	0.00244\\
0.216	0.0025\\
0.234	0.00257\\
0.252	0.00278\\
0.27	0.00245\\
0.288	0.0027\\
0.306	0.00267\\
0.324	0.00257\\
0.342	0.00235\\
0.36	0.00253\\
0.378	0.00254\\
0.396	0.00275\\
0.414	0.00274\\
0.432	0.00289\\
0.45	0.00303\\
0.468	0.00297\\
0.486	0.00267\\
0.504	0.00302\\
0.522	0.00305\\
0.54	0.00301\\
0.558	0.00335\\
0.576	0.00343\\
0.594	0.00321\\
0.612	0.00323\\
0.63	0.00333\\
0.648	0.0033\\
0.666	0.00366\\
0.684	0.00366\\
0.702	0.00364\\
0.72	0.00368\\
0.738	0.00354\\
0.756	0.00346\\
0.774	0.00381\\
0.792	0.00348\\
0.81	0.00338\\
0.828	0.00369\\
0.846	0.00341\\
0.864	0.00358\\
0.882	0.00352\\
0.9	0.0034\\
0.918	0.0031\\
0.936	0.00289\\
0.954	0.00296\\
0.972	0.00289\\
0.99	0.00267\\
1.008	0.00256\\
1.026	0.00247\\
1.044	0.00238\\
1.062	0.00238\\
1.08	0.00215\\
1.098	0.00214\\
1.116	0.00202\\
1.134	0.00186\\
1.152	0.00183\\
1.17	0.0015\\
1.188	0.00149\\
1.206	0.00143\\
1.224	0.00132\\
1.242	0.0012\\
1.26	0.00108\\
1.278	0.00103\\
1.296	0.00074\\
1.314	0.00089\\
1.332	0.00057\\
1.35	0.00057\\
1.368	0.00067\\
1.386	0.00065\\
1.404	0.00054\\
1.422	0.00039\\
1.44	0.00041\\
1.458	0.00038\\
1.476	0.00042\\
1.494	0.00032\\
1.512	0.00024\\
1.53	0.00017\\
1.548	0.00021\\
1.566	0.00016\\
1.584	0.00015\\
1.602	0.00015\\
1.62	0.00014\\
1.638	0.00012\\
1.656	0.00016\\
1.674	0.00015\\
1.692	8e-05\\
1.71	0.00011\\
1.728	4e-05\\
1.746	0.00011\\
1.764	4e-05\\
1.782	0\\
1.8	5e-05\\
};
\addlegendentry{$\hat I^\epsilon_N$}

\addplot [color=mycolor2, line width=1pt]
  table[row sep=crcr]{%
0.018	0.0025657701987118\\
0.036	0.0025757701987118\\
0.054	0.0025357701987118\\
0.072	0.0025957701987118\\
0.09	0.0025557701987118\\
0.108	0.0025857701987118\\
0.126	0.0025257701987118\\
0.144	0.0025257701987118\\
0.162	0.0026057701987118\\
0.18	0.0025257701987118\\
0.198	0.0026057701987118\\
0.216	0.0026557701987118\\
0.234	0.0025557701987118\\
0.252	0.0026057701987118\\
0.27	0.0027257701987118\\
0.288	0.0025157701987118\\
0.306	0.0026557701987118\\
0.324	0.0025957701987118\\
0.342	0.0027057701987118\\
0.36	0.0026757701987118\\
0.378	0.0027857701987118\\
0.396	0.0027457701987118\\
0.414	0.0029457701987118\\
0.432	0.0028257701987118\\
0.45	0.0031757701987118\\
0.468	0.0028057701987118\\
0.486	0.0029057701987118\\
0.504	0.0030557701987118\\
0.522	0.0030657701987118\\
0.54	0.0028357701987118\\
0.558	0.0034957701987118\\
0.576	0.0034157701987118\\
0.594	0.0033957701987118\\
0.612	0.0032857701987118\\
0.63	0.0035957701987118\\
0.648	0.0032357701987118\\
0.666	0.0035357701987118\\
0.684	0.0037057701987118\\
0.702	0.0035857701987118\\
0.72	0.0039557701987118\\
0.738	0.0036457701987118\\
0.756	0.0034557701987118\\
0.774	0.0038257701987118\\
0.792	0.0036757701987118\\
0.81	0.0036557701987118\\
0.828	0.0037657701987118\\
0.846	0.0034957701987118\\
0.864	0.0033857701987118\\
0.882	0.0033957701987118\\
0.9	0.0035057701987118\\
0.918	0.0030257701987118\\
0.936	0.0029657701987118\\
0.954	0.0030457701987118\\
0.972	0.0031857701987118\\
0.99	0.0028257701987118\\
1.008	0.0024957701987118\\
1.026	0.0025957701987118\\
1.044	0.0025657701987118\\
1.062	0.0020957701987118\\
1.08	0.0020457701987118\\
1.098	0.0021257701987118\\
1.116	0.0019457701987118\\
1.134	0.0017057701987118\\
1.152	0.0018157701987118\\
1.17	0.0015357701987118\\
1.188	0.0015657701987118\\
1.206	0.0014657701987118\\
1.224	0.0015857701987118\\
1.242	0.0011557701987118\\
1.26	0.0013757701987118\\
1.278	0.0009257701987118\\
1.296	0.0007357701987118\\
1.314	0.0007957701987118\\
1.332	0.0007657701987118\\
1.35	0.0005757701987118\\
1.368	0.0009257701987118\\
1.386	0.0007157701987118\\
1.404	0.0004857701987118\\
1.422	0.0005457701987118\\
1.44	0.0002457701987118\\
1.458	0.0004857701987118\\
1.476	0.0005157701987118\\
1.494	0.0002057701987118\\
1.512	0.0004457701987118\\
1.53	0.0002457701987118\\
1.548	0.0002057701987118\\
1.566	0.0004757701987118\\
1.584	-7.42298012882002e-05\\
1.602	0.0002557701987118\\
1.62	0.0002057701987118\\
1.638	5.57701987117997e-05\\
1.656	0.0002757701987118\\
1.674	0.0002857701987118\\
1.692	-3.42298012882001e-05\\
1.71	-6.42298012882001e-05\\
1.728	0.0002357701987118\\
1.746	0.0003457701987118\\
1.764	0.0002057701987118\\
1.782	6.57701987117998e-05\\
1.8	0.0001957701987118\\
};
\addlegendentry{$\hat J^\epsilon_N$}

\addplot [color=mycolor3, line width=1pt]
  table[row sep=crcr]{%
0.018	0.00256576971651696\\
0.036	0.0025757918301453\\
0.054	0.00253689578991656\\
0.072	0.00259449515043805\\
0.09	0.00255403727159403\\
0.108	0.00258020899961903\\
0.126	0.00251532370098432\\
0.144	0.00252774079714315\\
0.162	0.00260226243024068\\
0.18	0.00252632259847016\\
0.198	0.00260022274619307\\
0.216	0.00265121924348094\\
0.234	0.002556433620329\\
0.252	0.00261215527440808\\
0.27	0.0027148541520427\\
0.288	0.00253054208547033\\
0.306	0.0026567115316413\\
0.324	0.00259342995332602\\
0.342	0.00267827388457014\\
0.36	0.00265944991504932\\
0.378	0.00276022663110591\\
0.396	0.00274623298384163\\
0.414	0.00292471321464616\\
0.432	0.00283411239852186\\
0.45	0.00316239591436591\\
0.468	0.00282873352522829\\
0.486	0.00285700125915053\\
0.504	0.00304965767168299\\
0.522	0.00306284819717917\\
0.54	0.00287662667711049\\
0.558	0.00347389578552407\\
0.576	0.00341871306375416\\
0.594	0.00336040504206163\\
0.612	0.00327036006487476\\
0.63	0.00352368230380617\\
0.648	0.00325393333747497\\
0.666	0.00357432603662945\\
0.684	0.0036931538836962\\
0.702	0.00360512577347784\\
0.72	0.00387571743709104\\
0.738	0.00360600150126906\\
0.756	0.00345739104917438\\
0.774	0.00381992208645489\\
0.792	0.00359192287172263\\
0.81	0.00352510801289562\\
0.828	0.00372968216132447\\
0.846	0.00345235168194451\\
0.864	0.00348966593523882\\
0.882	0.00345931404325629\\
0.9	0.00345308579046409\\
0.918	0.00307091133609183\\
0.936	0.00291937099490221\\
0.954	0.00299129577025412\\
0.972	0.0030053654228802\\
0.99	0.00272141543689145\\
1.008	0.00253935147403297\\
1.026	0.00250844176886803\\
1.044	0.00243275987001898\\
1.062	0.0022986685788438\\
1.08	0.00212439115266781\\
1.098	0.00213638093314141\\
1.116	0.00200154380805056\\
1.134	0.00182333106225816\\
1.152	0.00182737185523831\\
1.17	0.00150490720390819\\
1.188	0.00149842841335025\\
1.206	0.00143490971417179\\
1.224	0.00135670921046821\\
1.242	0.00119376668639827\\
1.26	0.00111102064761241\\
1.278	0.00101836512529263\\
1.296	0.000739656621787141\\
1.314	0.000881559504020439\\
1.332	0.000582307900582225\\
1.35	0.000570222679430359\\
1.368	0.000692023490481377\\
1.386	0.000654440716992002\\
1.404	0.000537539031424803\\
1.422	0.000396418146458786\\
1.44	0.000403432150877811\\
1.458	0.000381252763898584\\
1.476	0.000422680522639955\\
1.494	0.000316617660520047\\
1.512	0.000243446317116071\\
1.53	0.000170902257567636\\
1.548	0.00020996788899166\\
1.566	0.000164168941452158\\
1.584	0.000146810027758427\\
1.602	0.000151277150925469\\
1.62	0.000141045727861814\\
1.638	0.000119762864505017\\
1.656	0.000160928810118562\\
1.674	0.000150539666864437\\
1.692	7.91544151300119e-05\\
1.71	0.000107468756966736\\
1.728	4.08201342412908e-05\\
1.746	0.000114030918210042\\
1.764	3.99933533161141e-05\\
1.782	0\\
1.8	5.1200318106467e-05\\
};
\addlegendentry{$\hat K^\epsilon_N$}

\addplot [color=black, dashed, line width=1.0pt]
  table[row sep=crcr]{%
0	0.0025657701987118\\
0.2	0.0025657701987118\\
0.4	0.0025657701987118\\
0.6	0.0025657701987118\\
0.8	0.0025657701987118\\
1	0.0025657701987118\\
1.2	0.0025657701987118\\
1.4	0.0025657701987118\\
1.6	0.0025657701987118\\
1.8	0.0025657701987118\\
};
\addlegendentry{$P^0_2$}

\end{axis}

\node[] at (2.85,0.65) {(a')};

\end{tikzpicture}%
        \captionlistentry{}
        \label{fig:P2_estimates_PSD2}
    \end{subfigure}%

    \centering
    \renewcommand{\thesubfigure}{\alph{subfigure}}
    \begin{subfigure}{.45\textwidth}
        \centering
%
%
\definecolor{mycolor1}{rgb}{0.00000,0.44700,0.74100}%
\definecolor{mycolor2}{rgb}{0.85000,0.32500,0.09800}%
\definecolor{mycolor3}{rgb}{0.92900,0.69400,0.12500}%
\begin{tikzpicture}[scale=0.95]

\begin{axis}[%
width=0.8\linewidth,
height=0.64\linewidth,
scale only axis,
xmin=0,
xmax=1.2,
xtick distance = 0.3,
xlabel = $\epsilon$,
y label style={at={(axis description cs:0.1,.5)},anchor=south},
axis background/.style={fill=white},
legend style={legend cell align=left, align=left, draw=white!15!black, minimum height=0.6cm, font={\tiny\arraycolsep=2pt}, at={(0.99,0.46)},anchor=east}
]
\addplot [color=mycolor1, line width=1pt]
  table[row sep=crcr]{%
0.012	0.002594234799\\
0.024	0.002432056156\\
0.036	0.002397225591\\
0.048	0.002395235199\\
0.06	0.002469869424\\
0.072	0.002476834711\\
0.084	0.002484794919\\
0.096	0.002498724975\\
0.108	0.002498724975\\
0.12	0.002567374524\\
0.132	0.002555436156\\
0.144	0.002485789936\\
0.156	0.002403196719\\
0.168	0.002464894159\\
0.18	0.002381302231\\
0.192	0.002510664711\\
0.204	0.002472854559\\
0.216	0.002457928704\\
0.228	0.002409167775\\
0.24	0.002397225591\\
0.252	0.002421109671\\
0.264	0.002418124224\\
0.276	0.002390259184\\
0.288	0.002279778775\\
0.3	0.002355425679\\
0.312	0.002299686975\\
0.324	0.002268828924\\
0.336	0.0023146176\\
0.348	0.002272810716\\
0.36	0.002226022639\\
0.372	0.002114509839\\
0.384	0.002181221404\\
0.396	0.002044801599\\
0.408	0.002120484375\\
0.42	0.002057748156\\
0.432	0.002041813884\\
0.444	0.0019262751\\
0.456	0.0018864279\\
0.468	0.001728003639\\
0.48	0.001763877711\\
0.492	0.001674187671\\
0.504	0.001595446396\\
0.516	0.001580494111\\
0.528	0.001525665216\\
0.54	0.001504728951\\
0.552	0.001410006256\\
0.564	0.001364134044\\
0.576	0.001277364159\\
0.588	0.001259409879\\
0.6	0.001231479711\\
0.612	0.001117747839\\
0.624	0.001104776764\\
0.636	0.0010688551\\
0.648	0.001026943216\\
0.66	0.000855267264\\
0.672	0.000892202551\\
0.684	0.000858262119\\
0.696	0.000807347136\\
0.708	0.000714488775\\
0.72	0.000708497319\\
0.732	0.000700508599\\
0.744	0.0006096279\\
0.756	0.000597642396\\
0.768	0.000536711631\\
0.78	0.000520728559\\
0.792	0.000446800191\\
0.804	0.000440805519\\
0.816	0.000395843184\\
0.828	0.000372860871\\
0.84	0.000323895024\\
0.852	0.000311902656\\
0.864	0.000303907584\\
0.876	0.000252935991\\
0.888	0.000237943356\\
0.9	0.000217952476\\
0.912	0.000222950271\\
0.924	0.0001799676\\
0.936	0.000185965404\\
0.948	0.000137980956\\
0.96	0.000147978096\\
0.972	0.000138980679\\
0.984	0.0001199856\\
0.996	0.000111987456\\
1.008	9.1991536e-05\\
1.02	9.6990591e-05\\
1.032	8.3992944e-05\\
1.044	6.99951e-05\\
1.056	5.7996636e-05\\
1.068	5.5996864e-05\\
1.08	4.99975e-05\\
1.092	3.8998479e-05\\
1.104	5.1997296e-05\\
1.116	4.2998151e-05\\
1.128	3.6998631e-05\\
1.14	2.7999216e-05\\
1.152	3.7998556e-05\\
1.164	3.0999039e-05\\
1.176	1.99996e-05\\
1.188	1.6999711e-05\\
1.2	2.0999559e-05\\
};
\addlegendentry{$\displaystyle \sigma^2_{\displaystyle \hat I^\epsilon_N}$}

\addplot [color=mycolor2, line width=1pt]
  table[row sep=crcr]{%
0.012	2.999991e-06\\
0.024	0\\
0.036	1.0999999e-05\\
0.048	1.6999919e-05\\
0.06	1.8999999e-05\\
0.072	2.5999996e-05\\
0.084	4.4999991e-05\\
0.096	4.99999e-05\\
0.108	6.4999975e-05\\
0.12	8.4999831e-05\\
0.132	0.000119999984\\
0.144	0.000131999516\\
0.156	0.000142999271\\
0.168	0.000148998911\\
0.18	0.000204999159\\
0.192	0.000188999711\\
0.204	0.000223998704\\
0.216	0.0002919951\\
0.228	0.000295997696\\
0.24	0.000336987231\\
0.252	0.000336991351\\
0.264	0.00040399\\
0.276	0.000407982576\\
0.288	0.000444974079\\
0.3	0.000501972444\\
0.312	0.000558962751\\
0.324	0.000604955479\\
0.336	0.0006459324\\
0.348	0.0006439271\\
0.36	0.000709912384\\
0.372	0.000674891759\\
0.384	0.000810851775\\
0.396	0.0008458556\\
0.408	0.000917795696\\
0.42	0.000901792064\\
0.432	0.001001721216\\
0.444	0.001064616839\\
0.456	0.001086552439\\
0.468	0.001144477271\\
0.48	0.001210380631\\
0.492	0.001269280896\\
0.504	0.001282248311\\
0.516	0.001379031744\\
0.528	0.001347051324\\
0.54	0.001420033711\\
0.552	0.001450884864\\
0.564	0.001559684391\\
0.576	0.001541729871\\
0.588	0.001644392176\\
0.6	0.001687254959\\
0.612	0.001719125839\\
0.624	0.001774894599\\
0.636	0.001798882975\\
0.648	0.001836740991\\
0.66	0.001932187671\\
0.672	0.001985137136\\
0.684	0.001965130364\\
0.696	0.002000916464\\
0.708	0.00204676\\
0.72	0.002043863559\\
0.732	0.002050752796\\
0.744	0.002170134844\\
0.756	0.002108305916\\
0.768	0.0022039599\\
0.78	0.002245772864\\
0.792	0.002165983984\\
0.804	0.002268458839\\
0.816	0.002334190751\\
0.828	0.002332312775\\
0.84	0.0022235479\\
0.852	0.002241446044\\
0.864	0.002339968951\\
0.876	0.0023289375\\
0.888	0.002351869775\\
0.9	0.002361824375\\
0.912	0.0024444775\\
0.924	0.002361824375\\
0.936	0.002424439836\\
0.948	0.002412411504\\
0.96	0.002409482199\\
0.972	0.002448268764\\
0.984	0.002478124224\\
0.996	0.0024004304\\
1.008	0.002483012191\\
1.02	0.002469090239\\
1.032	0.0024003356\\
1.044	0.002422220784\\
1.056	0.002495889216\\
1.068	0.002476963151\\
1.08	0.002536624375\\
1.092	0.002499789936\\
1.104	0.002538684831\\
1.116	0.002430162944\\
1.128	0.002436114524\\
1.14	0.002465918844\\
1.152	0.002512724975\\
1.164	0.002472904039\\
1.176	0.002523669744\\
1.188	0.0025196496\\
1.2	0.002478874375\\
};
\addlegendentry{$\displaystyle \sigma^2_{\displaystyle \hat J^\epsilon_N}$}

\addplot [color=mycolor3, line width=1pt]
  table[row sep=crcr]{%
0.012	2.9999909769735e-06\\
0.024	1.09581906492948e-13\\
0.036	1.09895562189223e-05\\
0.048	1.69298597222659e-05\\
0.06	1.89671725064546e-05\\
0.072	2.59417624207232e-05\\
0.084	4.47686036175597e-05\\
0.096	4.96417461113453e-05\\
0.108	6.45110499303074e-05\\
0.12	8.40705985899226e-05\\
0.132	0.000118681056227837\\
0.144	0.00012963857437947\\
0.156	0.000140030700143561\\
0.168	0.000145689518934613\\
0.18	0.000199326260612987\\
0.192	0.000184805923806402\\
0.204	0.000217271459815262\\
0.216	0.000279059076271046\\
0.228	0.000283973217824575\\
0.24	0.000316866200196374\\
0.252	0.000318641824430768\\
0.264	0.000378816774147483\\
0.276	0.000379143882418941\\
0.288	0.000407445444840131\\
0.3	0.00045782607727198\\
0.312	0.000502370750599856\\
0.324	0.000537973569180467\\
0.336	0.000566424282555982\\
0.348	0.000561999086098407\\
0.36	0.000609834763064169\\
0.372	0.000572027187146254\\
0.384	0.000671863109932728\\
0.396	0.000691244467323052\\
0.408	0.000735863692582899\\
0.42	0.000718852051742211\\
0.432	0.000774583546187748\\
0.444	0.00078681597365401\\
0.456	0.000785711970903844\\
0.468	0.000789472076917601\\
0.48	0.000820194956580817\\
0.492	0.000825804475211168\\
0.504	0.000814143241680098\\
0.516	0.000835697352681271\\
0.528	0.000809227533218067\\
0.54	0.000840709349001428\\
0.552	0.000814798624459823\\
0.564	0.000831424213762136\\
0.576	0.000802214736447659\\
0.588	0.000806561657788153\\
0.6	0.000802112102729323\\
0.612	0.000761553753815558\\
0.624	0.000758202928786443\\
0.636	0.000751312965535144\\
0.648	0.000735755189172483\\
0.66	0.000647325993409326\\
0.672	0.000678268174627936\\
0.684	0.000654796302427391\\
0.696	0.000625513445629039\\
0.708	0.000576574792860651\\
0.72	0.000577512763493329\\
0.732	0.000568892593951241\\
0.744	0.000510596298923169\\
0.756	0.000497590555374216\\
0.768	0.000461668619202793\\
0.78	0.000451228828234793\\
0.792	0.000392935158538622\\
0.804	0.000387767885347927\\
0.816	0.00035569763128869\\
0.828	0.000340538290460941\\
0.84	0.000295144826038035\\
0.852	0.000285203724667959\\
0.864	0.0002788662576901\\
0.876	0.000235211176040313\\
0.888	0.000223269952202184\\
0.9	0.000206131051595757\\
0.912	0.000211365546407494\\
0.924	0.000172683825839461\\
0.936	0.000177154361071926\\
0.948	0.000133074808739825\\
0.96	0.000142702552647227\\
0.972	0.000134308612124302\\
0.984	0.00011681513544041\\
0.996	0.000108795181663222\\
1.008	9.00141305438551e-05\\
1.02	9.47744356594237e-05\\
1.032	8.19539701929812e-05\\
1.044	6.86401143332562e-05\\
1.056	5.72689443897806e-05\\
1.068	5.52640565989057e-05\\
1.08	4.93471032569059e-05\\
1.092	3.85953677069584e-05\\
1.104	5.14944801837767e-05\\
1.116	4.25570368010837e-05\\
1.128	3.66584823300078e-05\\
1.14	2.77493895340346e-05\\
1.152	3.76220507571044e-05\\
1.164	3.07295749373442e-05\\
1.176	1.99333104968712e-05\\
1.188	1.69227343748228e-05\\
1.2	2.08974108348801e-05\\
};
\addlegendentry{$\displaystyle \sigma^2_{\displaystyle \hat K^\epsilon_N}$}

\end{axis}

\node[] at (2.85,0.65) {(b)};

\end{tikzpicture}%
        \captionlistentry{}
        \label{fig:P2_variances_PSD1}
    \end{subfigure}%
    \setcounter{subfigure}{1}
    \renewcommand{\thesubfigure}{\alph{subfigure}'}
    \begin{subfigure}{.45\textwidth}
        \centering
%
%
\definecolor{mycolor1}{rgb}{0.00000,0.44700,0.74100}%
\definecolor{mycolor2}{rgb}{0.85000,0.32500,0.09800}%
\definecolor{mycolor3}{rgb}{0.92900,0.69400,0.12500}%
\begin{tikzpicture}[scale=0.95]

\begin{axis}[%
width=0.8\linewidth,
height=0.64\linewidth,
scale only axis,
xmin=0,
xmax=1.8,
xtick distance = 0.45,
xlabel = $\epsilon$,
y label style={at={(axis description cs:0.1,.5)},anchor=south},
axis background/.style={fill=white},
legend style={legend cell align=left, align=left, draw=white!15!black, minimum height=0.1cm, font={\tiny\arraycolsep=2pt}, at={(0.99,0.46)},anchor=east}
]
\addplot [color=mycolor1, line width=1pt]
  table[row sep=crcr]{%
0.018	0.002630046231\\
0.036	0.002551456636\\
0.054	0.002564389959\\
0.072	0.002614130359\\
0.09	0.002627062044\\
0.108	0.0025335484\\
0.126	0.002539517884\\
0.144	0.002518624375\\
0.162	0.002636014551\\
0.18	0.002527578844\\
0.198	0.002527578844\\
0.216	0.0025136496\\
0.234	0.002491759996\\
0.252	0.002544492399\\
0.27	0.002588265975\\
0.288	0.002576328111\\
0.306	0.002566379671\\
0.324	0.002555436156\\
0.342	0.002636014551\\
0.36	0.002571353916\\
0.378	0.002659887111\\
0.396	0.002710612476\\
0.414	0.002709617911\\
0.432	0.002694699196\\
0.45	0.0028219911\\
0.468	0.002919426816\\
0.486	0.002922409239\\
0.504	0.002935332864\\
0.522	0.003081445719\\
0.54	0.003056599644\\
0.558	0.003174855775\\
0.576	0.00318976\\
0.594	0.003337784199\\
0.612	0.003403337775\\
0.63	0.003394399164\\
0.648	0.003486756999\\
0.666	0.003639655591\\
0.684	0.003664472316\\
0.702	0.003750824775\\
0.72	0.003746854879\\
0.738	0.003688295196\\
0.756	0.003665464959\\
0.774	0.003647597079\\
0.792	0.003473847804\\
0.81	0.003640648284\\
0.828	0.003509595516\\
0.846	0.003464910471\\
0.864	0.003508602559\\
0.882	0.003360629616\\
0.9	0.003207644476\\
0.918	0.003172868511\\
0.936	0.003081445719\\
0.954	0.002990005999\\
0.972	0.002753376879\\
0.99	0.002715585271\\
1.008	0.002586276351\\
1.026	0.002437031751\\
1.044	0.002341491591\\
1.062	0.002198146791\\
1.08	0.002046793399\\
1.098	0.001985043879\\
1.116	0.001870488124\\
1.134	0.001744944496\\
1.152	0.001658241079\\
1.17	0.0015376284\\
1.188	0.001446900399\\
1.206	0.001313270775\\
1.224	0.001273374375\\
1.242	0.001150672896\\
1.26	0.001093800975\\
1.278	0.000968061039\\
1.296	0.0009091719\\
1.314	0.000781388476\\
1.332	0.000755428464\\
1.35	0.000632599311\\
1.368	0.000615620544\\
1.386	0.000573670524\\
1.404	0.000536711631\\
1.422	0.000422821071\\
1.44	0.000407833536\\
1.458	0.000380854839\\
1.476	0.000343881664\\
1.494	0.000343881664\\
1.512	0.000264929775\\
1.53	0.000241941436\\
1.548	0.000220951159\\
1.566	0.000191963136\\
1.584	0.000178967959\\
1.602	0.000158974719\\
1.62	0.000132982311\\
1.638	0.000114986775\\
1.656	0.000103989184\\
1.674	9.4990975e-05\\
1.692	8.1993276e-05\\
1.71	8.7992256e-05\\
1.728	6.6995511e-05\\
1.746	5.0997399e-05\\
1.764	3.7998556e-05\\
1.782	3.6998631e-05\\
1.8	3.2998911e-05\\
};
\addlegendentry{$\displaystyle \sigma^2_{\displaystyle \hat I^\epsilon_N}$}

\addplot [color=mycolor2, line width=1pt]
  table[row sep=crcr]{%
0.018	6.999991e-06\\
0.036	1.6999975e-05\\
0.054	2.7999996e-05\\
0.072	4.3999996e-05\\
0.09	5.2999951e-05\\
0.108	7.7999984e-05\\
0.126	9.6999879e-05\\
0.144	0.000114999831\\
0.162	0.000127999804\\
0.18	0.000159999804\\
0.198	0.000214999639\\
0.216	0.000224999831\\
0.234	0.000269999984\\
0.252	0.000280999991\\
0.27	0.000344997599\\
0.288	0.0004159964\\
0.306	0.000460987679\\
0.324	0.000509996156\\
0.342	0.000620988119\\
0.36	0.000673983616\\
0.378	0.000717949824\\
0.396	0.000793924924\\
0.414	0.000834935991\\
0.432	0.000888925471\\
0.45	0.001045867504\\
0.468	0.001154860871\\
0.486	0.001256814239\\
0.504	0.001331767676\\
0.522	0.001465647164\\
0.54	0.001505690864\\
0.558	0.001708462711\\
0.576	0.001823394716\\
0.594	0.001944316071\\
0.612	0.002087317724\\
0.63	0.002186216775\\
0.648	0.002296021879\\
0.666	0.002447977879\\
0.684	0.002615716311\\
0.702	0.002736588656\\
0.72	0.002856397244\\
0.738	0.002919661351\\
0.756	0.002999571975\\
0.774	0.003132794396\\
0.792	0.003171997999\\
0.81	0.003266776764\\
0.828	0.0033070784\\
0.846	0.003375949375\\
0.864	0.003519168256\\
0.882	0.003404292719\\
0.9	0.003355472924\\
0.918	0.003499582684\\
0.936	0.003484636391\\
0.954	0.003526772471\\
0.972	0.003378843975\\
0.99	0.003516991719\\
1.008	0.003484991351\\
1.026	0.003424999039\\
1.044	0.003481949824\\
1.062	0.003469857116\\
1.08	0.003409761856\\
1.098	0.003403679644\\
1.116	0.003215695296\\
1.134	0.003279372736\\
1.152	0.003291200764\\
1.17	0.003233932911\\
1.188	0.003220837916\\
1.206	0.0031465116\\
1.224	0.003187333319\\
1.242	0.003082166684\\
1.26	0.002997098359\\
1.278	0.003033522524\\
1.296	0.003022267591\\
1.314	0.002917926991\\
1.332	0.002818180959\\
1.35	0.002790584896\\
1.368	0.0027985775\\
1.386	0.002723529231\\
1.404	0.002848866911\\
1.422	0.002708768751\\
1.44	0.002758458839\\
1.458	0.002737351664\\
1.476	0.002688398975\\
1.494	0.002730217031\\
1.512	0.002677950991\\
1.53	0.002635887879\\
1.548	0.002622719196\\
1.566	0.002534973436\\
1.584	0.002597603671\\
1.602	0.002604458684\\
1.62	0.002615235199\\
1.638	0.002557330839\\
1.656	0.002550316544\\
1.674	0.002567187079\\
1.692	0.002625789936\\
1.71	0.002551177431\\
1.728	0.002437585071\\
1.746	0.002496172604\\
1.764	0.002483109671\\
1.782	0.002568684831\\
1.8	0.002622348759\\
};
\addlegendentry{$\displaystyle \sigma^2_{\displaystyle \hat J^\epsilon_N}$}

\addplot [color=mycolor3, line width=1pt]
  table[row sep=crcr]{%
0.018	6.99052630392936e-06\\
0.036	1.69857776327274e-05\\
0.054	2.79339900685984e-05\\
0.072	4.38310847342101e-05\\
0.09	5.27977272502743e-05\\
0.108	7.74584882280243e-05\\
0.126	9.62676880293383e-05\\
0.144	0.000113383523193751\\
0.162	0.000126759114603619\\
0.18	0.000157877729339896\\
0.198	0.000211167588615043\\
0.216	0.000220503952595811\\
0.234	0.000262888561789129\\
0.252	0.000273396974443822\\
0.27	0.000336357792647124\\
0.288	0.000403385146932166\\
0.306	0.000448478928238298\\
0.324	0.000489847599003709\\
0.342	0.000595003768831523\\
0.36	0.000643418804127951\\
0.378	0.000692804631754414\\
0.396	0.000766054570975131\\
0.414	0.000800336044085405\\
0.432	0.00084960686213459\\
0.45	0.000998348056149715\\
0.468	0.00109457880100879\\
0.486	0.00118805827870196\\
0.504	0.00125781017038734\\
0.522	0.00138880668860067\\
0.54	0.0014150439260098\\
0.558	0.00161038314541348\\
0.576	0.00170936811213645\\
0.594	0.00181917208514394\\
0.612	0.00193208214666465\\
0.63	0.00201652645082243\\
0.648	0.00212388897745903\\
0.666	0.00225032884630771\\
0.684	0.00239714377425719\\
0.702	0.00250106527435832\\
0.72	0.00259978321250259\\
0.738	0.00261116428761496\\
0.756	0.00266832294244265\\
0.774	0.00272517081249512\\
0.792	0.00269402754423551\\
0.81	0.0028045858960657\\
0.828	0.00276546557240149\\
0.846	0.00280812411549309\\
0.864	0.00286332640276299\\
0.882	0.00275112386833151\\
0.9	0.00265791339854815\\
0.918	0.00269303939565712\\
0.936	0.00264621317070248\\
0.954	0.00260084665210616\\
0.972	0.00243457533409131\\
0.99	0.00239885515686798\\
1.008	0.00233122505764488\\
1.026	0.00221516721198333\\
1.044	0.00214382454418509\\
1.062	0.00203336270728388\\
1.08	0.0019117588247028\\
1.098	0.00185981885455892\\
1.116	0.0017511316014212\\
1.134	0.00164644222403646\\
1.152	0.00157601899049331\\
1.17	0.0014638833894114\\
1.188	0.00139160374161591\\
1.206	0.0012654683422476\\
1.224	0.00123268009068636\\
1.242	0.00111838118015373\\
1.26	0.00106151082573393\\
1.278	0.00094618979769368\\
1.296	0.00088995113662818\\
1.314	0.000766192988468373\\
1.332	0.000741617721674521\\
1.35	0.00062245729455594\\
1.368	0.000607824769802981\\
1.386	0.000565538812577396\\
1.404	0.000530555591383474\\
1.422	0.000419253035625156\\
1.44	0.000404566293916355\\
1.458	0.000377861508524146\\
1.476	0.000341956074556283\\
1.494	0.000341988739822102\\
1.512	0.000264072977338183\\
1.53	0.000240883505779331\\
1.548	0.000219727767779391\\
1.566	0.000191218089941885\\
1.584	0.000178372437168733\\
1.602	0.000158601163262874\\
1.62	0.000132779108408299\\
1.638	0.000114760929048136\\
1.656	0.00010386236047247\\
1.674	9.49152102263245e-05\\
1.692	8.19391432989073e-05\\
1.71	8.78924439459316e-05\\
1.728	6.69627933407967e-05\\
1.746	5.09781809198217e-05\\
1.764	3.79791583291356e-05\\
1.782	3.69849188961263e-05\\
1.8	3.29748710661344e-05\\
};
\addlegendentry{$\displaystyle \sigma^2_{\displaystyle \hat K^\epsilon_N}$}

\end{axis}

\node[] at (2.85,0.65) {(b')};

\end{tikzpicture}%
        \captionlistentry{}
        \label{fig:P2_variances_PSD2}
    \end{subfigure}%

    \centering
    \renewcommand{\thesubfigure}{\alph{subfigure}}
    \begin{subfigure}{.45\textwidth}
        \centering
%
%
\definecolor{mycolor1}{rgb}{0.00000,0.44700,0.74100}%
\definecolor{mycolor2}{rgb}{0.85000,0.32500,0.09800}%
\definecolor{mycolor3}{rgb}{0.92900,0.69400,0.12500}%
\begin{tikzpicture}[scale=0.95]

\begin{axis}[%
width=0.8\linewidth,
height=0.64\linewidth,
scale only axis,
xmin=0,
xmax=1.2,
xtick distance = 0.3,
xlabel = $\epsilon$,
y label style={at={(axis description cs:0.1,.5)},anchor=south},
axis background/.style={fill=white},
legend style={legend cell align=left, align=left, draw=white!15!black, minimum height=0.6cm, font={\tiny\arraycolsep=2pt}}
]
\addplot [color=black, line width=1pt]
  table[row sep=crcr]{%
0.036	0.000305265450525619\\
0.048	0.000352705410880539\\
0.06	0.000316119541774243\\
0.072	0.000360302255843377\\
0.084	0.000532959566875711\\
0.096	0.00051710152199318\\
0.108	0.000597324536391735\\
0.12	0.000700588321582688\\
0.132	0.000899098910816946\\
0.144	0.00090026787763521\\
0.156	0.000897632693227952\\
0.168	0.000867199517467933\\
0.18	0.0011073681145166\\
0.192	0.000962530853158344\\
0.204	0.00106505617556501\\
0.216	0.00129194016792151\\
0.228	0.00124549656940603\\
0.24	0.00132027583415156\\
0.252	0.00126445168424908\\
0.264	0.00143491202328592\\
0.276	0.00137370971890921\\
0.288	0.00141474112791712\\
0.3	0.00152608692423993\\
0.312	0.00161016266217903\\
0.324	0.001660412250557\\
0.336	0.00168578655522614\\
0.348	0.00161493990258163\\
0.36	0.00169398545295603\\
0.372	0.00153770749232864\\
0.384	0.00174964351544981\\
0.396	0.00174556683667437\\
0.408	0.00180358748182083\\
0.42	0.00171155250414812\\
0.432	0.00179301746802719\\
0.444	0.00177210804877029\\
0.456	0.00172305256777159\\
0.468	0.00168690614725983\\
0.48	0.0017087394928767\\
0.492	0.0016784643805105\\
0.504	0.0016153635747621\\
0.516	0.0016195685129482\\
0.528	0.0015326279038221\\
0.54	0.00155686916481746\\
0.552	0.00147608446460113\\
0.564	0.00147415640737967\\
0.576	0.00139273391744385\\
0.588	0.00137170349963971\\
0.6	0.00133685350454887\\
0.612	0.00124436887878359\\
0.624	0.00121506879613212\\
0.636	0.00118130969423765\\
0.648	0.00113542467464889\\
0.66	0.0009807969597111\\
0.672	0.00100932764081538\\
0.684	0.00095730453571256\\
0.696	0.000898726214984252\\
0.708	0.000814371176356852\\
0.72	0.000802101060407401\\
0.732	0.000777175674796777\\
0.744	0.000686285348015012\\
0.756	0.000658188565309809\\
0.768	0.00060113101458697\\
0.78	0.000578498497736914\\
0.792	0.000496130250680078\\
0.804	0.000482298364860605\\
0.816	0.000435903959912611\\
0.828	0.000411278128576015\\
0.84	0.000351362888140518\\
0.852	0.000334746155713567\\
0.864	0.000322761872326505\\
0.876	0.000268505908721818\\
0.888	0.000251430126353811\\
0.9	0.000229034501773063\\
0.912	0.000231760467552077\\
0.924	0.000186887257402015\\
0.936	0.000189267479777698\\
0.948	0.000140374270822601\\
0.96	0.000148648492340861\\
0.972	0.000138177584490023\\
0.984	0.000118714568537002\\
0.996	0.000109232110103637\\
1.008	8.92997326823959e-05\\
1.02	9.29161133915918e-05\\
1.032	7.94127618149043e-05\\
1.044	6.57472359513948e-05\\
1.056	5.4231954914565e-05\\
1.068	5.17453713472899e-05\\
1.08	4.56917622749129e-05\\
1.092	3.53437433213905e-05\\
1.104	4.6643550891102e-05\\
1.116	3.81335455206843e-05\\
1.128	3.24986545478793e-05\\
1.14	2.43415697666971e-05\\
1.152	3.26580301710976e-05\\
1.164	2.63999784685088e-05\\
1.176	1.69500939599245e-05\\
1.188	1.4244725904733e-05\\
1.2	1.74145090290668e-05\\
};
\addlegendentry{$\displaystyle \sigma^2_{\displaystyle \hat K^\epsilon_N}/\epsilon$}

\addplot [color=black,dashed, line width=0.5pt]
  table[row sep=crcr]{%
0.036	0.00847959584793386\\
0.048	0.00734802939334456\\
0.06	0.00526865902957072\\
0.072	0.00500419799782468\\
0.084	0.00634475674852037\\
0.096	0.00538647418742896\\
0.108	0.00553078274436792\\
0.12	0.00583823601318907\\
0.132	0.00681135538497686\\
0.144	0.00625186026135562\\
0.156	0.00575405572582021\\
0.168	0.00516190188969008\\
0.18	0.00615204508064776\\
0.192	0.00501318152686637\\
0.204	0.00522086360571083\\
0.216	0.0059812044811181\\
0.228	0.00546270425178084\\
0.24	0.00550114930896482\\
0.252	0.00501766541368683\\
0.264	0.00543527281547697\\
0.276	0.00497720912648264\\
0.288	0.00491229558304556\\
0.3	0.00508695641413311\\
0.312	0.00516077776339432\\
0.324	0.00512472916838579\\
0.336	0.00501722189055398\\
0.348	0.0046406319039702\\
0.36	0.00470551514710007\\
0.372	0.00413362229120602\\
0.384	0.00455636332148389\\
0.396	0.00440799706230902\\
0.408	0.00442055755348243\\
0.42	0.00407512500987648\\
0.432	0.0041505033982111\\
0.444	0.00399123434407724\\
0.456	0.00377862405213067\\
0.468	0.00360450031465776\\
0.48	0.00355987394349313\\
0.492	0.00341151296851728\\
0.504	0.00320508645786131\\
0.516	0.00313869866850426\\
0.528	0.00290270436329942\\
0.54	0.00288309104595826\\
0.552	0.00267406605906002\\
0.564	0.00261375249535403\\
0.576	0.00241794082889558\\
0.588	0.00233282908101992\\
0.6	0.00222808917424812\\
0.612	0.00203328248167254\\
0.624	0.00194722563482712\\
0.636	0.00185740517961894\\
0.648	0.00175219857198903\\
0.66	0.00148605599956227\\
0.672	0.00150197565597527\\
0.684	0.00139956803466749\\
0.696	0.00129127329739117\\
0.708	0.00115024177451533\\
0.72	0.00111402925056583\\
0.732	0.00106171540272784\\
0.744	0.000922426543030931\\
0.756	0.000870619795383345\\
0.768	0.00078272267524345\\
0.78	0.000741664740688351\\
0.792	0.000626427084192018\\
0.804	0.000599873588135081\\
0.816	0.00053419602930467\\
0.828	0.000496712715671516\\
0.84	0.000418289152548236\\
0.852	0.000392894548959585\\
0.864	0.00037356698185938\\
0.876	0.0003065135944313\\
0.888	0.00028314203418222\\
0.9	0.000254482779747848\\
0.912	0.000254123319684295\\
0.924	0.000202258936582267\\
0.936	0.000202208845916345\\
0.948	0.000148074125340296\\
0.96	0.000154842179521731\\
0.972	0.000142158008734592\\
0.984	0.000120644886724595\\
0.996	0.000109670793276744\\
1.008	8.85910046452341e-05\\
1.02	9.10942288152861e-05\\
1.032	7.69503505958375e-05\\
1.044	6.29762796469299e-05\\
1.056	5.13560179115199e-05\\
1.068	4.84507222352902e-05\\
1.08	4.2307187291586e-05\\
1.092	3.23660653126286e-05\\
1.104	4.2249593198462e-05\\
1.116	3.41698436565272e-05\\
1.128	2.88108639608859e-05\\
1.14	2.13522541813132e-05\\
1.152	2.83489845235222e-05\\
1.164	2.26803938732893e-05\\
1.176	1.44133452040174e-05\\
1.188	1.1990510020819e-05\\
1.2	1.45120908575557e-05\\
};
\addlegendentry{$\displaystyle \sigma^2_{\displaystyle \hat K^\epsilon_N}/\epsilon^2$}

\end{axis}

\node[] at (2.85,0.65) {(c)};

\end{tikzpicture}%
        \captionlistentry{}
        \label{fig:P2_bounds_PSD1}
    \end{subfigure}%
    \setcounter{subfigure}{2}
    \renewcommand{\thesubfigure}{\alph{subfigure}'}
    \begin{subfigure}{.45\textwidth}
        \centering
%
%
\definecolor{mycolor1}{rgb}{0.00000,0.44700,0.74100}%
\definecolor{mycolor2}{rgb}{0.85000,0.32500,0.09800}%
\definecolor{mycolor3}{rgb}{0.92900,0.69400,0.12500}%
\begin{tikzpicture}[scale=0.95]

\begin{axis}[%
width=0.8\linewidth,
height=0.64\linewidth,
scale only axis,
xmin=0,
xmax=1.8,
xtick distance = 0.45,
xlabel = $\epsilon$,
y label style={at={(axis description cs:0.1,.5)},anchor=south},
axis background/.style={fill=white},
legend style={legend cell align=left, align=left, draw=white!15!black, minimum height=0.6cm, font={\tiny\arraycolsep=2pt}}
]
\addplot [color=black, line width=1pt]
  table[row sep=crcr]{%
0.036	0.00047182715646465\\
0.054	0.000517296112381451\\
0.072	0.000608765065752918\\
0.09	0.000586641413891937\\
0.108	0.000717208224333558\\
0.126	0.000764029270074113\\
0.144	0.000787385577734383\\
0.162	0.00078246367039271\\
0.18	0.000877098496332754\\
0.198	0.00106650297280325\\
0.216	0.00102085163238801\\
0.234	0.00112345539226123\\
0.252	0.00108490862874533\\
0.27	0.00124576960239676\\
0.288	0.00140064287129224\\
0.306	0.00146561741254346\\
0.324	0.001511875305567\\
0.342	0.00173977710184656\\
0.36	0.00178727445591097\\
0.378	0.00183281648612279\\
0.396	0.00193448123983619\\
0.414	0.00193317885044784\\
0.432	0.00196668255123748\\
0.45	0.00221855123588826\\
0.468	0.00233884359189913\\
0.486	0.00244456435946905\\
0.504	0.00249565509997488\\
0.522	0.00266054921187869\\
0.54	0.00262045171483296\\
0.558	0.00288599130002416\\
0.576	0.00296765297245912\\
0.594	0.00306257926791909\\
0.612	0.00315699697167426\\
0.63	0.00320083563622608\\
0.648	0.00327760644669603\\
0.666	0.00337887214160316\\
0.684	0.00350459616119472\\
0.702	0.00356277104609447\\
0.72	0.00361081001736471\\
0.738	0.00353816299134818\\
0.756	0.00352952770164371\\
0.774	0.00352089252260351\\
0.792	0.00340154992959029\\
0.81	0.0034624517235379\\
0.828	0.00333993426618537\\
0.846	0.00331929564479089\\
0.864	0.00331403518838309\\
0.882	0.00311918805933278\\
0.9	0.00295323710949794\\
0.918	0.00293359411291625\\
0.936	0.00282715082340008\\
0.954	0.00272625435231254\\
0.972	0.00250470713383879\\
0.99	0.00242308601703837\\
1.008	0.00231272327147309\\
1.026	0.00215903237035412\\
1.044	0.0020534717856179\\
1.062	0.00191465414998482\\
1.08	0.00177014705991\\
1.098	0.00169382409340521\\
1.116	0.00156911433819104\\
1.134	0.00145188908645191\\
1.152	0.00136807204035878\\
1.17	0.00125118238411231\\
1.188	0.00117138362088881\\
1.206	0.00104931039987364\\
1.224	0.00100709157735813\\
1.242	0.000900467938932151\\
1.26	0.000842468909312641\\
1.278	0.00074036760382917\\
1.296	0.000686690691842731\\
1.314	0.000583099686810025\\
1.332	0.000556770061317208\\
1.35	0.000461079477448844\\
1.368	0.000444316352195161\\
1.386	0.000408036661311252\\
1.404	0.000377888597851477\\
1.422	0.000294833358386186\\
1.44	0.000280948815219691\\
1.458	0.000259164271964435\\
1.476	0.000231677557287455\\
1.494	0.000228908125717605\\
1.512	0.000174651440038481\\
1.53	0.000157440199855772\\
1.548	0.000141943002441467\\
1.566	0.000122106059988432\\
1.584	0.000112608861848948\\
1.602	9.90019745710825e-05\\
1.62	8.19624125977156e-05\\
1.638	7.00616172455041e-05\\
1.656	6.2718816710429e-05\\
1.674	5.66996476859764e-05\\
1.692	4.84273896565646e-05\\
1.71	5.13990900268606e-05\\
1.728	3.87516165166647e-05\\
1.746	2.91971253836321e-05\\
1.764	2.15301351072197e-05\\
1.782	2.07547244086006e-05\\
1.8	1.83193728145191e-05\\
};
\addlegendentry{$\displaystyle \sigma^2_{\displaystyle \hat K^\epsilon_N}/\epsilon$}

\addplot [color=black, dashed, line width=0.5pt]
  table[row sep=crcr]{%
0.036	0.0131063099017958\\
0.054	0.00957955763669354\\
0.072	0.00845507035767941\\
0.09	0.00651823793213263\\
0.108	0.00664081689197739\\
0.126	0.00606372436566756\\
0.144	0.00546795540093321\\
0.162	0.00483002265674512\\
0.18	0.00487276942407085\\
0.198	0.00538637865052144\\
0.216	0.00472616496475932\\
0.234	0.00480109141991981\\
0.252	0.00430519297121161\\
0.27	0.00461396149035836\\
0.288	0.00486334330309807\\
0.306	0.00478959938739692\\
0.324	0.00466628180730556\\
0.342	0.00508706754925894\\
0.36	0.00496465126641937\\
0.378	0.0048487208627587\\
0.396	0.00488505363594997\\
0.414	0.00466951413151652\\
0.432	0.00455250590564231\\
0.45	0.00493011385752946\\
0.468	0.0049975290425195\\
0.486	0.00502996781783756\\
0.504	0.00495169662693429\\
0.522	0.00509683757064883\\
0.54	0.00485268836080178\\
0.558	0.00517202741939814\\
0.576	0.00515217529940819\\
0.594	0.00515585735339915\\
0.612	0.00515849178378147\\
0.63	0.00508069148607314\\
0.648	0.005058034639963\\
0.666	0.00507338159399874\\
0.684	0.0051236785982379\\
0.702	0.005075172430334\\
0.72	0.00501501391300654\\
0.738	0.00479425879586474\\
0.756	0.00466868743603666\\
0.774	0.00454895674754975\\
0.792	0.00429488627473521\\
0.81	0.0042746317574542\\
0.828	0.00403373703645576\\
0.846	0.00392351731062753\\
0.864	0.00383568887544339\\
0.882	0.00353649439833648\\
0.9	0.00328137456610883\\
0.918	0.0031956362885798\\
0.936	0.00302046028141034\\
0.954	0.0028577089646882\\
0.972	0.00257685919119217\\
0.99	0.00244756163337209\\
1.008	0.0022943683248741\\
1.026	0.00210432004907809\\
1.044	0.0019669269977183\\
1.062	0.00180287584744334\\
1.08	0.00163902505547222\\
1.098	0.00154264489381166\\
1.116	0.00140601643207082\\
1.134	0.00128032547306165\\
1.152	0.00118756253503366\\
1.17	0.00106938665308744\\
1.188	0.000986013148896304\\
1.206	0.00087007495843585\\
1.224	0.000822787236403705\\
1.242	0.000725014443584663\\
1.26	0.000668626118502096\\
1.278	0.00057931737388824\\
1.296	0.000529853928890996\\
1.314	0.00044375927458906\\
1.332	0.000417995541529435\\
1.35	0.00034154035366581\\
1.368	0.000324792655113422\\
1.386	0.000294398745534814\\
1.404	0.000269151422971138\\
1.422	0.000207337101537402\\
1.44	0.000195103343902563\\
1.458	0.000177753272952287\\
1.476	0.000156963114693397\\
1.494	0.000153218290306295\\
1.512	0.000115510211665662\\
1.53	0.000102902091409001\\
1.548	9.16944460216197e-05\\
1.566	7.79732183834178e-05\\
1.584	7.10914531874669e-05\\
1.602	6.17989853752075e-05\\
1.62	5.05940818504417e-05\\
1.638	4.27726601010404e-05\\
1.656	3.78736815884233e-05\\
1.674	3.38707572795558e-05\\
1.692	2.86213886859129e-05\\
1.71	3.0057947384129e-05\\
1.728	2.24257040026995e-05\\
1.746	1.67222940341535e-05\\
1.764	1.22052920108955e-05\\
1.782	1.16468711608309e-05\\
1.8	1.01774293413995e-05\\
};
\addlegendentry{$\displaystyle \sigma^2_{\displaystyle \hat K^\epsilon_N}/\epsilon^2$}

\end{axis}

\node[] at (2.85,0.65) {(c')};

\end{tikzpicture}%
        \captionlistentry{}
        \label{fig:P2_bounds_PSD2}
    \end{subfigure}%

\caption{Monte Carlo estimation of $P^\epsilon_2$ (for $a=0.5$ and $\Delta_f=1$) using $\hat I_N^\eps$ (standard MC), $\hat J_N^\eps$ (simple control variate) and $K_N^\eps$ (practical optimal control variate).
    For $\hat J_N^\eps$ and $K_N^\eps$, the expectation of the control variate $P^0_2$ is obtained by solving the PDE \eqref{eq:pdeBEPO2}. This is done numerically by finite differences shown in Section \ref{sec:fdm_pde}. The driving coloured noise is related to PSD1 for the left column (a)-(b)-(c), and PSD2 for the right column (a')-(b')-(c'). The time discretization of the KBEPO dynamics together with the driving noise (coloured or white) is given in \eqref{eq:OU_EM_BEPO}-\eqref{eq:colorednoise_EM_BEPO}-\eqref{eq:whitenoise_EM_BEPO} in the appendix.}
\label{fig:P2_PSD1_PSD2}
\end{figure}

\begin{figure}
    \centering
    \begin{subfigure}{0.45\textwidth}
        \centering
%
%
\definecolor{mycolor1}{rgb}{0.00000,0.44700,0.74100}%
\definecolor{mycolor2}{rgb}{0.85000,0.32500,0.09800}%
\definecolor{mycolor3}{rgb}{0.92900,0.69400,0.12500}%
\begin{tikzpicture}[scale=0.95]

\begin{axis}[%
width=0.8\linewidth,
height=0.64\linewidth,
scale only axis,
xmin=0,
xmax=1.2,
xtick distance = 0.3,
xlabel = $\epsilon$,
y label style={at={(axis description cs:0.1,.5)},anchor=south},
axis background/.style={fill=white},
legend style={legend cell align=left, align=left, draw=white!15!black, minimum height=0.5cm, font={\tiny\arraycolsep=2pt}},
]
\addplot [color=mycolor1, line width=1pt]
  table[row sep=crcr]{%
0.012	0.00459\\
0.024	0.00421\\
0.036	0.00466\\
0.048	0.0043\\
0.06	0.00461\\
0.072	0.00418\\
0.084	0.00472\\
0.096	0.0046\\
0.108	0.00469\\
0.12	0.0046\\
0.132	0.00426\\
0.144	0.00467\\
0.156	0.00419\\
0.168	0.00441\\
0.18	0.00492\\
0.192	0.00431\\
0.204	0.0046\\
0.216	0.005\\
0.228	0.00461\\
0.24	0.00454\\
0.252	0.00449\\
0.264	0.00425\\
0.276	0.00432\\
0.288	0.00456\\
0.3	0.00453\\
0.312	0.00428\\
0.324	0.00446\\
0.336	0.00432\\
0.348	0.00435\\
0.36	0.00423\\
0.372	0.00436\\
0.384	0.00363\\
0.396	0.00393\\
0.408	0.00367\\
0.42	0.00376\\
0.432	0.00349\\
0.444	0.00345\\
0.456	0.00352\\
0.468	0.00325\\
0.48	0.00355\\
0.492	0.00348\\
0.504	0.00317\\
0.516	0.00311\\
0.528	0.003\\
0.54	0.00293\\
0.552	0.0029\\
0.564	0.0027\\
0.576	0.0028\\
0.588	0.00285\\
0.6	0.00254\\
0.612	0.00215\\
0.624	0.0025\\
0.636	0.00226\\
0.648	0.00195\\
0.66	0.00197\\
0.672	0.00219\\
0.684	0.0017\\
0.696	0.00162\\
0.708	0.00176\\
0.72	0.00154\\
0.732	0.0015\\
0.744	0.00127\\
0.756	0.00137\\
0.768	0.00129\\
0.78	0.00132\\
0.792	0.00107\\
0.804	0.00118\\
0.816	0.00094\\
0.828	0.00088\\
0.84	0.00085\\
0.852	0.0008\\
0.864	0.00086\\
0.876	0.0006\\
0.888	0.00087\\
0.9	0.00067\\
0.912	0.00053\\
0.924	0.0005\\
0.936	0.00048\\
0.948	0.00062\\
0.96	0.00048\\
0.972	0.00036\\
0.984	0.0004\\
0.996	0.00029\\
1.008	0.00032\\
1.02	0.00031\\
1.032	0.00026\\
1.044	0.00026\\
1.056	0.00028\\
1.068	0.0002\\
1.08	0.00028\\
1.092	0.00013\\
1.104	9e-05\\
1.116	0.00011\\
1.128	0.0001\\
1.14	0.00016\\
1.152	7e-05\\
1.164	0.00011\\
1.176	0.00011\\
1.188	0.00018\\
1.2	7e-05\\
};
\addlegendentry{$\hat I^\epsilon_N$}

\addplot [color=mycolor2, line width=1pt]
  table[row sep=crcr]{%
0.012	0.004593570119343\\
0.024	0.004583570119343\\
0.036	0.004563570119343\\
0.048	0.004583570119343\\
0.06	0.004583570119343\\
0.072	0.004623570119343\\
0.084	0.004553570119343\\
0.096	0.004563570119343\\
0.108	0.004503570119343\\
0.12	0.004553570119343\\
0.132	0.004553570119343\\
0.144	0.004533570119343\\
0.156	0.004503570119343\\
0.168	0.004563570119343\\
0.18	0.004513570119343\\
0.192	0.004423570119343\\
0.204	0.004433570119343\\
0.216	0.004403570119343\\
0.228	0.004503570119343\\
0.24	0.004493570119343\\
0.252	0.004303570119343\\
0.264	0.004373570119343\\
0.276	0.004533570119343\\
0.288	0.004593570119343\\
0.3	0.004273570119343\\
0.312	0.004233570119343\\
0.324	0.004233570119343\\
0.336	0.004293570119343\\
0.348	0.004363570119343\\
0.36	0.004093570119343\\
0.372	0.004253570119343\\
0.384	0.003873570119343\\
0.396	0.003733570119343\\
0.408	0.003823570119343\\
0.42	0.003793570119343\\
0.432	0.003713570119343\\
0.444	0.003553570119343\\
0.456	0.003413570119343\\
0.468	0.003313570119343\\
0.48	0.003433570119343\\
0.492	0.003353570119343\\
0.504	0.003163570119343\\
0.516	0.003263570119343\\
0.528	0.003073570119343\\
0.54	0.003233570119343\\
0.552	0.002923570119343\\
0.564	0.002513570119343\\
0.576	0.002703570119343\\
0.588	0.002613570119343\\
0.6	0.002713570119343\\
0.612	0.002263570119343\\
0.624	0.002243570119343\\
0.636	0.002093570119343\\
0.648	0.002303570119343\\
0.66	0.001643570119343\\
0.672	0.002093570119343\\
0.684	0.001943570119343\\
0.696	0.001553570119343\\
0.708	0.001593570119343\\
0.72	0.001443570119343\\
0.732	0.001343570119343\\
0.744	0.001103570119343\\
0.756	0.001163570119343\\
0.768	0.001363570119343\\
0.78	0.001603570119343\\
0.792	0.001273570119343\\
0.804	0.001373570119343\\
0.816	0.001023570119343\\
0.828	0.000783570119343\\
0.84	0.000673570119343\\
0.852	0.001073570119343\\
0.864	0.000903570119343\\
0.876	0.000673570119343\\
0.888	0.000683570119343\\
0.9	0.000273570119343001\\
0.912	0.000923570119343\\
0.924	0.000583570119343\\
0.936	0.000263570119343\\
0.948	0.000353570119343\\
0.96	0.000343570119343\\
0.972	0.000403570119343001\\
0.984	0.000163570119343\\
0.996	0.000263570119343\\
1.008	0.000383570119343001\\
1.02	0.000263570119343\\
1.032	2.35701193430004e-05\\
1.044	0.000603570119343\\
1.056	2.35701193430004e-05\\
1.068	0.000653570119343\\
1.08	0.000423570119343001\\
1.092	-0.000236429880656999\\
1.104	0.000123570119343\\
1.116	2.35701193430004e-05\\
1.128	0.000103570119343001\\
1.14	-9.64298806569999e-05\\
1.152	8.35701193430006e-05\\
1.164	-7.64298806569998e-05\\
1.176	0.000253570119343001\\
1.188	-0.000266429880657\\
1.2	-0.000456429880657\\
};
\addlegendentry{$\hat J^\epsilon_N$}

\addplot [color=mycolor3, line width=1.0pt]
  table[row sep=crcr]{%
0.012	0.00459357008346756\\
0.024	0.00458355768416288\\
0.036	0.00456398240452509\\
0.048	0.00458224019184342\\
0.06	0.00458368531765123\\
0.072	0.00462245530802018\\
0.084	0.00455532632216016\\
0.096	0.00456404470901134\\
0.108	0.00450905442004693\\
0.12	0.00455477770486041\\
0.132	0.00454876231146919\\
0.144	0.0045376296297373\\
0.156	0.00449178824699171\\
0.168	0.00455974218228112\\
0.18	0.00453238161101781\\
0.192	0.00441720774501739\\
0.204	0.00444305133063215\\
0.216	0.00443590846573337\\
0.228	0.00451108576185537\\
0.24	0.00449638685700924\\
0.252	0.00432276175955115\\
0.264	0.0043619070205045\\
0.276	0.00451738026734204\\
0.288	0.00459067918348414\\
0.3	0.00430437197301076\\
0.312	0.00424020311247778\\
0.324	0.00426613117935117\\
0.336	0.0042974214317467\\
0.348	0.00436154466171473\\
0.36	0.00412109301016186\\
0.372	0.00427017895482084\\
0.384	0.00381850177639295\\
0.396	0.00378377915654995\\
0.408	0.00378573272097267\\
0.42	0.00378507321718393\\
0.432	0.00365093106274998\\
0.444	0.00352070134627864\\
0.456	0.00345015296931733\\
0.468	0.0032907637513394\\
0.48	0.00347251200996238\\
0.492	0.00340034175545054\\
0.504	0.00316598151142788\\
0.516	0.0032045788973081\\
0.528	0.0030440219715172\\
0.54	0.00310932099524273\\
0.552	0.00291317236989896\\
0.564	0.00260586297685777\\
0.576	0.0027497595704145\\
0.588	0.00272548341945212\\
0.6	0.00262455178060784\\
0.612	0.00219416160200772\\
0.624	0.00238618177317203\\
0.636	0.00219630502915249\\
0.648	0.0020928431446732\\
0.66	0.00186174715221938\\
0.672	0.00215253541246321\\
0.684	0.00177793287894089\\
0.696	0.00160044488939941\\
0.708	0.00171206361693571\\
0.72	0.00151220215158939\\
0.732	0.00145679735659532\\
0.744	0.00123289519269608\\
0.756	0.00132005263556182\\
0.768	0.00130563575146137\\
0.78	0.00138916567275606\\
0.792	0.00111039388369288\\
0.804	0.00122140079939258\\
0.816	0.000954285482564871\\
0.828	0.000865801210175967\\
0.84	0.000824826033274077\\
0.852	0.000839303333034925\\
0.864	0.000866036004463269\\
0.876	0.000608968259790556\\
0.888	0.000841495236301594\\
0.9	0.000627866785569331\\
0.912	0.000573178906595176\\
0.924	0.000506487295874654\\
0.936	0.000462434817075133\\
0.948	0.00058814945988599\\
0.96	0.000468449157623973\\
0.972	0.00036306926958527\\
0.984	0.000383829710677984\\
0.996	0.000288798160171103\\
1.008	0.000323511573156753\\
1.02	0.000307697280790328\\
1.032	0.000250204000165132\\
1.044	0.000276184983217215\\
1.056	0.000269956865785903\\
1.068	0.000216449895651135\\
1.08	0.00028742994569937\\
1.092	0.000122608625221865\\
1.104	9.04416710756947e-05\\
1.116	0.000108707952656133\\
1.128	0.000100054457546273\\
1.14	0.000153652628100238\\
1.152	7.02078573994656e-05\\
1.164	0.000106486227275654\\
1.176	0.000111932917213751\\
1.188	0.000171160901397129\\
1.2	6.38244552063899e-05\\
};
\addlegendentry{$\hat K^\epsilon_N$}

\addplot [color=black, dashed, line width=1.0pt]
  table[row sep=crcr]{%
0	0.004583570119343\\
0.133333333333333	0.004583570119343\\
0.266666666666667	0.004583570119343\\
0.4	0.004583570119343\\
0.533333333333333	0.004583570119343\\
0.666666666666667	0.004583570119343\\
0.8	0.004583570119343\\
0.933333333333333	0.004583570119343\\
1.06666666666667	0.004583570119343\\
1.2	0.004583570119343\\
};
\addlegendentry{$P^0_1$}

\end{axis}

\node[] at (2.85,0.65) {(a)};

\end{tikzpicture}%
        \captionlistentry{}
        \label{fig:a0_estimates_PSD1}
    \end{subfigure}%
    \setcounter{subfigure}{0}
    \renewcommand{\thesubfigure}{\alph{subfigure}'}
    \begin{subfigure}{.45\textwidth}
        \centering
%
%
\definecolor{mycolor1}{rgb}{0.00000,0.44700,0.74100}%
\definecolor{mycolor2}{rgb}{0.85000,0.32500,0.09800}%
\definecolor{mycolor3}{rgb}{0.92900,0.69400,0.12500}%
\begin{tikzpicture}[scale=0.95]

\begin{axis}[%
width=0.8\linewidth,
height=0.64\linewidth,
scale only axis,
xmin=0,
xmax=1.8,
xtick distance = 0.45,
xlabel = $\epsilon$,
y label style={at={(axis description cs:0.1,.5)},anchor=south},
axis background/.style={fill=white},
legend style={legend cell align=left, align=left, draw=white!15!black, minimum height=0.5cm, font={\tiny\arraycolsep=2pt}},
]
\addplot [color=mycolor1, line width=1pt]
  table[row sep=crcr]{%
0.018	0.00454\\
0.036	0.00427\\
0.054	0.00458\\
0.072	0.00415\\
0.09	0.00459\\
0.108	0.00498\\
0.126	0.0045\\
0.144	0.00455\\
0.162	0.0045\\
0.18	0.00463\\
0.198	0.00434\\
0.216	0.00457\\
0.234	0.00466\\
0.252	0.00452\\
0.27	0.00444\\
0.288	0.00465\\
0.306	0.00449\\
0.324	0.00485\\
0.342	0.00476\\
0.36	0.00479\\
0.378	0.00505\\
0.396	0.00529\\
0.414	0.00514\\
0.432	0.00459\\
0.45	0.00503\\
0.468	0.00476\\
0.486	0.00498\\
0.504	0.00534\\
0.522	0.00527\\
0.54	0.00534\\
0.558	0.00574\\
0.576	0.00568\\
0.594	0.00601\\
0.612	0.00618\\
0.63	0.00588\\
0.648	0.00617\\
0.666	0.00625\\
0.684	0.00601\\
0.702	0.00624\\
0.72	0.00646\\
0.738	0.00589\\
0.756	0.00712\\
0.774	0.0068\\
0.792	0.00684\\
0.81	0.00639\\
0.828	0.00615\\
0.846	0.00574\\
0.864	0.00585\\
0.882	0.00612\\
0.9	0.00605\\
0.918	0.00603\\
0.936	0.00576\\
0.954	0.00572\\
0.972	0.00528\\
0.99	0.00508\\
1.008	0.0045\\
1.026	0.00443\\
1.044	0.00472\\
1.062	0.00442\\
1.08	0.00438\\
1.098	0.00388\\
1.116	0.00399\\
1.134	0.00359\\
1.152	0.00345\\
1.17	0.00335\\
1.188	0.00322\\
1.206	0.00273\\
1.224	0.00255\\
1.242	0.00265\\
1.26	0.00232\\
1.278	0.00212\\
1.296	0.00194\\
1.314	0.00197\\
1.332	0.00187\\
1.35	0.00165\\
1.368	0.00154\\
1.386	0.00141\\
1.404	0.00135\\
1.422	0.00119\\
1.44	0.00102\\
1.458	0.00094\\
1.476	0.00099\\
1.494	0.00103\\
1.512	0.00094\\
1.53	0.00081\\
1.548	0.00073\\
1.566	0.00059\\
1.584	0.00066\\
1.602	0.00043\\
1.62	0.00044\\
1.638	0.00047\\
1.656	0.00028\\
1.674	0.00041\\
1.692	0.00024\\
1.71	0.00022\\
1.728	0.00027\\
1.746	0.00023\\
1.764	0.00028\\
1.782	0.00016\\
1.8	0.00011\\
};
\addlegendentry{$\hat I^\epsilon_N$}

\addplot [color=mycolor2, line width=1pt]
  table[row sep=crcr]{%
0.018	0.004603570119343\\
0.036	0.004583570119343\\
0.054	0.004583570119343\\
0.072	0.004543570119343\\
0.09	0.004593570119343\\
0.108	0.004613570119343\\
0.126	0.004613570119343\\
0.144	0.004483570119343\\
0.162	0.004563570119343\\
0.18	0.004583570119343\\
0.198	0.004543570119343\\
0.216	0.004633570119343\\
0.234	0.004643570119343\\
0.252	0.004563570119343\\
0.27	0.004563570119343\\
0.288	0.004793570119343\\
0.306	0.004723570119343\\
0.324	0.004633570119343\\
0.342	0.004893570119343\\
0.36	0.004803570119343\\
0.378	0.004963570119343\\
0.396	0.004873570119343\\
0.414	0.005003570119343\\
0.432	0.005003570119343\\
0.45	0.005003570119343\\
0.468	0.005033570119343\\
0.486	0.005263570119343\\
0.504	0.005323570119343\\
0.522	0.005523570119343\\
0.54	0.005473570119343\\
0.558	0.005543570119343\\
0.576	0.005813570119343\\
0.594	0.005863570119343\\
0.612	0.005843570119343\\
0.63	0.006023570119343\\
0.648	0.005823570119343\\
0.666	0.006453570119343\\
0.684	0.005863570119343\\
0.702	0.006673570119343\\
0.72	0.006683570119343\\
0.738	0.006293570119343\\
0.756	0.006983570119343\\
0.774	0.007013570119343\\
0.792	0.006553570119343\\
0.81	0.006273570119343\\
0.828	0.006063570119343\\
0.846	0.005923570119343\\
0.864	0.005753570119343\\
0.882	0.006263570119343\\
0.9	0.005963570119343\\
0.918	0.005973570119343\\
0.936	0.005663570119343\\
0.954	0.006223570119343\\
0.972	0.005493570119343\\
0.99	0.005243570119343\\
1.008	0.004473570119343\\
1.026	0.004213570119343\\
1.044	0.004553570119343\\
1.062	0.004483570119343\\
1.08	0.004453570119343\\
1.098	0.003933570119343\\
1.116	0.003983570119343\\
1.134	0.003573570119343\\
1.152	0.003613570119343\\
1.17	0.003603570119343\\
1.188	0.003233570119343\\
1.206	0.003043570119343\\
1.224	0.002233570119343\\
1.242	0.002273570119343\\
1.26	0.002573570119343\\
1.278	0.001843570119343\\
1.296	0.001763570119343\\
1.314	0.002193570119343\\
1.332	0.002223570119343\\
1.35	0.001673570119343\\
1.368	0.001563570119343\\
1.386	0.001313570119343\\
1.404	0.001563570119343\\
1.422	0.001003570119343\\
1.44	0.000693570119343\\
1.458	0.000683570119343\\
1.476	0.000683570119343\\
1.494	0.000823570119343\\
1.512	0.000833570119343\\
1.53	0.000913570119343\\
1.548	0.000883570119343\\
1.566	0.000553570119343001\\
1.584	0.000883570119343\\
1.602	0.000353570119343\\
1.62	0.000703570119343\\
1.638	0.000183570119343\\
1.656	7.35701193430001e-05\\
1.674	0.000553570119343001\\
1.692	0.000343570119343\\
1.71	0.000193570119343\\
1.728	0.000453570119343\\
1.746	0.000213570119343\\
1.764	0.000173570119343\\
1.782	0.000233570119343001\\
1.8	-0.000106429880656999\\
};
\addlegendentry{$\hat J^\epsilon_N$}

\addplot [color=mycolor3, line width=1pt]
  table[row sep=crcr]{%
0.018	0.00460356878507529\\
0.036	0.0045835628678989\\
0.054	0.00458357011933302\\
0.072	0.00453979841585814\\
0.09	0.00459354659081006\\
0.108	0.00461582292926211\\
0.126	0.00461229030819346\\
0.144	0.00448571641702138\\
0.162	0.00456173469529365\\
0.18	0.00458467835730413\\
0.198	0.00453424011973275\\
0.216	0.00463074126296066\\
0.234	0.0046444322831677\\
0.252	0.00456010032176568\\
0.27	0.00455383156037564\\
0.288	0.00478704323152982\\
0.306	0.00470465946795887\\
0.324	0.00465442414563103\\
0.342	0.00488207106468528\\
0.36	0.00480231424964876\\
0.378	0.00497215662196509\\
0.396	0.00491220872892416\\
0.414	0.00502047235853868\\
0.432	0.004952588435846\\
0.45	0.00500738268672497\\
0.468	0.00498818134961288\\
0.486	0.00521171166647989\\
0.504	0.00532623761650712\\
0.522	0.00547862753038251\\
0.54	0.00545144081702836\\
0.558	0.00558463943355769\\
0.576	0.00578838022008639\\
0.594	0.00589673006254815\\
0.612	0.00591753060575617\\
0.63	0.00598730935255494\\
0.648	0.0059158113077765\\
0.666	0.00640183649755864\\
0.684	0.00590575041846773\\
0.702	0.00654255788987507\\
0.72	0.00660996385815131\\
0.738	0.00616197973742295\\
0.756	0.00702630037553548\\
0.774	0.00694138125689633\\
0.792	0.00666525375000263\\
0.81	0.0063210576747613\\
0.828	0.00609865601070067\\
0.846	0.00582903642146011\\
0.864	0.00580440238251064\\
0.882	0.00619576946093849\\
0.9	0.00600850229271799\\
0.918	0.00600199540977215\\
0.936	0.00571563575407093\\
0.954	0.00596743029476259\\
0.972	0.0053628042563593\\
0.99	0.00514495784563388\\
1.008	0.00449027037391775\\
1.026	0.00436028512207225\\
1.044	0.00465742049601406\\
1.062	0.00444062723154619\\
1.08	0.00440015948928354\\
1.098	0.00389416532430727\\
1.116	0.00398813998485595\\
1.134	0.00358582515768205\\
1.152	0.00348028501542966\\
1.17	0.00339972777945635\\
1.188	0.00322243201339809\\
1.206	0.00278223946325278\\
1.224	0.00250343809987778\\
1.242	0.00259074978513497\\
1.26	0.00236234413544007\\
1.278	0.00209201115750587\\
1.296	0.00192060557983767\\
1.314	0.00199170329103696\\
1.332	0.00190542976083805\\
1.35	0.00165188217956631\\
1.368	0.00154219633795265\\
1.386	0.00140247704376294\\
1.404	0.00136689047371324\\
1.422	0.00117687081800047\\
1.44	0.0010009599122695\\
1.458	0.000925867863866984\\
1.476	0.000972043168278596\\
1.494	0.00101852182212669\\
1.512	0.000935084556100334\\
1.53	0.000813863523674479\\
1.548	0.000734065796586181\\
1.566	0.000588991752594251\\
1.584	0.000663971919781526\\
1.602	0.000428385215784087\\
1.62	0.000446623813693162\\
1.638	0.00046540711843904\\
1.656	0.000275727773560173\\
1.674	0.000414488028005116\\
1.692	0.000241136146899639\\
1.71	0.000219775452932495\\
1.728	0.000272045444441453\\
1.746	0.000229824385396365\\
1.764	0.000278433953710363\\
1.782	0.000160643637414913\\
1.8	0.000109117790119287\\
};
\addlegendentry{$\hat K^\epsilon_N$}

\addplot [color=black, dashed, line width=1.0pt]
  table[row sep=crcr]{%
0	0.004583570119343\\
0.133333333333333	0.004583570119343\\
0.266666666666667	0.004583570119343\\
0.4	0.004583570119343\\
0.533333333333333	0.004583570119343\\
0.666666666666667	0.004583570119343\\
0.8	0.004583570119343\\
0.933333333333333	0.004583570119343\\
1.06666666666667	0.004583570119343\\
1.2	0.004583570119343\\
};
\addlegendentry{$P^0_1$}

\end{axis}

\node[] at (2.85,0.65) {(a')};

\end{tikzpicture}%
        \captionlistentry{}
        \label{fig:a0_estimates_PSD2}
    \end{subfigure}%

    \centering
    \renewcommand{\thesubfigure}{\alph{subfigure}}
    \begin{subfigure}{.45\textwidth}
        \centering
%
%
\definecolor{mycolor1}{rgb}{0.00000,0.44700,0.74100}%
\definecolor{mycolor2}{rgb}{0.85000,0.32500,0.09800}%
\definecolor{mycolor3}{rgb}{0.92900,0.69400,0.12500}%
\begin{tikzpicture}[scale=0.95]

\begin{axis}[%
width=0.8\linewidth,
height=0.64\linewidth,
scale only axis,
xmin=0,
xmax=1.2,
xtick distance = 0.3,
xlabel = $\epsilon$,
y label style={at={(axis description cs:0.1,.5)},anchor=south},
axis background/.style={fill=white},
legend style={legend cell align=left, align=left, draw=white!15!black, minimum height=0.6cm, font={\tiny\arraycolsep=2pt}, at={(0.98,0.47)},anchor=east}
]
\addplot [color=mycolor1, line width=1pt]
  table[row sep=crcr]{%
0.012	0.004538215519\\
0.024	0.004527315696\\
0.036	0.004625405391\\
0.048	0.004604600124\\
0.06	0.0044995696\\
0.072	0.004480740999\\
0.084	0.004647200439\\
0.096	0.0045292975\\
0.108	0.0046184704\\
0.12	0.004537224636\\
0.132	0.004606581616\\
0.144	0.004534251975\\
0.156	0.004323147036\\
0.168	0.0044896599\\
0.18	0.004597664839\\
0.192	0.004457947516\\
0.204	0.004425241975\\
0.216	0.004486686951\\
0.228	0.004506506271\\
0.24	0.004468848879\\
0.252	0.004323147036\\
0.264	0.004391543079\\
0.276	0.004407401671\\
0.288	0.004343964231\\
0.3	0.004235903484\\
0.312	0.004171452279\\
0.324	0.004165502511\\
0.336	0.004169469031\\
0.348	0.004186326384\\
0.36	0.004173435519\\
0.372	0.004029629884\\
0.384	0.003935389599\\
0.396	0.004042524519\\
0.408	0.003854030839\\
0.42	0.003748839831\\
0.432	0.003662487024\\
0.444	0.003662487024\\
0.456	0.003574133431\\
0.468	0.003465903516\\
0.48	0.003379501119\\
0.492	0.003309970959\\
0.504	0.003274208775\\
0.516	0.003066538224\\
0.528	0.003076476604\\
0.54	0.002864745871\\
0.552	0.002837900284\\
0.564	0.0027324924\\
0.576	0.002646956284\\
0.588	0.002697682975\\
0.6	0.002530563631\\
0.612	0.002391254391\\
0.624	0.002311631511\\
0.636	0.002273806159\\
0.648	0.0020358384\\
0.66	0.002003967936\\
0.672	0.001989027951\\
0.684	0.001801741975\\
0.696	0.001750923484\\
0.708	0.001651264284\\
0.72	0.0015675351\\
0.732	0.001475815516\\
0.744	0.001381087311\\
0.756	0.001293322975\\
0.768	0.001265394711\\
0.78	0.001263399775\\
0.792	0.001111761231\\
0.804	0.001067857239\\
0.816	0.001021953471\\
0.828	0.000890206119\\
0.84	0.000934125775\\
0.852	0.000847280896\\
0.864	0.000787379056\\
0.876	0.00069951\\
0.888	0.000685529404\\
0.9	0.0006395904\\
0.912	0.000585656604\\
0.924	0.000520728559\\
0.936	0.000523725424\\
0.948	0.000460787479\\
0.96	0.000398840799\\
0.972	0.000403836784\\
0.984	0.000367864576\\
0.996	0.000332889111\\
1.008	0.000294912975\\
1.02	0.000293913564\\
1.032	0.000292914151\\
1.044	0.000231946176\\
1.056	0.000231946176\\
1.068	0.000221950716\\
1.08	0.000197960796\\
1.092	0.000164972775\\
1.104	0.000156975351\\
1.116	0.000131982576\\
1.128	0.000123984624\\
1.14	0.0001499775\\
1.152	0.000100989799\\
1.164	0.0001099879\\
1.176	9.3991164e-05\\
1.188	9.0991719e-05\\
1.2	6.7995376e-05\\
};
\addlegendentry{$\displaystyle \sigma^2_{\displaystyle \hat I^\epsilon_N}$}

\addplot [color=mycolor2, line width=1pt]
  table[row sep=crcr]{%
0.012	5.999984e-06\\
0.024	0\\
0.036	1.3999984e-05\\
0.048	2.1999984e-05\\
0.06	2.9999984e-05\\
0.072	5.4999975e-05\\
0.084	7.6999975e-05\\
0.096	8.8999775e-05\\
0.108	0.000120999471\\
0.12	0.000146999775\\
0.132	0.0001819999\\
0.144	0.000208999039\\
0.156	0.000229998704\\
0.168	0.000279999216\\
0.18	0.000335999424\\
0.192	0.000369987456\\
0.204	0.000390994071\\
0.216	0.000437991164\\
0.228	0.0005199879\\
0.24	0.000541979836\\
0.252	0.000588988119\\
0.264	0.000683964656\\
0.276	0.000676977199\\
0.288	0.000842930831\\
0.3	0.000842914151\\
0.312	0.000903860124\\
0.324	0.0009778844\\
0.336	0.001071833536\\
0.348	0.001123833536\\
0.36	0.001141808156\\
0.372	0.001245780976\\
0.384	0.001396560431\\
0.396	0.001492604359\\
0.408	0.0015054224\\
0.42	0.001640319375\\
0.432	0.001756285975\\
0.444	0.001734258679\\
0.456	0.001795023856\\
0.468	0.001967605239\\
0.48	0.001971614671\\
0.492	0.002156376924\\
0.504	0.0021982576\\
0.516	0.002302900399\\
0.528	0.002256963671\\
0.54	0.002434326775\\
0.552	0.002485849375\\
0.564	0.002649476871\\
0.576	0.002651278959\\
0.588	0.002788111216\\
0.6	0.0028418384\\
0.612	0.003089932999\\
0.624	0.002984856176\\
0.636	0.003046515036\\
0.648	0.003138704919\\
0.66	0.003315072576\\
0.672	0.0032750831\\
0.684	0.003369243775\\
0.696	0.003451438524\\
0.708	0.003446745871\\
0.72	0.003487023984\\
0.732	0.003627358975\\
0.744	0.003749476464\\
0.756	0.003652791975\\
0.768	0.003686599375\\
0.78	0.003750208775\\
0.792	0.003833986844\\
0.804	0.003851296759\\
0.816	0.003884447151\\
0.828	0.004166292775\\
0.84	0.004107892464\\
0.852	0.004082779559\\
0.864	0.004135239036\\
0.876	0.004159578671\\
0.888	0.004166680604\\
0.9	0.004293887804\\
0.912	0.004193199375\\
0.924	0.004137373791\\
0.936	0.004485475584\\
0.948	0.004202394375\\
0.96	0.004253337276\\
0.972	0.004249532636\\
0.984	0.004291569375\\
0.996	0.0043798524\\
1.008	0.004348860919\\
1.02	0.004363920496\\
1.032	0.004493666391\\
1.044	0.0044171644\\
1.056	0.00438951\\
1.068	0.004299183159\\
1.08	0.004475631199\\
1.092	0.004419859375\\
1.104	0.004397964231\\
1.116	0.004598233751\\
1.128	0.004373216444\\
1.14	0.0045161084\\
1.152	0.0044900191\\
1.164	0.004520848879\\
1.176	0.004508848879\\
1.188	0.004487224191\\
1.2	0.004519605744\\
};
\addlegendentry{$\displaystyle \sigma^2_{\displaystyle \hat J^\epsilon_N}$}

\addplot [color=mycolor3, line width=1pt]
  table[row sep=crcr]{%
0.012	5.99975534678263e-06\\
0.024	3.53590237929422e-16\\
0.036	1.39825592780837e-05\\
0.048	2.19634175443146e-05\\
0.06	2.99622871957798e-05\\
0.072	5.47996384905509e-05\\
0.084	7.66390488053656e-05\\
0.096	8.84062917536443e-05\\
0.108	0.000119885857114492\\
0.12	0.000145560898425414\\
0.132	0.000180005506149946\\
0.144	0.000205852051229135\\
0.156	0.000225950120591541\\
0.168	0.000274757960838277\\
0.18	0.000328997303660201\\
0.192	0.000357329469629434\\
0.204	0.000378866437618464\\
0.216	0.000422591883254\\
0.228	0.000498559245689174\\
0.24	0.000516703065896513\\
0.252	0.00056157728070653\\
0.264	0.000642604008748617\\
0.276	0.000639491456531514\\
0.288	0.000776808937593755\\
0.3	0.000771970873635738\\
0.312	0.000814444245993907\\
0.324	0.000881881636915672\\
0.336	0.000952716253036897\\
0.348	0.000996647478075313\\
0.36	0.0010070515837264\\
0.372	0.00108314329634332\\
0.384	0.00116693404080396\\
0.396	0.00125268429734241\\
0.408	0.00122854568520004\\
0.42	0.00130947073735677\\
0.432	0.00138240470071981\\
0.444	0.00136346057411437\\
0.456	0.00137230291910079\\
0.468	0.00143640681908598\\
0.48	0.0014297968673387\\
0.492	0.00151696456114431\\
0.504	0.00152714535233885\\
0.516	0.00152622393912654\\
0.528	0.00150618955409728\\
0.54	0.00151729632353497\\
0.552	0.00150522686086062\\
0.564	0.00154168891365986\\
0.576	0.00150861756977285\\
0.588	0.00157867652383263\\
0.6	0.00154187158314802\\
0.612	0.00155761604997756\\
0.624	0.00148242284099422\\
0.636	0.00147779656673652\\
0.648	0.00138804438003121\\
0.66	0.00141185261842564\\
0.672	0.00139179243005813\\
0.684	0.00130881590155677\\
0.696	0.0012814563672536\\
0.708	0.00124353963017284\\
0.72	0.00118833405957093\\
0.732	0.0011574087349462\\
0.744	0.00110940276132926\\
0.756	0.00104422762510001\\
0.768	0.001030309814039\\
0.78	0.00103313452566321\\
0.792	0.00092691241268852\\
0.804	0.000908794145842389\\
0.816	0.000866219512840968\\
0.828	0.0007818874066684\\
0.84	0.000814415511125473\\
0.852	0.000746387466617167\\
0.864	0.000701160622928237\\
0.876	0.000628280208457113\\
0.888	0.000619741961020695\\
0.9	0.000587965960922113\\
0.912	0.000537759762858534\\
0.924	0.000481374174996636\\
0.936	0.000486753828873429\\
0.948	0.000427634783559894\\
0.96	0.000378231249199883\\
0.972	0.000383818808315394\\
0.984	0.000348052501299229\\
0.996	0.000317820821212101\\
1.008	0.000282243227884203\\
1.02	0.000282460119981725\\
1.032	0.000281221566843661\\
1.044	0.000224592611052026\\
1.056	0.000225005896849857\\
1.068	0.000215183371065517\\
1.08	0.00019303996387715\\
1.092	0.00016110271582323\\
1.104	0.000153260329636864\\
1.116	0.000129823517304815\\
1.128	0.000121974704807859\\
1.14	0.000147277508097579\\
1.152	9.95642448488308e-05\\
1.164	0.00010846507430137\\
1.176	9.28047476237754e-05\\
1.188	9.01791311550044e-05\\
1.2	6.73157366632167e-05\\
};
\addlegendentry{$\displaystyle \sigma^2_{\displaystyle \hat K^\epsilon_N}$}

\end{axis}

\node[] at (2.85,0.65) {(b)};

\end{tikzpicture}%
        \captionlistentry{}
        \label{fig:a0_variances_PSD1}
    \end{subfigure}%
    \setcounter{subfigure}{1}
    \renewcommand{\thesubfigure}{\alph{subfigure}'}
    \begin{subfigure}{.45\textwidth}
        \centering
%
%
\definecolor{mycolor1}{rgb}{0.00000,0.44700,0.74100}%
\definecolor{mycolor2}{rgb}{0.85000,0.32500,0.09800}%
\definecolor{mycolor3}{rgb}{0.92900,0.69400,0.12500}%
\begin{tikzpicture}[scale=0.95]

\begin{axis}[%
width=0.8\linewidth,
height=0.64\linewidth,
scale only axis,
xmin=0,
xmax=1.8,
xtick distance = 0.45,
xlabel = $\epsilon$,
y label style={at={(axis description cs:0.1,.5)},anchor=south},
axis background/.style={fill=white},
legend style={legend cell align=left, align=left, draw=white!15!black, minimum height=0.1cm, font={\tiny}, at={(0.98,0.47)},anchor=east}
]
\addplot [color=mycolor1, line width=1pt]
  table[row sep=crcr]{%
0.018	0.004552087671\\
0.036	0.004627386799\\
0.054	0.004579830799\\
0.072	0.0044599296\\
0.09	0.004592711004\\
0.108	0.004594692544\\
0.126	0.004593701775\\
0.144	0.004602618624\\
0.162	0.004535242864\\
0.18	0.004704655471\\
0.198	0.004533261084\\
0.216	0.004659088239\\
0.234	0.004699702716\\
0.252	0.004696731039\\
0.27	0.004513442844\\
0.288	0.004660078876\\
0.306	0.004645219111\\
0.324	0.004809642111\\
0.342	0.004831428975\\
0.36	0.004721494464\\
0.378	0.004971039984\\
0.396	0.004963119856\\
0.414	0.004941338844\\
0.432	0.005107642044\\
0.45	0.005128425975\\
0.468	0.005176918384\\
0.486	0.005268941791\\
0.504	0.005254100476\\
0.522	0.005516230791\\
0.54	0.005553807775\\
0.558	0.005611156551\\
0.576	0.005839496124\\
0.594	0.005850366775\\
0.612	0.005928430704\\
0.63	0.006004506319\\
0.648	0.006047972775\\
0.666	0.0060825456\\
0.684	0.006143782876\\
0.702	0.006255372975\\
0.72	0.006320537679\\
0.738	0.006278082876\\
0.756	0.006363975975\\
0.774	0.006269196519\\
0.792	0.006333372124\\
0.81	0.006402474864\\
0.828	0.006316588551\\
0.846	0.006206987484\\
0.864	0.006086496624\\
0.882	0.006145758144\\
0.9	0.005813800896\\
0.918	0.005772290364\\
0.936	0.005673441564\\
0.954	0.005424253884\\
0.972	0.005246184924\\
0.99	0.004980939964\\
1.008	0.004997769471\\
1.026	0.004800729024\\
1.044	0.004691778204\\
1.062	0.004501551516\\
1.08	0.004279527196\\
1.098	0.003973087879\\
1.116	0.003837162096\\
1.134	0.0035671836\\
1.152	0.003332817664\\
1.17	0.003313944375\\
1.188	0.0030108796\\
1.206	0.002813041959\\
1.224	0.002714590716\\
1.242	0.002497729984\\
1.26	0.002301677751\\
1.278	0.002211089344\\
1.296	0.002050776975\\
1.314	0.001885431679\\
1.332	0.001762881244\\
1.35	0.001645284096\\
1.368	0.001528656039\\
1.386	0.001380090076\\
1.404	0.001331223111\\
1.422	0.001170626416\\
1.44	0.001039916319\\
1.458	0.001050893296\\
1.476	0.000978041559\\
1.494	0.000880223839\\
1.512	0.000781388476\\
1.53	0.000768408639\\
1.548	0.0006795376\\
1.566	0.000598641199\\
1.584	0.000566678511\\
1.602	0.000492756951\\
1.62	0.000471777216\\
1.638	0.00039984\\
1.656	0.000375858624\\
1.674	0.000346879591\\
1.692	0.000280921039\\
1.71	0.000275923824\\
1.728	0.000272925471\\
1.746	0.000233945244\\
1.764	0.000202958791\\
1.782	0.000184965775\\
1.8	0.0001699711\\
};
\addlegendentry{$\displaystyle \sigma^2_{\displaystyle \hat I^\epsilon_N}$}

\addplot [color=mycolor2, line width=1pt]
  table[row sep=crcr]{%
0.018	2.999999e-06\\
0.036	1.8999999e-05\\
0.054	3.6999991e-05\\
0.072	4.2999951e-05\\
0.09	8.4999975e-05\\
0.108	0.000114999775\\
0.126	0.000147999936\\
0.144	0.000223999964\\
0.162	0.0002519996\\
0.18	0.000286999039\\
0.198	0.000348998911\\
0.216	0.0004179996\\
0.234	0.000470994071\\
0.252	0.000588994071\\
0.27	0.000606996751\\
0.288	0.000695976896\\
0.306	0.000795973756\\
0.324	0.000862969375\\
0.342	0.0009879159\\
0.36	0.001077956736\\
0.378	0.001209919344\\
0.396	0.001329873264\\
0.414	0.001475830256\\
0.432	0.001591753984\\
0.45	0.0017077884\\
0.468	0.001923656604\\
0.486	0.002104500151\\
0.504	0.002141461244\\
0.522	0.002418155439\\
0.54	0.002586963676\\
0.558	0.002837857239\\
0.576	0.003050376924\\
0.594	0.003239307399\\
0.612	0.003411163975\\
0.63	0.003580000604\\
0.648	0.003807695676\\
0.666	0.003999323504\\
0.684	0.0042833756\\
0.702	0.004482227775\\
0.72	0.004722562684\\
0.738	0.0047385775\\
0.756	0.004998555264\\
0.774	0.005142716656\\
0.792	0.0053083519\\
0.81	0.005509325111\\
0.828	0.00560911\\
0.846	0.005805427184\\
0.864	0.005807316956\\
0.882	0.005975522524\\
0.9	0.005881048391\\
0.918	0.0060023871\\
0.936	0.006137586279\\
0.954	0.006126162775\\
0.972	0.006226563079\\
0.99	0.006100853311\\
1.008	0.006191809904\\
1.026	0.006191992944\\
1.044	0.006232996279\\
1.062	0.006247950716\\
1.08	0.006168899511\\
1.098	0.006161574896\\
1.116	0.005966468559\\
1.134	0.006014837916\\
1.152	0.005941628759\\
1.17	0.006062106624\\
1.188	0.0059734079\\
1.206	0.005853947991\\
1.224	0.005882\\
1.242	0.005691573184\\
1.26	0.005534040471\\
1.278	0.005656345116\\
1.296	0.005506203551\\
1.314	0.005437822959\\
1.332	0.005277876439\\
1.35	0.005473642519\\
1.368	0.005381234375\\
1.386	0.0052436959\\
1.404	0.0052512416\\
1.422	0.005146494336\\
1.44	0.005159960679\\
1.458	0.005156176439\\
1.476	0.005081140604\\
1.494	0.005104719159\\
1.512	0.004931952496\\
1.53	0.005053688431\\
1.548	0.004963672775\\
1.566	0.004954426959\\
1.584	0.004935369916\\
1.602	0.004860165391\\
1.62	0.004697983996\\
1.638	0.004844140924\\
1.656	0.004822758559\\
1.674	0.004690744284\\
1.692	0.004665971484\\
1.71	0.004695613056\\
1.728	0.004626292736\\
1.746	0.004671354876\\
1.764	0.004778001216\\
1.782	0.004637841871\\
1.8	0.004741857856\\
};
\addlegendentry{$\displaystyle \sigma^2_{\displaystyle \hat J^\epsilon_N}$}

\addplot [color=mycolor3, line width=1pt]
  table[row sep=crcr]{%
0.018	2.99977725897227e-06\\
0.036	1.89784132494357e-05\\
0.054	3.69367447377856e-05\\
0.072	4.28603832729753e-05\\
0.09	8.46508187371222e-05\\
0.108	0.000114452395331841\\
0.126	0.000146930288970066\\
0.144	0.000221413954873867\\
0.162	0.000249014929421102\\
0.18	0.000283485673006456\\
0.198	0.000343441566108314\\
0.216	0.000409455568973155\\
0.234	0.00046256956826527\\
0.252	0.000574770541291364\\
0.27	0.000589997388441523\\
0.288	0.000679487397080357\\
0.306	0.000773463416369524\\
0.324	0.00083732448364583\\
0.342	0.000960908316871799\\
0.36	0.00103586839720939\\
0.378	0.00116394507931152\\
0.396	0.00127808375552823\\
0.414	0.00141295634626999\\
0.432	0.00152615638678955\\
0.45	0.00162388521580934\\
0.468	0.00182550103721407\\
0.486	0.00199656502887469\\
0.504	0.00203094821419521\\
0.522	0.00229465870442503\\
0.54	0.00244980434431877\\
0.558	0.00266393019014886\\
0.576	0.00287549089708311\\
0.594	0.00303056139131482\\
0.612	0.00317755687986155\\
0.63	0.00332177464608494\\
0.648	0.00351530408877014\\
0.666	0.00368191017494185\\
0.684	0.00388774201911226\\
0.702	0.00404610514343445\\
0.72	0.00425759505555791\\
0.738	0.00426311179942692\\
0.756	0.00444626011117641\\
0.774	0.00451589903316763\\
0.792	0.00465073890403333\\
0.81	0.00478501982099198\\
0.828	0.00477702592496159\\
0.846	0.00484236954243785\\
0.864	0.00482607468100239\\
0.882	0.00491187448787067\\
0.9	0.00473937235761761\\
0.918	0.00475555124214893\\
0.936	0.00476938733448403\\
0.954	0.00461834951137523\\
0.972	0.00453611381046402\\
0.99	0.0043223504188182\\
1.008	0.0043752366878636\\
1.026	0.00421440226126575\\
1.044	0.0041763284091539\\
1.062	0.00403288677980789\\
1.08	0.00388151497261053\\
1.098	0.00365326512628296\\
1.116	0.00351324090322601\\
1.134	0.0033088860306675\\
1.152	0.00313501119461811\\
1.17	0.0031147656097509\\
1.188	0.00286390154501739\\
1.206	0.00268830838378821\\
1.224	0.00258969328612013\\
1.242	0.00239197820155276\\
1.26	0.00221077340475969\\
1.278	0.00214157262905317\\
1.296	0.00197514316670759\\
1.314	0.00183102862130781\\
1.332	0.00171698410752818\\
1.35	0.00160613830061088\\
1.368	0.00149568882852241\\
1.386	0.00135272770108772\\
1.404	0.00130665806742658\\
1.422	0.00115291678000515\\
1.44	0.0010258558038487\\
1.458	0.00103699926503632\\
1.476	0.000967339063022038\\
1.494	0.000870995135671299\\
1.512	0.000774246851003259\\
1.53	0.000760872588800824\\
1.548	0.000674948748935526\\
1.566	0.000594042588431886\\
1.584	0.00056316981903439\\
1.602	0.000490404804931682\\
1.62	0.000468865094971356\\
1.638	0.000398444882463805\\
1.656	0.000374136385886356\\
1.674	0.000345834277113952\\
1.692	0.000280102693435706\\
1.71	0.00027508479108467\\
1.728	0.000272277189557959\\
1.746	0.000233459094656522\\
1.764	0.000202631060217212\\
1.782	0.00018453642591373\\
1.8	0.000169747351838113\\
};
\addlegendentry{$\displaystyle \sigma^2_{\displaystyle \hat K^\epsilon_N}$}

\end{axis}

\node[] at (2.85,0.65) {(b')};

\end{tikzpicture}%
        \captionlistentry{}
        \label{fig:a0_variances_PSD2}
    \end{subfigure}%

    \centering
    \renewcommand{\thesubfigure}{\alph{subfigure}}
    \begin{subfigure}{.45\textwidth}
        \centering
%
%
\definecolor{mycolor1}{rgb}{0.00000,0.44700,0.74100}%
\definecolor{mycolor2}{rgb}{0.85000,0.32500,0.09800}%
\definecolor{mycolor3}{rgb}{0.92900,0.69400,0.12500}%
\begin{tikzpicture}[scale=0.95]

\begin{axis}[%
width=0.8\linewidth,
height=0.64\linewidth,
scale only axis,
xmin=0,
xmax=1.2,
xtick distance = 0.3,
xlabel = $\epsilon$,
y label style={at={(axis description cs:0.1,.5)},anchor=south},
axis background/.style={fill=white},
legend style={legend cell align=left, align=left, draw=white!15!black, minimum height=0.6cm, font={\tiny\arraycolsep=2pt}}
]
\addplot [color=black, line width=1pt]
  table[row sep=crcr]{%
0.036	0.000388404424391214\\
0.048	0.000457571198839888\\
0.06	0.000499371453262996\\
0.072	0.000761106090146541\\
0.084	0.000912369628635305\\
0.096	0.000920898872433794\\
0.108	0.00111005423254159\\
0.12	0.00121300748687845\\
0.132	0.00136367807689353\\
0.144	0.00142952813353566\\
0.156	0.00144839820892014\\
0.168	0.00163546405260879\\
0.18	0.00182776279811223\\
0.192	0.0018610909876533\\
0.204	0.00185718841969835\\
0.216	0.0019564439039537\\
0.228	0.00218666335828585\\
0.24	0.00215292944123547\\
0.252	0.00222848127264496\\
0.264	0.00243410609374476\\
0.276	0.00231699803091128\\
0.288	0.00269725325553387\\
0.3	0.00257323624545246\\
0.312	0.00261039822433945\\
0.324	0.00272185690406072\\
0.336	0.00283546503880029\\
0.348	0.00286392953469918\\
0.36	0.00279736551035111\\
0.372	0.00291167552780462\\
0.384	0.0030388907312603\\
0.396	0.0031633441852081\\
0.408	0.00301114138529422\\
0.42	0.00311778746989707\\
0.432	0.00320001088129585\\
0.444	0.00307085714890625\\
0.456	0.00300943622609822\\
0.468	0.0030692453399273\\
0.48	0.00297874347362228\\
0.492	0.00308326130313885\\
0.504	0.00303005030225963\\
0.516	0.00295779833164058\\
0.528	0.00285263173124484\\
0.54	0.00280980800654624\\
0.552	0.00272686025518229\\
0.564	0.00273349098166641\\
0.576	0.00261912772530008\\
0.588	0.00268482402012352\\
0.6	0.00256978597191337\\
0.612	0.00254512426466921\\
0.624	0.00237567762979844\\
0.636	0.00232357950744736\\
0.648	0.00214204379634446\\
0.66	0.00213917063397824\\
0.672	0.0020711196875865\\
0.684	0.00191347354028768\\
0.696	0.00184117294145632\\
0.708	0.00175641190702379\\
0.72	0.00165046397162629\\
0.732	0.00158115947397022\\
0.744	0.00149113274372212\\
0.756	0.00138125347235451\\
0.768	0.00134154923702995\\
0.78	0.00132453144315796\\
0.792	0.0011703439554148\\
0.804	0.00113034097741591\\
0.816	0.00106154352063844\\
0.828	0.000944308462159904\\
0.84	0.000969542275149372\\
0.852	0.000876041627484938\\
0.864	0.000811528498759534\\
0.876	0.000717214849836887\\
0.888	0.000697907613762044\\
0.9	0.000653295512135681\\
0.912	0.000589648862783481\\
0.924	0.000520967721857831\\
0.936	0.000520036141958791\\
0.948	0.00045109154383955\\
0.96	0.000393990884583212\\
0.972	0.000394875317196907\\
0.984	0.000353711891564258\\
0.996	0.000319097210052311\\
1.008	0.000280003202266075\\
1.02	0.000276921686256593\\
1.032	0.000272501518259362\\
1.044	0.000215127022080484\\
1.056	0.000213073765956304\\
1.068	0.00020148255717745\\
1.08	0.000178740707293657\\
1.092	0.000147529959545083\\
1.104	0.000138822762352232\\
1.116	0.000116329316581375\\
1.128	0.000108133603553066\\
1.14	0.000129190796576824\\
1.152	8.64272958757212e-05\\
1.164	9.31830535235134e-05\\
1.176	7.89156017208975e-05\\
1.188	7.59083595580845e-05\\
1.2	5.60964472193473e-05\\
};
\addlegendentry{$\displaystyle \sigma^2_{\displaystyle \hat K^\epsilon_N}/\epsilon$}

\addplot [color=black,dashed, line width=0.5pt]
  table[row sep=crcr]{%
0.036	0.0107890117886448\\
0.048	0.00953273330916434\\
0.06	0.00832285755438327\\
0.072	0.010570917918702\\
0.084	0.0108615431980393\\
0.096	0.00959269658785203\\
0.108	0.0102782799309406\\
0.12	0.0101083957239871\\
0.132	0.0103308945219207\\
0.144	0.00992727870510874\\
0.156	0.0092846039033342\\
0.168	0.00973490507505233\\
0.18	0.0101542377672902\\
0.192	0.00969318222736094\\
0.204	0.00910386480244291\\
0.216	0.00905761066645233\\
0.228	0.00959062876441162\\
0.24	0.00897053933848112\\
0.252	0.00884317965335301\\
0.264	0.00922009883994228\\
0.276	0.00839492040185247\\
0.288	0.00936546269282594\\
0.3	0.0085774541515082\\
0.312	0.00836666097544695\\
0.324	0.00840079291376765\\
0.336	0.00843888404404847\\
0.348	0.00822968257097464\\
0.36	0.00777045975097529\\
0.372	0.00782708475216295\\
0.384	0.00791377794599037\\
0.396	0.00798824289193964\\
0.408	0.00738024849336818\\
0.42	0.00742330349975492\\
0.432	0.00740743259559225\\
0.444	0.00691634492996903\\
0.456	0.00659964084670662\\
0.468	0.0065582165383062\\
0.48	0.00620571557004642\\
0.492	0.00626679126654237\\
0.504	0.00601200456797545\\
0.516	0.00573216730938098\\
0.528	0.00540271161220613\\
0.54	0.00520334816027081\\
0.552	0.0049399642304027\\
0.564	0.00484661521572059\\
0.576	0.00454709674531264\\
0.588	0.00456602724510802\\
0.6	0.00428297661985561\\
0.612	0.00415869977887125\\
0.624	0.00380717568877955\\
0.636	0.00365342689850214\\
0.648	0.00330562314250689\\
0.66	0.00324116762723976\\
0.672	0.00308202334462277\\
0.684	0.0027974759360931\\
0.696	0.00264536342163265\\
0.708	0.00248080777828218\\
0.72	0.00229231107170318\\
0.732	0.00216005392618883\\
0.744	0.00200421067704586\\
0.756	0.00182705485761179\\
0.768	0.00174680890238275\\
0.78	0.0016981172348179\\
0.792	0.00147770701441262\\
0.804	0.00140589673807949\\
0.816	0.00130091117725299\\
0.828	0.00114046915719795\\
0.84	0.00115421699422544\\
0.852	0.0010282178726349\\
0.864	0.000939269095786497\\
0.876	0.000818738413055807\\
0.888	0.00078593199747978\\
0.9	0.000725883902372979\\
0.912	0.000646544805683641\\
0.924	0.000563817880798518\\
0.936	0.000555594168759392\\
0.948	0.000475834961856066\\
0.96	0.000410407171440846\\
0.972	0.000406250326334267\\
0.984	0.000359463304435221\\
0.996	0.000320378724952119\\
1.008	0.000277780954629042\\
1.02	0.00027149184927117\\
1.032	0.000264051858778451\\
1.044	0.000206060365977475\\
1.056	0.000201774399579833\\
1.068	0.000188654079754167\\
1.08	0.000165500654901535\\
1.092	0.000135100695554105\\
1.104	0.000125745255753834\\
1.116	0.0001042377388722\\
1.128	9.58631237172572e-05\\
1.14	0.000113325260155109\\
1.152	7.50236943365635e-05\\
1.164	8.005416969374e-05\\
1.176	6.71051035041645e-05\\
1.188	6.38959255539432e-05\\
1.2	4.67470393494561e-05\\
};
\addlegendentry{$\displaystyle \sigma^2_{\displaystyle \hat K^\epsilon_N}/\epsilon^2$}

\end{axis}

\node[] at (2.85,0.4) {(c)};

\end{tikzpicture}%
        \captionlistentry{}
        \label{fig:a0_bounds_PSD1}
    \end{subfigure}%
    \setcounter{subfigure}{2}
    \renewcommand{\thesubfigure}{\alph{subfigure}'}
    \begin{subfigure}{.45\textwidth}
        \centering
%
%
\definecolor{mycolor1}{rgb}{0.00000,0.44700,0.74100}%
\definecolor{mycolor2}{rgb}{0.85000,0.32500,0.09800}%
\definecolor{mycolor3}{rgb}{0.92900,0.69400,0.12500}%
\begin{tikzpicture}[scale=0.95]

\begin{axis}[%
width=0.8\linewidth,
height=0.64\linewidth,
scale only axis,
xmin=0,
xmax=1.8,
xtick distance = 0.45,
xlabel = $\epsilon$,
y label style={at={(axis description cs:0.1,.5)},anchor=south},
axis background/.style={fill=white},
legend style={legend cell align=left, align=left, draw=white!15!black, minimum height=0.6cm, font={\tiny\arraycolsep=2pt}}
]
\addplot [color=black, line width=1pt]
  table[row sep=crcr]{%
0.036	0.000527178145817657\\
0.054	0.000684013791440474\\
0.072	0.000595283101013545\\
0.09	0.000940564652634691\\
0.108	0.00105974440122075\\
0.126	0.00116611340452433\\
0.144	0.0015375969088463\\
0.162	0.00153712919395742\\
0.18	0.00157492040559142\\
0.198	0.0017345533641834\\
0.216	0.00189562763413498\\
0.234	0.00197679302677466\\
0.252	0.00228083548131494\\
0.27	0.00218517551274638\\
0.288	0.00235933123986235\\
0.306	0.00252765822342982\\
0.324	0.00258433482606738\\
0.342	0.00280967344114561\\
0.36	0.00287741221447054\\
0.378	0.00307921978653841\\
0.396	0.0032274842311319\\
0.414	0.00341293803446859\\
0.432	0.00353276941386469\\
0.45	0.00360863381290964\\
0.468	0.0039006432419104\\
0.486	0.00410815849562694\\
0.504	0.00402965915514922\\
0.522	0.00439589790119738\\
0.54	0.00453667471170143\\
0.558	0.00477406844112699\\
0.576	0.00499217169632485\\
0.594	0.0051019552042337\\
0.612	0.00519208640500253\\
0.63	0.0052726581683888\\
0.648	0.0054248519888428\\
0.666	0.00552839365606884\\
0.684	0.00568383336127523\\
0.702	0.00576368254050492\\
0.72	0.00591332646605265\\
0.738	0.00577657425396601\\
0.756	0.00588129644335504\\
0.774	0.00583449487489358\\
0.792	0.00587214508085016\\
0.81	0.00590743187776788\\
0.828	0.00576935498183767\\
0.846	0.0057238410667114\\
0.864	0.0055857345844935\\
0.882	0.00556901869373092\\
0.9	0.00526596928624179\\
0.918	0.00518033904373522\\
0.936	0.00509549928897866\\
0.954	0.00484103722366376\\
0.972	0.00466678375562142\\
0.99	0.00436601052405879\\
1.008	0.00434051258716627\\
1.026	0.004107604543144\\
1.044	0.00400031456815507\\
1.062	0.00379744517872683\\
1.08	0.00359399534500975\\
1.098	0.00332719956856371\\
1.116	0.00314806532547133\\
1.134	0.00291788891593254\\
1.152	0.00272136388421711\\
1.17	0.00266219282884692\\
1.188	0.00241069153620993\\
1.206	0.00222911142934346\\
1.224	0.00211576248866024\\
1.242	0.00192590837484119\\
1.26	0.00175458206726959\\
1.278	0.00167572193196649\\
1.296	0.00152403022122499\\
1.314	0.00139347688075176\\
1.332	0.00128902710775389\\
1.35	0.00118973207452657\\
1.368	0.00109333978693159\\
1.386	0.000975994012328799\\
1.404	0.000930668139192723\\
1.422	0.000810771293955804\\
1.44	0.00071239986378382\\
1.458	0.000711247781232041\\
1.476	0.000655378768985121\\
1.494	0.000582995405402476\\
1.512	0.000512068023150304\\
1.53	0.000497302345621453\\
1.548	0.00043601340370512\\
1.566	0.000379337540505674\\
1.584	0.00035553650191565\\
1.602	0.000306120352641499\\
1.62	0.000289422898130466\\
1.638	0.000243250843994997\\
1.656	0.000225927769255046\\
1.674	0.000206591563389457\\
1.692	0.000165545327089661\\
1.71	0.000160868298879924\\
1.728	0.000157567818031226\\
1.746	0.000133710821681857\\
1.764	0.000114870215542637\\
1.782	0.000103555794564382\\
1.8	9.4304084354507e-05\\
};
\addlegendentry{$\displaystyle \sigma^2_{\displaystyle \hat K^\epsilon_N}/\epsilon$}

\addplot [color=black, dashed, line width=0.5pt]
  table[row sep=crcr]{%
0.036	0.0146438373838238\\
0.054	0.0126669220637125\\
0.072	0.00826782084741035\\
0.09	0.0104507183626077\\
0.108	0.00981244815945141\\
0.126	0.00925486828987564\\
0.144	0.0106777563114326\\
0.162	0.00948845181455197\\
0.18	0.00874955780884123\\
0.198	0.00876037052617881\\
0.216	0.00877605386173601\\
0.234	0.00844783344775495\\
0.252	0.00905093444966244\\
0.27	0.00809324263980141\\
0.288	0.00819212236063316\\
0.306	0.0082603209916007\\
0.324	0.00797634205576351\\
0.342	0.00821541941855442\\
0.36	0.0079928117068626\\
0.378	0.00814608409137146\\
0.396	0.00815021270487853\\
0.414	0.00824381167746035\\
0.432	0.00817770697653864\\
0.45	0.0080191862509103\\
0.468	0.00833470778185984\\
0.486	0.00845300101980851\\
0.504	0.00799535546656592\\
0.522	0.00842126034712141\\
0.54	0.00840124946611376\\
0.558	0.00855567820990499\\
0.576	0.00866696475056397\\
0.594	0.00858915017547762\\
0.612	0.00848380131536361\\
0.63	0.00836929867998222\\
0.648	0.00837168516796729\\
0.666	0.00830089137547874\\
0.684	0.00830969789660121\\
0.702	0.00821037398932324\\
0.72	0.00821295342507313\\
0.738	0.00782733638748782\\
0.756	0.00777949264994053\\
0.774	0.00753810707350592\\
0.792	0.00741432459703303\\
0.81	0.00729312577502207\\
0.828	0.006967820026374\\
0.846	0.00676576958240119\\
0.864	0.00646497058390452\\
0.882	0.00631408015162236\\
0.9	0.00585107698471309\\
0.918	0.00564307085374206\\
0.936	0.00544390949677208\\
0.954	0.00507446249859933\\
0.972	0.00480121785557759\\
0.99	0.00441011164046342\\
1.008	0.00430606407456971\\
1.026	0.00400351319994542\\
1.044	0.00383171893501444\\
1.062	0.0035757487558633\\
1.08	0.00332777346760162\\
1.098	0.003030236401242\\
1.116	0.00282084706583453\\
1.134	0.00257309428212746\\
1.152	0.00236229503838291\\
1.17	0.00227537848619395\\
1.188	0.00202920162980634\\
1.206	0.00184835110227484\\
1.224	0.00172856412472242\\
1.242	0.00155065086541158\\
1.26	0.00139252545021396\\
1.278	0.00131120651953559\\
1.296	0.00117594924477237\\
1.314	0.00106048468854776\\
1.332	0.000967738068884298\\
1.35	0.000881283018167833\\
1.368	0.000799224990447067\\
1.386	0.000704180384075612\\
1.404	0.000662869045009062\\
1.422	0.000570162653977358\\
1.44	0.000494722127627653\\
1.458	0.000487824266962991\\
1.476	0.000444023556222982\\
1.494	0.000390224501608083\\
1.512	0.00033866932748036\\
1.53	0.000325034212824479\\
1.548	0.000281662405494264\\
1.566	0.000242233423055986\\
1.584	0.000224454862320486\\
1.602	0.000191086362447877\\
1.62	0.000178656109957078\\
1.638	0.000148504788763734\\
1.656	0.000136429812352081\\
1.674	0.000123411925561205\\
1.692	9.78400278307689e-05\\
1.71	9.40750285847508e-05\\
1.728	9.1185079879182e-05\\
1.746	7.65812266219109e-05\\
1.764	6.51191698087511e-05\\
1.782	5.81121181618308e-05\\
1.8	5.23911579747261e-05\\
};
\addlegendentry{$\displaystyle \sigma^2_{\displaystyle \hat K^\epsilon_N}/\epsilon^2$}

\end{axis}

\node[] at (2.85,0.4) {(c')};

\end{tikzpicture}%
        \captionlistentry{}
        \label{fig:a0_bounds_PSD2}
    \end{subfigure}%
    
    \caption{Monte Carlo estimation of $P^\epsilon_1$ (for $a=0$ and $X_f=2$) using $\hat I_N^\eps$ (standard MC), $\hat J_N^\eps$ (simple control variate) and $K_N^\eps$ (practical optimal control variate). Results have been obtained with the same methodology as in Figure \ref{fig:P1_PSD1_PSD2}. The left column (a)-(b)-(c) corresponds to PSD1, and the right column (a')-(b')-(c') corresponds to PSD2.}
    \label{fig:a0_PSD1_PSD2}
\end{figure}

\begin{figure}[h!]
    \centering
    \begin{subfigure}{0.45\textwidth}
        \centering
%
%
\definecolor{mycolor1}{rgb}{0.00000,0.44700,0.74100}%
\definecolor{mycolor2}{rgb}{0.85000,0.32500,0.09800}%
\definecolor{mycolor3}{rgb}{0.92900,0.69400,0.12500}%
\begin{tikzpicture}[scale=0.95]

\begin{axis}[%
width=0.8\linewidth,
height=0.64\linewidth,
scale only axis,
xmin=0,
xmax=1.2,
xtick distance = 0.3,
xlabel = $\epsilon$,
y label style={at={(axis description cs:0.1,.5)},anchor=south},
axis background/.style={fill=white},
legend style={legend cell align=left, align=left, draw=white!15!black, minimum height=0.5cm, font={\tiny\arraycolsep=2pt}},
]
\addplot [color=mycolor1, line width=1pt]
  table[row sep=crcr]{%
0.012	0.00161\\
0.024	0.0014\\
0.036	0.00155\\
0.048	0.0015\\
0.06	0.0014\\
0.072	0.00141\\
0.084	0.00134\\
0.096	0.00165\\
0.108	0.00122\\
0.12	0.00149\\
0.132	0.00142\\
0.144	0.00149\\
0.156	0.00107\\
0.168	0.00157\\
0.18	0.00127\\
0.192	0.00139\\
0.204	0.00119\\
0.216	0.00136\\
0.228	0.00156\\
0.24	0.00123\\
0.252	0.0013\\
0.264	0.00117\\
0.276	0.00129\\
0.288	0.00101\\
0.3	0.00156\\
0.312	0.00124\\
0.324	0.00126\\
0.336	0.00129\\
0.348	0.00133\\
0.36	0.00118\\
0.372	0.00107\\
0.384	0.00121\\
0.396	0.00113\\
0.408	0.0011\\
0.42	0.00085\\
0.432	0.00117\\
0.444	0.00094\\
0.456	0.00108\\
0.468	0.00092\\
0.48	0.00093\\
0.492	0.00078\\
0.504	0.00072\\
0.516	0.00092\\
0.528	0.00078\\
0.54	0.0007\\
0.552	0.00073\\
0.564	0.00076\\
0.576	0.00085\\
0.588	0.00056\\
0.6	0.0007\\
0.612	0.00055\\
0.624	0.00044\\
0.636	0.00053\\
0.648	0.00039\\
0.66	0.00042\\
0.672	0.00045\\
0.684	0.00034\\
0.696	0.00029\\
0.708	0.00032\\
0.72	0.00035\\
0.732	0.00032\\
0.744	0.00017\\
0.756	0.00024\\
0.768	0.00023\\
0.78	0.00028\\
0.792	0.00017\\
0.804	0.00015\\
0.816	0.00014\\
0.828	0.00015\\
0.84	9e-05\\
0.852	7e-05\\
0.864	0.0001\\
0.876	0.00014\\
0.888	4e-05\\
0.9	0.0001\\
0.912	3e-05\\
0.924	7e-05\\
0.936	7e-05\\
0.948	0.0001\\
0.96	6e-05\\
0.972	2e-05\\
0.984	6e-05\\
0.996	2e-05\\
1.008	3e-05\\
1.02	2e-05\\
1.032	2e-05\\
1.044	2e-05\\
1.056	1e-05\\
1.068	1e-05\\
1.08	1e-05\\
1.092	0\\
1.104	1e-05\\
1.116	1e-05\\
1.128	0\\
1.14	0\\
1.152	1e-05\\
1.164	2e-05\\
1.176	0\\
1.188	0\\
1.2	0\\
};
\addlegendentry{$\hat I^\epsilon_N$}

\addplot [color=mycolor2, line width=1pt]
  table[row sep=crcr]{%
0.012	0.001419135783723\\
0.024	0.001419135783723\\
0.036	0.001419135783723\\
0.048	0.001429135783723\\
0.06	0.001409135783723\\
0.072	0.001419135783723\\
0.084	0.001399135783723\\
0.096	0.001459135783723\\
0.108	0.001409135783723\\
0.12	0.001419135783723\\
0.132	0.001439135783723\\
0.144	0.001429135783723\\
0.156	0.001389135783723\\
0.168	0.001409135783723\\
0.18	0.001369135783723\\
0.192	0.001379135783723\\
0.204	0.001429135783723\\
0.216	0.001359135783723\\
0.228	0.001319135783723\\
0.24	0.001359135783723\\
0.252	0.001359135783723\\
0.264	0.001369135783723\\
0.276	0.001239135783723\\
0.288	0.001359135783723\\
0.3	0.001379135783723\\
0.312	0.001259135783723\\
0.324	0.001289135783723\\
0.336	0.001259135783723\\
0.348	0.001279135783723\\
0.36	0.001219135783723\\
0.372	0.001199135783723\\
0.384	0.001139135783723\\
0.396	0.001129135783723\\
0.408	0.001109135783723\\
0.42	0.001069135783723\\
0.432	0.001049135783723\\
0.444	0.000899135783723\\
0.456	0.000979135783723\\
0.468	0.000899135783723\\
0.48	0.001039135783723\\
0.492	0.000879135783723\\
0.504	0.000949135783723\\
0.516	0.001009135783723\\
0.528	0.000799135783723\\
0.54	0.000569135783723\\
0.552	0.000829135783723\\
0.564	0.000569135783723\\
0.576	0.000639135783723\\
0.588	0.000749135783723\\
0.6	0.000449135783723\\
0.612	0.000539135783723\\
0.624	0.000549135783723\\
0.636	0.000539135783723\\
0.648	0.000439135783723\\
0.66	0.000389135783723\\
0.672	0.000529135783723\\
0.684	0.000409135783723\\
0.696	0.000549135783723\\
0.708	0.000409135783723\\
0.72	0.000149135783723\\
0.732	0.000419135783723\\
0.744	0.000409135783723\\
0.756	0.000189135783723\\
0.768	0.000139135783723\\
0.78	0.000119135783723\\
0.792	0.000199135783723\\
0.804	0.000109135783723\\
0.816	0.000149135783723\\
0.828	0.000219135783723\\
0.84	0.000179135783723\\
0.852	-2.0864216277e-05\\
0.864	-5.08642162769999e-05\\
0.876	5.9135783723e-05\\
0.888	1.91357837230001e-05\\
0.9	-4.08642162769998e-05\\
0.912	0.000129135783723\\
0.924	-6.08642162769999e-05\\
0.936	0.000189135783723\\
0.948	0.000269135783723\\
0.96	3.91357837230002e-05\\
0.972	-7.08642162769999e-05\\
0.984	5.9135783723e-05\\
0.996	9.91357837230001e-05\\
1.008	-1.0864216277e-05\\
1.02	0.000239135783723\\
1.032	5.9135783723e-05\\
1.044	0.000119135783723\\
1.056	8.91357837230001e-05\\
1.068	-4.08642162769998e-05\\
1.08	2.91357837230001e-05\\
1.092	-3.08642162769998e-05\\
1.104	-0.000150864216277\\
1.116	0.000119135783723\\
1.128	0.000139135783723\\
1.14	8.91357837230001e-05\\
1.152	9.13578372300008e-06\\
1.164	-0.000150864216277\\
1.176	8.91357837230001e-05\\
1.188	3.91357837230002e-05\\
1.2	-7.08642162769999e-05\\
};
\addlegendentry{$\hat J^\epsilon_N$}

\addplot [color=mycolor3, line width=1.0pt]
  table[row sep=crcr]{%
0.012	0.00141913577871091\\
0.024	0.00141913723182749\\
0.036	0.001429136732609\\
0.048	0.00140907099039294\\
0.06	0.00141913578318146\\
0.072	0.00139826598981822\\
0.084	0.00145914775594261\\
0.096	0.0014060530937544\\
0.108	0.00141961232977067\\
0.12	0.00143899851969907\\
0.132	0.00143078395870069\\
0.144	0.00137459498885898\\
0.156	0.00141117625615043\\
0.168	0.00136237186641644\\
0.18	0.00137959184342119\\
0.192	0.00141289082592898\\
0.204	0.0013592210573807\\
0.216	0.00134237370617913\\
0.228	0.00134510921333974\\
0.24	0.00135391422865291\\
0.252	0.00135279726062162\\
0.264	0.00124709675306222\\
0.276	0.00131665797949582\\
0.288	0.00139837946634412\\
0.3	0.0012557169507351\\
0.312	0.00128473162543922\\
0.324	0.00126467316455292\\
0.336	0.0012870987906225\\
0.348	0.00121005608319334\\
0.36	0.00116708791398739\\
0.372	0.00116102626689837\\
0.384	0.00112934884670425\\
0.396	0.00110634859543312\\
0.408	0.000985104186428847\\
0.42	0.00109389617354426\\
0.432	0.000915932598746204\\
0.444	0.00101365048916008\\
0.456	0.000907975981690061\\
0.468	0.000994130445622756\\
0.48	0.000830307626121775\\
0.492	0.000843208895208028\\
0.504	0.000970920148244058\\
0.516	0.000789292604827519\\
0.528	0.00064597420483883\\
0.54	0.000776552510794154\\
0.552	0.00068177841620534\\
0.564	0.000753000358761817\\
0.576	0.000635331326245325\\
0.588	0.000614411584300449\\
0.6	0.000546734443605705\\
0.612	0.000472485046150316\\
0.624	0.000533044712139517\\
0.636	0.000401473506909824\\
0.648	0.00041191268845096\\
0.66	0.000473027194176929\\
0.672	0.000354847876474504\\
0.684	0.000341361827963829\\
0.696	0.000336748498564165\\
0.708	0.000317781444014879\\
0.72	0.000341024675191642\\
0.732	0.000200392353575051\\
0.744	0.000232388700509665\\
0.756	0.000217365143911265\\
0.768	0.000255571774643951\\
0.78	0.000172933662283494\\
0.792	0.000146642528857003\\
0.804	0.000140907099039294\\
0.816	0.000157168919121798\\
0.828	9.53606262896156e-05\\
0.84	6.57877746295524e-05\\
0.852	9.23157850006798e-05\\
0.864	0.000132452706098184\\
0.876	3.94204386122055e-05\\
0.888	9.09703575801151e-05\\
0.9	3.15010534702897e-05\\
0.912	6.40900685341448e-05\\
0.924	7.54973216271414e-05\\
0.936	0.000108111557850339\\
0.948	5.91306579183082e-05\\
0.96	1.93991635298116e-05\\
0.972	5.99695785038695e-05\\
0.984	2.11811255729619e-05\\
0.996	2.91603252848628e-05\\
1.008	2.36521167394282e-05\\
1.02	2.05671846409812e-05\\
1.032	2.15020461310484e-05\\
1.044	9.99920758402871e-06\\
1.056	1.00005093900697e-05\\
1.068	1.01366841336493e-05\\
1.08	0\\
1.092	8.98188874992959e-06\\
1.104	1.08330900012187e-05\\
1.116	0\\
1.128	0\\
1.14	9.99391396988349e-06\\
1.152	1.78508041001608e-05\\
1.164	0\\
1.176	0\\
1.188	0\\
1.2	0\\
};
\addlegendentry{$\hat K^\epsilon_N$}

\addplot [color=black, dashed, line width=1.0pt]
  table[row sep=crcr]{%
0	0.001419135783723\\
0.133333333333333	0.001419135783723\\
0.266666666666667	0.001419135783723\\
0.4	0.001419135783723\\
0.533333333333333	0.001419135783723\\
0.666666666666667	0.001419135783723\\
0.8	0.001419135783723\\
0.933333333333333	0.001419135783723\\
1.06666666666667	0.001419135783723\\
1.2	0.001419135783723\\
};
\addlegendentry{$P^0_1$}

\end{axis}

\node[] at (2.85,0.65) {(a)};

\end{tikzpicture}%
        \captionlistentry{}
        \label{fig:a1_estimates_PSD1}
    \end{subfigure}%
    \setcounter{subfigure}{0}
    \renewcommand{\thesubfigure}{\alph{subfigure}'}
    \begin{subfigure}{.45\textwidth}
        \centering
%
%
\definecolor{mycolor1}{rgb}{0.00000,0.44700,0.74100}%
\definecolor{mycolor2}{rgb}{0.85000,0.32500,0.09800}%
\definecolor{mycolor3}{rgb}{0.92900,0.69400,0.12500}%
\begin{tikzpicture}[scale=0.95]

\begin{axis}[%
width=0.8\linewidth,
height=0.64\linewidth,
scale only axis,
xmin=0,
xmax=1.8,
xtick distance = 0.45,
xlabel = $\epsilon$,
y label style={at={(axis description cs:0.1,.5)},anchor=south},
axis background/.style={fill=white},
legend style={legend cell align=left, align=left, draw=white!15!black, minimum height=0.5cm, font={\tiny\arraycolsep=2pt}},
]
\addplot [color=mycolor1, line width=1pt]
  table[row sep=crcr]{%
0.018	0.00135\\
0.036	0.00133\\
0.054	0.00169\\
0.072	0.00134\\
0.09	0.00147\\
0.108	0.00145\\
0.126	0.00127\\
0.144	0.00161\\
0.162	0.0014\\
0.18	0.00146\\
0.198	0.00143\\
0.216	0.00143\\
0.234	0.0014\\
0.252	0.00144\\
0.27	0.00161\\
0.288	0.00121\\
0.306	0.00141\\
0.324	0.00141\\
0.342	0.00128\\
0.36	0.00138\\
0.378	0.00154\\
0.396	0.00168\\
0.414	0.00171\\
0.432	0.00168\\
0.45	0.00152\\
0.468	0.00149\\
0.486	0.00169\\
0.504	0.00167\\
0.522	0.00188\\
0.54	0.00165\\
0.558	0.00191\\
0.576	0.00195\\
0.594	0.00176\\
0.612	0.00189\\
0.63	0.00216\\
0.648	0.00201\\
0.666	0.00219\\
0.684	0.00236\\
0.702	0.00231\\
0.72	0.00249\\
0.738	0.00193\\
0.756	0.00221\\
0.774	0.00221\\
0.792	0.00196\\
0.81	0.00209\\
0.828	0.00196\\
0.846	0.00196\\
0.864	0.00213\\
0.882	0.002\\
0.9	0.00181\\
0.918	0.00151\\
0.936	0.00185\\
0.954	0.00157\\
0.972	0.00133\\
0.99	0.00137\\
1.008	0.00135\\
1.026	0.00148\\
1.044	0.00128\\
1.062	0.00103\\
1.08	0.00109\\
1.098	0.00095\\
1.116	0.00085\\
1.134	0.00091\\
1.152	0.00087\\
1.17	0.00076\\
1.188	0.00088\\
1.206	0.00055\\
1.224	0.00054\\
1.242	0.00045\\
1.26	0.00045\\
1.278	0.0004\\
1.296	0.0004\\
1.314	0.00033\\
1.332	0.0002\\
1.35	0.00027\\
1.368	0.00029\\
1.386	0.00026\\
1.404	0.0002\\
1.422	0.00016\\
1.44	0.00015\\
1.458	0.0001\\
1.476	0.0001\\
1.494	9e-05\\
1.512	5e-05\\
1.53	9e-05\\
1.548	9e-05\\
1.566	4e-05\\
1.584	2e-05\\
1.602	3e-05\\
1.62	8e-05\\
1.638	6e-05\\
1.656	1e-05\\
1.674	0\\
1.692	2e-05\\
1.71	1e-05\\
1.728	1e-05\\
1.746	3e-05\\
1.764	3e-05\\
1.782	2e-05\\
1.8	0\\
};
\addlegendentry{$\hat I^\epsilon_N$}

\addplot [color=mycolor2, line width=1pt]
  table[row sep=crcr]{%
0.018	0.001419135783723\\
0.036	0.001409135783723\\
0.054	0.001419135783723\\
0.072	0.001419135783723\\
0.09	0.001419135783723\\
0.108	0.001409135783723\\
0.126	0.001429135783723\\
0.144	0.001409135783723\\
0.162	0.001449135783723\\
0.18	0.001439135783723\\
0.198	0.001389135783723\\
0.216	0.001419135783723\\
0.234	0.001469135783723\\
0.252	0.001509135783723\\
0.27	0.001459135783723\\
0.288	0.001369135783723\\
0.306	0.001519135783723\\
0.324	0.001499135783723\\
0.342	0.001469135783723\\
0.36	0.001559135783723\\
0.378	0.001569135783723\\
0.396	0.001559135783723\\
0.414	0.001629135783723\\
0.432	0.001599135783723\\
0.45	0.001539135783723\\
0.468	0.001649135783723\\
0.486	0.001669135783723\\
0.504	0.001709135783723\\
0.522	0.001899135783723\\
0.54	0.001729135783723\\
0.558	0.001969135783723\\
0.576	0.001929135783723\\
0.594	0.001839135783723\\
0.612	0.001889135783723\\
0.63	0.001869135783723\\
0.648	0.002049135783723\\
0.666	0.002069135783723\\
0.684	0.002229135783723\\
0.702	0.002419135783723\\
0.72	0.002359135783723\\
0.738	0.001879135783723\\
0.756	0.002209135783723\\
0.774	0.002219135783723\\
0.792	0.002079135783723\\
0.81	0.002159135783723\\
0.828	0.001929135783723\\
0.846	0.001689135783723\\
0.864	0.002099135783723\\
0.882	0.001909135783723\\
0.9	0.001869135783723\\
0.918	0.001519135783723\\
0.936	0.001629135783723\\
0.954	0.001609135783723\\
0.972	0.001349135783723\\
0.99	0.001469135783723\\
1.008	0.001359135783723\\
1.026	0.001419135783723\\
1.044	0.001369135783723\\
1.062	0.000959135783723\\
1.08	0.001139135783723\\
1.098	0.000919135783723\\
1.116	0.000819135783723\\
1.134	0.000879135783723\\
1.152	0.000669135783723\\
1.17	0.000669135783723\\
1.188	0.000779135783723\\
1.206	0.000799135783723\\
1.224	0.000489135783723\\
1.242	0.000299135783723\\
1.26	0.000569135783723\\
1.278	0.000169135783723\\
1.296	0.000269135783723\\
1.314	0.000409135783723\\
1.332	0.000269135783723\\
1.35	0.000179135783723\\
1.368	0.000229135783723\\
1.386	0.000199135783723\\
1.404	-1.0864216277e-05\\
1.422	-2.0864216277e-05\\
1.44	0.000239135783723\\
1.458	0.000139135783723\\
1.476	-6.08642162769999e-05\\
1.494	0.000189135783723\\
1.512	5.9135783723e-05\\
1.53	5.9135783723e-05\\
1.548	0.000139135783723\\
1.566	-7.08642162769999e-05\\
1.584	-4.08642162769998e-05\\
1.602	-8.08642162769999e-05\\
1.62	0.000159135783723\\
1.638	0.000109135783723\\
1.656	-6.08642162769999e-05\\
1.674	-8.64216276999941e-07\\
1.692	0.000119135783723\\
1.71	-0.000170864216277\\
1.728	0.000129135783723\\
1.746	6.9135783723e-05\\
1.764	3.91357837230002e-05\\
1.782	6.9135783723e-05\\
1.8	5.9135783723e-05\\
};
\addlegendentry{$\hat J^\epsilon_N$}

\addplot [color=mycolor3, line width=1pt]
  table[row sep=crcr]{%
0.018	0.00141913553861355\\
0.036	0.00140854485060196\\
0.054	0.00141914756205771\\
0.072	0.00141913541338844\\
0.09	0.00141948239723147\\
0.108	0.00141025662453439\\
0.126	0.0014240727867043\\
0.144	0.00141410631639471\\
0.162	0.00144769763390503\\
0.18	0.00143971660508506\\
0.198	0.00139137700929961\\
0.216	0.00141966836094398\\
0.234	0.00146554242542793\\
0.252	0.00150605246365884\\
0.27	0.0014697305899233\\
0.288	0.00135143684342217\\
0.306	0.00151078203455861\\
0.324	0.00149241728628643\\
0.342	0.0014460277359376\\
0.36	0.00153741003996179\\
0.378	0.00156682247196796\\
0.396	0.00158116287603671\\
0.414	0.0016404910914623\\
0.432	0.00161210836872698\\
0.45	0.00153625908824417\\
0.468	0.00162506947844528\\
0.486	0.00167305869444184\\
0.504	0.00170117281705532\\
0.522	0.00189543093514761\\
0.54	0.00170841341640751\\
0.558	0.00195734691207408\\
0.576	0.00193378962340889\\
0.594	0.00181781334462937\\
0.612	0.00188936169381174\\
0.63	0.00195276699093271\\
0.648	0.00203576379236025\\
0.666	0.00210380118106374\\
0.684	0.00226137568537306\\
0.702	0.00238982911489223\\
0.72	0.00239562065685719\\
0.738	0.00189925768901851\\
0.756	0.00220947776976101\\
0.774	0.00221549494481564\\
0.792	0.00202766947918641\\
0.81	0.00213087990266018\\
0.828	0.00194492582306803\\
0.846	0.00183691671967037\\
0.864	0.00211535740882858\\
0.882	0.00195799604776506\\
0.9	0.00182905098991306\\
0.918	0.00151329528353175\\
0.936	0.00176407933090221\\
0.954	0.00158697752510192\\
0.972	0.00133627079868034\\
0.99	0.00140295269743251\\
1.008	0.00135232348319076\\
1.026	0.00146155681758772\\
1.044	0.00130270255630329\\
1.062	0.00101864173420038\\
1.08	0.00110000247772103\\
1.098	0.000945126554439562\\
1.116	0.000844057635467771\\
1.134	0.000906191146149386\\
1.152	0.00084036982726321\\
1.17	0.000743797435553604\\
1.188	0.000868126345363062\\
1.206	0.000571180289369113\\
1.224	0.000535869220780849\\
1.242	0.000436594176045392\\
1.26	0.000455451991595196\\
1.278	0.000384676609316122\\
1.296	0.000392442147122066\\
1.314	0.000336478757499922\\
1.332	0.00020306299854536\\
1.35	0.000265805974855635\\
1.368	0.0002883702638742\\
1.386	0.000256721022380069\\
1.404	0.000192267907459384\\
1.422	0.000158896794348777\\
1.44	0.000153342023872419\\
1.458	0.000100280065236494\\
1.476	9.69569443796598e-05\\
1.494	9.07430817794546e-05\\
1.512	5.01293111165573e-05\\
1.53	8.97896161971864e-05\\
1.548	9.03547184743034e-05\\
1.566	3.855302688215e-05\\
1.584	2.00012190855207e-05\\
1.602	3.00033309961537e-05\\
1.62	7.99936606722297e-05\\
1.638	5.9997047813684e-05\\
1.656	9.52440283833413e-06\\
1.674	0\\
1.692	2.07500304047655e-05\\
1.71	1.00018115035324e-05\\
1.728	1.09164190871198e-05\\
1.746	2.99988243053345e-05\\
1.764	2.99997255395153e-05\\
1.782	1.99990159378947e-05\\
1.8	0\\
};
\addlegendentry{$\hat K^\epsilon_N$}

\addplot [color=black, dashed, line width=1.0pt]
  table[row sep=crcr]{%
0	0.001419135783723\\
0.133333333333333	0.001419135783723\\
0.266666666666667	0.001419135783723\\
0.4	0.001419135783723\\
0.533333333333333	0.001419135783723\\
0.666666666666667	0.001419135783723\\
0.8	0.001419135783723\\
0.933333333333333	0.001419135783723\\
1.06666666666667	0.001419135783723\\
1.2	0.001419135783723\\
};
\addlegendentry{$P^0_1$}

\end{axis}

\node[] at (2.85,0.65) {(a')};

\end{tikzpicture}%
        \captionlistentry{}
        \label{fig:a1_estimates_PSD2}
    \end{subfigure}%

    \centering
    \renewcommand{\thesubfigure}{\alph{subfigure}}
    \begin{subfigure}{.45\textwidth}
        \centering
%
%
\definecolor{mycolor1}{rgb}{0.00000,0.44700,0.74100}%
\definecolor{mycolor2}{rgb}{0.85000,0.32500,0.09800}%
\definecolor{mycolor3}{rgb}{0.92900,0.69400,0.12500}%
\begin{tikzpicture}[scale=0.95]

\begin{axis}[%
width=0.8\linewidth,
height=0.64\linewidth,
scale only axis,
xmin=0,
xmax=1.2,
xtick distance = 0.3,
xlabel = $\epsilon$,
y label style={at={(axis description cs:0.1,.5)},anchor=south},
axis background/.style={fill=white},
legend style={legend cell align=left, align=left, draw=white!15!black, minimum height=0.6cm, font={\tiny\arraycolsep=2pt}, at={(0.98,0.47)},anchor=east}
]
\addplot [color=mycolor1, line width=1pt]
  table[row sep=crcr]{%
0.012	0.001421972224\\
0.024	0.001404023164\\
0.036	0.001413994944\\
0.048	0.001433937904\\
0.06	0.0014379264\\
0.072	0.001428952239\\
0.084	0.001403025975\\
0.096	0.001373109375\\
0.108	0.0013382044\\
0.12	0.00139804\\
0.132	0.001396045596\\
0.144	0.001384079004\\
0.156	0.001326236416\\
0.168	0.001446900399\\
0.18	0.001374106624\\
0.192	0.001403025975\\
0.204	0.001393053975\\
0.216	0.0013282311\\
0.228	0.001392056764\\
0.24	0.001331223111\\
0.252	0.001336209756\\
0.264	0.001255419951\\
0.276	0.001256417436\\
0.288	0.001222501824\\
0.3	0.001325239071\\
0.312	0.0012783616\\
0.324	0.001285343631\\
0.336	0.0012284871\\
0.348	0.001246442496\\
0.36	0.001246442496\\
0.372	0.001216516476\\
0.384	0.001153665975\\
0.396	0.001099787799\\
0.408	0.001066859376\\
0.42	0.001109765679\\
0.432	0.001034926704\\
0.444	0.001014967744\\
0.456	0.001017961639\\
0.468	0.000981035676\\
0.48	0.000935123904\\
0.492	0.000901186396\\
0.504	0.000858262119\\
0.516	0.000807347136\\
0.528	0.000786380631\\
0.54	0.000755428464\\
0.552	0.000704502975\\
0.564	0.000707498736\\
0.576	0.000660563079\\
0.588	0.000623610624\\
0.6	0.000592648351\\
0.612	0.000572671671\\
0.624	0.000498750999\\
0.636	0.000542705151\\
0.648	0.000480768639\\
0.66	0.000437808156\\
0.672	0.000433811644\\
0.684	0.000382853311\\
0.696	0.000377857116\\
0.708	0.000343881664\\
0.72	0.000305906364\\
0.732	0.000282919911\\
0.744	0.0002499375\\
0.756	0.000246938991\\
0.768	0.000213954204\\
0.78	0.000222950271\\
0.792	0.00019996\\
0.804	0.0001399804\\
0.816	0.000185965404\\
0.828	0.000163973104\\
0.84	0.000115986544\\
0.852	0.000121985116\\
0.864	0.000133982044\\
0.876	0.000103989184\\
0.888	7.2994671e-05\\
0.9	8.8992079e-05\\
0.912	6.8995239e-05\\
0.924	6.5995644e-05\\
0.936	5.8996519e-05\\
0.948	5.99964e-05\\
0.96	5.0997399e-05\\
0.972	3.5998704e-05\\
0.984	4.2998151e-05\\
0.996	2.2999471e-05\\
1.008	2.3999424e-05\\
1.02	3.1998976e-05\\
1.032	3.0999039e-05\\
1.044	2.3999424e-05\\
1.056	1.3999804e-05\\
1.068	1.99996e-05\\
1.08	1.0999879e-05\\
1.092	1.0999879e-05\\
1.104	1.1999856e-05\\
1.116	6.999951e-06\\
1.128	8.999919e-06\\
1.14	5.999964e-06\\
1.152	7.999936e-06\\
1.164	7.999936e-06\\
1.176	2.999991e-06\\
1.188	2.999991e-06\\
1.2	2.999991e-06\\
};
\addlegendentry{$\displaystyle \sigma^2_{\displaystyle \hat I^\epsilon_N}$}

\addplot [color=mycolor2, line width=1pt]
  table[row sep=crcr]{%
0.012	9.99999e-07\\
0.024	9.99999e-07\\
0.036	9.999996e-06\\
0.048	1.2999951e-05\\
0.06	1.2999999e-05\\
0.072	2e-05\\
0.084	2.7999936e-05\\
0.096	3.3999936e-05\\
0.108	3.6999991e-05\\
0.12	4.5999984e-05\\
0.132	5.7999964e-05\\
0.144	5.9999964e-05\\
0.156	8.2999711e-05\\
0.168	8.8999991e-05\\
0.18	9.7998844e-05\\
0.192	0.000150998631\\
0.204	0.000165998844\\
0.216	0.000192997599\\
0.228	0.000178999039\\
0.24	0.000214996519\\
0.252	0.000186996279\\
0.264	0.000215992256\\
0.276	0.000260991719\\
0.288	0.000300988551\\
0.3	0.000278991351\\
0.312	0.000314985359\\
0.324	0.000376973431\\
0.336	0.000371969024\\
0.348	0.000375971776\\
0.36	0.000420964279\\
0.372	0.000382964279\\
0.384	0.000470951159\\
0.396	0.000506905751\\
0.408	0.000524904519\\
0.42	0.000556893071\\
0.432	0.000619854076\\
0.444	0.000639809904\\
0.456	0.000608863839\\
0.468	0.0006817884\\
0.48	0.000707761856\\
0.492	0.000749737856\\
0.504	0.000783679644\\
0.516	0.000804662439\\
0.528	0.000809625456\\
0.54	0.000855561756\\
0.552	0.000874528031\\
0.564	0.0009214524\\
0.576	0.000946380631\\
0.588	0.000950367975\\
0.6	0.000970325959\\
0.612	0.001040213231\\
0.624	0.000990279199\\
0.636	0.001047197184\\
0.648	0.001134005991\\
0.66	0.001156930844\\
0.672	0.001105051324\\
0.684	0.0011728336\\
0.696	0.001120037639\\
0.708	0.001207827111\\
0.72	0.0011808336\\
0.732	0.001166816256\\
0.744	0.001241595775\\
0.756	0.001204750076\\
0.768	0.001275523775\\
0.78	0.001283557599\\
0.792	0.001270564796\\
0.804	0.001257547975\\
0.816	0.001268496924\\
0.828	0.001339353911\\
0.84	0.001326315196\\
0.852	0.0013602576\\
0.864	0.001367228439\\
0.876	0.001404084544\\
0.888	0.001327307399\\
0.9	0.001408073456\\
0.912	0.001384117616\\
0.924	0.0014539551\\
0.936	0.001355239071\\
0.948	0.001349239071\\
0.96	0.001373163975\\
0.972	0.001472859631\\
0.984	0.001433972224\\
0.996	0.001439949376\\
1.008	0.0014538975\\
1.02	0.001329270775\\
1.032	0.001429966524\\
1.044	0.001401042799\\
1.056	0.001379114871\\
1.068	0.001342215104\\
1.08	0.001346188284\\
1.092	0.001432952239\\
1.104	0.001429977916\\
1.116	0.001363136775\\
1.128	0.001397048391\\
1.14	0.001458871319\\
1.152	0.001401031591\\
1.164	0.001463856704\\
1.176	0.001402028784\\
1.188	0.001396045596\\
1.2	0.001436929279\\
};
\addlegendentry{$\displaystyle \sigma^2_{\displaystyle \hat J^\epsilon_N}$}

\addplot [color=mycolor3, line width=1pt]
  table[row sep=crcr]{%
0.012	9.99298245614865e-07\\
0.024	9.992892679614e-07\\
0.036	9.97459610704001e-06\\
0.048	1.29936017463353e-05\\
0.06	1.29659597844313e-05\\
0.072	1.99300186547519e-05\\
0.084	2.77706004954437e-05\\
0.096	3.36809587497931e-05\\
0.108	3.67018699804131e-05\\
0.12	4.55544016848449e-05\\
0.132	5.72699783202526e-05\\
0.144	5.92169423979823e-05\\
0.156	8.11401734946894e-05\\
0.168	8.75408480366203e-05\\
0.18	9.49096128519982e-05\\
0.192	0.000144869145753557\\
0.204	0.000158997737136563\\
0.216	0.000182377694992344\\
0.228	0.000171257674080302\\
0.24	0.00020151043050809\\
0.252	0.000176005317791367\\
0.264	0.000198818203333865\\
0.276	0.000238030571035947\\
0.288	0.000269723861738149\\
0.3	0.000254627958419265\\
0.312	0.000281069093145734\\
0.324	0.000326712672306293\\
0.336	0.000318593511923012\\
0.348	0.000323740581092205\\
0.36	0.000356250964441095\\
0.372	0.000324855538652631\\
0.384	0.000383981446495042\\
0.396	0.000389341548400153\\
0.408	0.000398707274166005\\
0.42	0.000421128619654688\\
0.432	0.000442975240611921\\
0.444	0.000440647983075601\\
0.456	0.00043670820246369\\
0.468	0.000455884332497406\\
0.48	0.000456861478250647\\
0.492	0.000468400953318573\\
0.504	0.000464251259940152\\
0.516	0.000459235818681888\\
0.528	0.000448645653460785\\
0.54	0.000449727531094663\\
0.552	0.00043680077812\\
0.564	0.00044508493903601\\
0.576	0.000427871353150378\\
0.588	0.000413905112428532\\
0.6	0.000403231990797364\\
0.612	0.000404489952742183\\
0.624	0.000363101777123292\\
0.636	0.00039143827405024\\
0.648	0.0003661481412904\\
0.66	0.000341952672072768\\
0.672	0.000337813819676287\\
0.684	0.000305830323648799\\
0.696	0.00030819082548833\\
0.708	0.000288662458404527\\
0.72	0.000259081810974456\\
0.732	0.000239928375921673\\
0.744	0.000215963617721745\\
0.756	0.000216808317764144\\
0.768	0.000190563768205592\\
0.78	0.000199991913259763\\
0.792	0.000180993621930385\\
0.804	0.000130505589720935\\
0.816	0.000166951356675399\\
0.828	0.000151404133593338\\
0.84	0.000108785488967114\\
0.852	0.000114925355866628\\
0.864	0.000124972334716112\\
0.876	9.81873788197428e-05\\
0.888	7.04663248365901e-05\\
0.9	8.48807183606471e-05\\
0.912	6.63323589792547e-05\\
0.924	6.41221569500351e-05\\
0.936	5.76029492908927e-05\\
0.948	5.83387166507526e-05\\
0.96	4.98043095320328e-05\\
0.972	3.53995636791213e-05\\
0.984	4.2066766946431e-05\\
0.996	2.2777294551201e-05\\
1.008	2.37008050979775e-05\\
1.02	3.15723189871788e-05\\
1.032	3.04618990170674e-05\\
1.044	2.36598695009759e-05\\
1.056	1.39278859244247e-05\\
1.068	1.98111934178042e-05\\
1.08	1.09263070310662e-05\\
1.092	1.09438240108672e-05\\
1.104	1.19658048106815e-05\\
1.116	6.96428571428581e-06\\
1.128	8.95447979517808e-06\\
1.14	5.9829341521165e-06\\
1.152	7.95464209780298e-06\\
1.164	7.96671095504769e-06\\
1.176	2.99360341151386e-06\\
1.188	2.99357601713062e-06\\
1.2	2.99375866851595e-06\\
};
\addlegendentry{$\displaystyle \sigma^2_{\displaystyle \hat K^\epsilon_N}$}

\end{axis}

\node[] at (2.85,0.65) {(b)};

\end{tikzpicture}%
        \captionlistentry{}
        \label{fig:a1_variances_PSD1}
    \end{subfigure}%
    \setcounter{subfigure}{1}
    \renewcommand{\thesubfigure}{\alph{subfigure}'}
    \begin{subfigure}{.45\textwidth}
        \centering
%
%
\definecolor{mycolor1}{rgb}{0.00000,0.44700,0.74100}%
\definecolor{mycolor2}{rgb}{0.85000,0.32500,0.09800}%
\definecolor{mycolor3}{rgb}{0.92900,0.69400,0.12500}%
\begin{tikzpicture}[scale=0.95]

\begin{axis}[%
width=0.8\linewidth,
height=0.64\linewidth,
scale only axis,
xmin=0,
xmax=1.8,
xtick distance = 0.45,
xlabel = $\epsilon$,
y label style={at={(axis description cs:0.1,.5)},anchor=south},
axis background/.style={fill=white},
legend style={legend cell align=left, align=left, draw=white!15!black, minimum height=0.1cm, font={\tiny}, at={(0.98,0.47)},anchor=east}
]
\addplot [color=mycolor1, line width=1pt]
  table[row sep=crcr]{%
0.018	0.001438923519\\
0.036	0.001414992111\\
0.054	0.001428952239\\
0.072	0.001422969375\\
0.09	0.001391059551\\
0.108	0.001445903296\\
0.126	0.001365131311\\
0.144	0.001412997775\\
0.162	0.001394051184\\
0.18	0.001411003431\\
0.198	0.001403025975\\
0.216	0.001455874236\\
0.234	0.001410006256\\
0.252	0.001475815516\\
0.27	0.001466842039\\
0.288	0.001399037199\\
0.306	0.001471827324\\
0.324	0.001440917751\\
0.342	0.001496752999\\
0.36	0.001580494111\\
0.378	0.001594449591\\
0.396	0.001589465536\\
0.414	0.001563547644\\
0.432	0.001616378839\\
0.45	0.001628339839\\
0.468	0.001628339839\\
0.486	0.0017070759\\
0.504	0.001748930496\\
0.522	0.001803734751\\
0.54	0.001783806631\\
0.558	0.001975083559\\
0.576	0.001898382396\\
0.594	0.001897386199\\
0.612	0.001977075639\\
0.63	0.001996\\
0.648	0.002102560551\\
0.666	0.002180225775\\
0.684	0.002108535231\\
0.702	0.002110526775\\
0.72	0.002255887879\\
0.738	0.002176243239\\
0.756	0.002122475871\\
0.774	0.002206111479\\
0.792	0.002078661111\\
0.81	0.002119488624\\
0.828	0.002031854704\\
0.846	0.002012931711\\
0.864	0.001984047856\\
0.882	0.001922290524\\
0.9	0.001898382396\\
0.918	0.0018166876\\
0.936	0.001704086151\\
0.954	0.001643290684\\
0.972	0.001489773936\\
0.99	0.001479803676\\
1.008	0.001400034396\\
1.026	0.001276366716\\
1.044	0.001227489559\\
1.062	0.001069852959\\
1.08	0.001007981919\\
1.098	0.000953089884\\
1.116	0.0009790396\\
1.134	0.000812339031\\
1.152	0.000792371151\\
1.17	0.000707498736\\
1.188	0.000660563079\\
1.206	0.000603635184\\
1.224	0.000523725424\\
1.242	0.000468780039\\
1.26	0.000498750999\\
1.278	0.000381854076\\
1.296	0.000334887775\\
1.314	0.000302908191\\
1.332	0.0002799216\\
1.35	0.000252935991\\
1.368	0.000227948016\\
1.386	0.000213954204\\
1.404	0.000206957151\\
1.422	0.000162973431\\
1.44	0.000167971776\\
1.458	0.000138980679\\
1.476	0.000104988975\\
1.494	0.000101989596\\
1.512	8.5992604e-05\\
1.53	9.2991351e-05\\
1.548	9.0991719e-05\\
1.566	6.2996031e-05\\
1.584	5.1997296e-05\\
1.602	4.99975e-05\\
1.62	4.8997599e-05\\
1.638	3.3998844e-05\\
1.656	2.7999216e-05\\
1.674	2.4999375e-05\\
1.692	1.7999676e-05\\
1.71	1.6999711e-05\\
1.728	1.5999744e-05\\
1.746	1.7999676e-05\\
1.764	1.2999831e-05\\
1.782	1.2999831e-05\\
1.8	8.999919e-06\\
};
\addlegendentry{$\displaystyle \sigma^2_{\displaystyle \hat I^\epsilon_N}$}

\addplot [color=mycolor2, line width=1pt]
  table[row sep=crcr]{%
0.018	4.999999e-06\\
0.036	1.4999951e-05\\
0.054	6.999999e-06\\
0.072	1.6e-05\\
0.09	2.2999879e-05\\
0.108	4.3999984e-05\\
0.126	5.59999e-05\\
0.144	7.4999999e-05\\
0.162	7.5999676e-05\\
0.18	0.000106999999\\
0.198	0.000119999984\\
0.216	0.000150999559\\
0.234	0.000177999516\\
0.252	0.0001859991\\
0.27	0.0002379984\\
0.288	0.000272997191\\
0.306	0.000292997599\\
0.324	0.000329989184\\
0.342	0.000382995239\\
0.36	0.000428988975\\
0.378	0.000434982311\\
0.396	0.000517986076\\
0.414	0.000525958384\\
0.432	0.00059596\\
0.45	0.000662934975\\
0.468	0.000691946176\\
0.486	0.000816905751\\
0.504	0.000818880975\\
0.522	0.000936859375\\
0.54	0.000999804636\\
0.558	0.001092776271\\
0.576	0.001161808156\\
0.594	0.001212720159\\
0.612	0.0012536975\\
0.63	0.001324655431\\
0.648	0.001478508599\\
0.666	0.001530601839\\
0.684	0.001587529404\\
0.702	0.001633428464\\
0.72	0.001781431484\\
0.738	0.001758423919\\
0.756	0.0018895239\\
0.774	0.001888374319\\
0.792	0.001923487344\\
0.81	0.001971501564\\
0.828	0.001990636391\\
0.846	0.002038687519\\
0.864	0.002068746991\\
0.882	0.002101799296\\
0.9	0.002118756951\\
0.918	0.001999804636\\
0.936	0.002059917056\\
0.954	0.002099963136\\
0.972	0.001964986311\\
0.99	0.002034995511\\
1.008	0.002093996636\\
1.026	0.001969971776\\
1.044	0.001885966876\\
1.062	0.001840903279\\
1.08	0.001885825276\\
1.098	0.001824780039\\
1.116	0.001903782844\\
1.134	0.001852594231\\
1.152	0.001808560431\\
1.17	0.001761526656\\
1.188	0.001739400924\\
1.206	0.001706374319\\
1.224	0.001728148071\\
1.242	0.001724017919\\
1.26	0.001679175536\\
1.278	0.001635852959\\
1.296	0.001598798784\\
1.314	0.001569761231\\
1.332	0.001553675199\\
1.35	0.001546714044\\
1.368	0.001580531056\\
1.386	0.001543523775\\
1.404	0.0016282576\\
1.422	0.001535348775\\
1.44	0.001509414919\\
1.458	0.001531249671\\
1.476	0.001520134044\\
1.494	0.001492225776\\
1.512	0.001432304796\\
1.53	0.001413404831\\
1.548	0.001469244375\\
1.566	0.001453163975\\
1.584	0.001465092839\\
1.602	0.001473053975\\
1.62	0.001391286519\\
1.638	0.001482969375\\
1.656	0.001512842039\\
1.674	0.001486906191\\
1.692	0.001416051184\\
1.71	0.001416073456\\
1.728	0.001402106624\\
1.746	0.00143004\\
1.764	0.0014300119\\
1.782	0.001423037199\\
1.8	0.001462900399\\
};
\addlegendentry{$\displaystyle \sigma^2_{\displaystyle \hat J^\epsilon_N}$}

\addplot [color=mycolor3, line width=1pt]
  table[row sep=crcr]{%
0.018	4.99375466292863e-06\\
0.036	1.49885313114235e-05\\
0.054	6.98881780275502e-06\\
0.072	1.59550236279689e-05\\
0.09	2.2794123493581e-05\\
0.108	4.37224148570617e-05\\
0.126	5.52086205995436e-05\\
0.144	7.39788550474478e-05\\
0.162	7.53874831642414e-05\\
0.18	0.000104934952227542\\
0.198	0.000117268455242087\\
0.216	0.000148052440259989\\
0.234	0.000173613007663848\\
0.252	0.000181786661627903\\
0.27	0.000231122008943508\\
0.288	0.000263997134025086\\
0.306	0.000282525804991412\\
0.324	0.000320416626828638\\
0.342	0.00036571178791967\\
0.36	0.000411172177880472\\
0.378	0.000419344772196808\\
0.396	0.00049076168466537\\
0.414	0.000506835022087366\\
0.432	0.000568213463928571\\
0.45	0.00063254484265881\\
0.468	0.00065397353359975\\
0.486	0.000770336598919735\\
0.504	0.000778739690001839\\
0.522	0.000881428683746646\\
0.54	0.000941605109443905\\
0.558	0.00102857456533646\\
0.576	0.0010718481327061\\
0.594	0.00112698958785358\\
0.612	0.00116659962559183\\
0.63	0.00122772171345627\\
0.648	0.00137018504937004\\
0.666	0.00139952074002824\\
0.684	0.00144416789234201\\
0.702	0.00149075914131688\\
0.72	0.00160507720789628\\
0.738	0.00158160394169257\\
0.756	0.00163781158457901\\
0.774	0.00167494762127009\\
0.792	0.00165538176937932\\
0.81	0.00168763119029159\\
0.828	0.00165321209710264\\
0.846	0.00166172714477966\\
0.864	0.0016544891990088\\
0.882	0.0016376324559162\\
0.9	0.00164818691833316\\
0.918	0.00155812901760668\\
0.936	0.00150541610089467\\
0.954	0.00147274529435412\\
0.972	0.00134298682688727\\
0.99	0.00134961397511755\\
1.008	0.00129996490048086\\
1.026	0.00117889587544144\\
1.044	0.00112754380028094\\
1.062	0.00100265998536033\\
1.08	0.000955464678721623\\
1.098	0.000900007686236179\\
1.116	0.000932372331885694\\
1.134	0.000783647040588176\\
1.152	0.000759429434014195\\
1.17	0.000686764952802482\\
1.188	0.00063868693324049\\
1.206	0.000588509656232095\\
1.224	0.000513719180118043\\
1.242	0.000461739086669286\\
1.26	0.000489775456469527\\
1.278	0.000375174440022265\\
1.296	0.000330122288992753\\
1.314	0.00029908025299983\\
1.332	0.000275707565198755\\
1.35	0.000251431496676952\\
1.368	0.000226681697867151\\
1.386	0.000212292540098583\\
1.404	0.000205205145854416\\
1.422	0.000162038658989266\\
1.44	0.00016674794153682\\
1.458	0.000138198261183433\\
1.476	0.0001044983258\\
1.494	0.000101685898197262\\
1.512	8.57074535074281e-05\\
1.53	9.2781089313453e-05\\
1.548	9.07658498841139e-05\\
1.566	6.28783145035627e-05\\
1.584	5.19416235733776e-05\\
1.602	4.99291935324052e-05\\
1.62	4.89621511270039e-05\\
1.638	3.39881322696284e-05\\
1.656	2.79827701400997e-05\\
1.674	2.49886887846737e-05\\
1.692	1.79652253633172e-05\\
1.71	1.69969277083413e-05\\
1.728	1.59969301635693e-05\\
1.746	1.79969227620054e-05\\
1.764	1.29970678648169e-05\\
1.782	1.29991485825172e-05\\
1.8	8.99925003549551e-06\\
};
\addlegendentry{$\displaystyle \sigma^2_{\displaystyle \hat K^\epsilon_N}$}

\end{axis}

\node[] at (2.85,0.65) {(b')};

\end{tikzpicture}%
        \captionlistentry{}
        \label{fig:a1_variances_PSD2}
    \end{subfigure}%

    \centering
    \renewcommand{\thesubfigure}{\alph{subfigure}}
    \begin{subfigure}{.45\textwidth}
        \centering
%
%
\definecolor{mycolor1}{rgb}{0.00000,0.44700,0.74100}%
\definecolor{mycolor2}{rgb}{0.85000,0.32500,0.09800}%
\definecolor{mycolor3}{rgb}{0.92900,0.69400,0.12500}%
\begin{tikzpicture}[scale=0.95]

\begin{axis}[%
width=0.8\linewidth,
height=0.64\linewidth,
scale only axis,
xmin=0,
xmax=1.2,
xtick distance = 0.3,
xlabel = $\epsilon$,
y label style={at={(axis description cs:0.1,.5)},anchor=south},
axis background/.style={fill=white},
legend style={legend cell align=left, align=left, draw=white!15!black, minimum height=0.6cm, font={\tiny\arraycolsep=2pt}}
]
\addplot [color=black, line width=1pt]
  table[row sep=crcr]{%
0.012	8.32748538012387e-05\\
0.024	4.1637052831725e-05\\
0.036	0.000277072114084445\\
0.048	0.000270700036381985\\
0.06	0.000216099329740521\\
0.072	0.000276805814649332\\
0.084	0.00033060238685052\\
0.096	0.000350843320310345\\
0.108	0.00033983212944827\\
0.12	0.000379620014040374\\
0.132	0.000433863472123126\\
0.144	0.000411228766652655\\
0.156	0.00052012931727365\\
0.168	0.000521076476408454\\
0.18	0.000527275626955545\\
0.192	0.000754526800799777\\
0.204	0.000779400672238053\\
0.216	0.00084434118052011\\
0.228	0.000751130149475009\\
0.24	0.000839626793783707\\
0.252	0.000698433800759395\\
0.264	0.000753099255052518\\
0.276	0.000862429605202708\\
0.288	0.000936541186590793\\
0.3	0.000848759861397551\\
0.312	0.000900862478031197\\
0.324	0.00100837244538979\\
0.336	0.000948194975961346\\
0.348	0.000930289026127026\\
0.36	0.000989586012336376\\
0.372	0.000873267577023202\\
0.384	0.00099995168358084\\
0.396	0.000983185728283214\\
0.408	0.000977223711191188\\
0.42	0.00100268718965402\\
0.432	0.00102540564956463\\
0.444	0.000992450412332434\\
0.456	0.00095769342645546\\
0.468	0.000974111821575653\\
0.48	0.000951794746355514\\
0.492	0.000952034457964578\\
0.504	0.000921133452262206\\
0.516	0.000889991896670325\\
0.528	0.000849707677009063\\
0.54	0.000832828761286414\\
0.552	0.000791305757463768\\
0.564	0.000789157693326259\\
0.576	0.000742832210330517\\
0.588	0.000703920259232197\\
0.6	0.000672053317995606\\
0.612	0.000660931295330364\\
0.624	0.00058189387359502\\
0.636	0.000615468984355724\\
0.648	0.000565043427917284\\
0.66	0.000518110109201163\\
0.672	0.000502699136423046\\
0.684	0.000447120356211695\\
0.696	0.000442802910184382\\
0.708	0.000407715336729558\\
0.72	0.000359835848575633\\
0.732	0.000327771005357477\\
0.744	0.000290273679733529\\
0.756	0.00028678348910601\\
0.768	0.000248129906517698\\
0.78	0.000256399888794568\\
0.792	0.000228527300417152\\
0.804	0.000162320385225044\\
0.816	0.000204597250827695\\
0.828	0.000182855233808379\\
0.84	0.000129506534484659\\
0.852	0.000134888915336418\\
0.864	0.000144643905921425\\
0.876	0.000112086048880985\\
0.888	7.93539694105745e-05\\
0.9	9.43119092896079e-05\\
0.912	7.27328497579547e-05\\
0.924	6.93962737554492e-05\\
0.936	6.15416124902699e-05\\
0.948	6.15387306442538e-05\\
0.96	5.18794890958675e-05\\
0.972	3.64193041966269e-05\\
0.984	4.27507794171047e-05\\
0.996	2.28687696297199e-05\\
1.008	2.35127034702158e-05\\
1.02	3.09532539089988e-05\\
1.032	2.95173440087863e-05\\
1.044	2.26627102499769e-05\\
1.056	1.31892859132809e-05\\
1.068	1.85498065709777e-05\\
1.08	1.01169509546909e-05\\
1.092	1.00218168597685e-05\\
1.104	1.08385913140231e-05\\
1.116	6.24039938556076e-06\\
1.128	7.93836861274653e-06\\
1.14	5.24818785273378e-06\\
1.152	6.90507126545397e-06\\
1.164	6.84425339780729e-06\\
1.176	2.54558113223968e-06\\
1.188	2.51984513226483e-06\\
1.2	2.49479889042996e-06\\
};
\addlegendentry{$\displaystyle \sigma^2_{\displaystyle \hat K^\epsilon_N}/\epsilon$}

\addplot [color=black,dashed, line width=0.5pt]
  table[row sep=crcr]{%
0.012	0.00693957115010323\\
0.024	0.00173487720132188\\
0.036	0.0076964476134568\\
0.048	0.00563958409129136\\
0.06	0.00360165549567535\\
0.072	0.00384452520346295\\
0.084	0.00393574270060143\\
0.096	0.00365461791989943\\
0.108	0.00314659379118768\\
0.12	0.00316350011700312\\
0.132	0.00328684448578126\\
0.144	0.00285575532397677\\
0.156	0.0033341622902157\\
0.168	0.00310164569290747\\
0.18	0.00292930903864192\\
0.192	0.00392982708749884\\
0.204	0.00382059153057869\\
0.216	0.00390898694685236\\
0.228	0.00329443048015355\\
0.24	0.00349844497409878\\
0.252	0.00277156270142617\\
0.264	0.00285264869338075\\
0.276	0.00312474494638662\\
0.288	0.00325187912010692\\
0.3	0.00282919953799184\\
0.312	0.00288737973727948\\
0.324	0.00311226063391911\\
0.336	0.00282200885702782\\
0.348	0.00267324432795122\\
0.36	0.00274885003426771\\
0.372	0.00234749348662151\\
0.384	0.00260404084265844\\
0.396	0.00248279224313943\\
0.408	0.00239515615488036\\
0.42	0.00238735045155719\\
0.432	0.00237362418880702\\
0.444	0.0022352486764244\\
0.456	0.00210020488257776\\
0.468	0.00208143551618729\\
0.48	0.00198290572157399\\
0.492	0.00193502938610687\\
0.504	0.00182764573861549\\
0.516	0.00172479049742311\\
0.528	0.0016092948428202\\
0.54	0.00154227548386373\\
0.552	0.00143352492294161\\
0.564	0.00139921576830897\\
0.576	0.00128963925404604\\
0.588	0.00119714329801394\\
0.6	0.00112008886332601\\
0.612	0.00107995309694504\\
0.624	0.000932522233325353\\
0.636	0.000967718528861201\\
0.648	0.000871980598637784\\
0.66	0.000785015316971459\\
0.672	0.000748064191105723\\
0.684	0.000653684731303647\\
0.696	0.000636211077851123\\
0.708	0.000575869119674518\\
0.72	0.000499772011910601\\
0.732	0.000447774597482893\\
0.744	0.000390152795340764\\
0.756	0.000379343239558214\\
0.768	0.000323085815778252\\
0.78	0.000328717806146882\\
0.792	0.000288544571233778\\
0.804	0.000201891026399308\\
0.816	0.00025073192503394\\
0.828	0.000220839654357946\\
0.84	0.000154174445815071\\
0.852	0.00015832032316481\\
0.864	0.000167411928149798\\
0.876	0.000127952110594732\\
0.888	8.93625781650613e-05\\
0.9	0.000104791010321787\\
0.912	7.97509317521434e-05\\
0.924	7.51041923760273e-05\\
0.936	6.57495859938781e-05\\
0.948	6.49142728314913e-05\\
0.96	5.4041134474862e-05\\
0.972	3.74684199553774e-05\\
0.984	4.34459140417731e-05\\
0.996	2.2960612078032e-05\\
1.008	2.33260947125157e-05\\
1.02	3.03463273617635e-05\\
1.032	2.86020775278937e-05\\
1.044	2.1707576867794e-05\\
1.056	1.24898540845463e-05\\
1.068	1.73687327443611e-05\\
1.08	9.36754718026939e-06\\
1.092	9.177487966821e-06\\
1.104	9.81756459603542e-06\\
1.116	5.59175572182864e-06\\
1.128	7.03756082690295e-06\\
1.14	4.60367355502963e-06\\
1.152	5.99398547348435e-06\\
1.164	5.87994278162139e-06\\
1.176	2.16460980632626e-06\\
1.188	2.12108176116568e-06\\
1.2	2.0789990753583e-06\\
};
\addlegendentry{$\displaystyle \sigma^2_{\displaystyle \hat K^\epsilon_N}/\epsilon^2$}

\end{axis}

\node[] at (2.85,0.4) {(c)};

\end{tikzpicture}%
        \captionlistentry{}
        \label{fig:a1_bounds_PSD1}
    \end{subfigure}%
    \setcounter{subfigure}{2}
    \renewcommand{\thesubfigure}{\alph{subfigure}'}
    \begin{subfigure}{.45\textwidth}
        \centering
%
%
\definecolor{mycolor1}{rgb}{0.00000,0.44700,0.74100}%
\definecolor{mycolor2}{rgb}{0.85000,0.32500,0.09800}%
\definecolor{mycolor3}{rgb}{0.92900,0.69400,0.12500}%
\begin{tikzpicture}[scale=0.95]

\begin{axis}[%
width=0.8\linewidth,
height=0.64\linewidth,
scale only axis,
xmin=0,
xmax=1.8,
xtick distance = 0.45,
xlabel = $\epsilon$,
y label style={at={(axis description cs:0.1,.5)},anchor=south},
axis background/.style={fill=white},
legend style={legend cell align=left, align=left, draw=white!15!black, minimum height=0.6cm, font={\tiny\arraycolsep=2pt}}
]
\addplot [color=black, line width=1pt]
  table[row sep=crcr]{%
0.018	0.000277430814607146\\
0.036	0.000416348091983987\\
0.054	0.000129422551902871\\
0.072	0.000221597550388457\\
0.09	0.000253268038817567\\
0.108	0.000404837174602423\\
0.126	0.000438163655551933\\
0.144	0.00051374204894061\\
0.162	0.000465354834347169\\
0.18	0.000582971956819678\\
0.198	0.000592264925465088\\
0.216	0.000685427964166615\\
0.234	0.000741935930187386\\
0.252	0.000721375641380566\\
0.27	0.000856007440531512\\
0.288	0.000916656715364881\\
0.306	0.000923286944416378\\
0.324	0.000988940206261227\\
0.342	0.00106933271321541\\
0.36	0.00114214493855687\\
0.378	0.00110937770422436\\
0.396	0.00123929718349841\\
0.414	0.00122423918378591\\
0.432	0.0013153089442791\\
0.45	0.00140565520590847\\
0.468	0.00139737934529861\\
0.486	0.00158505473028752\\
0.504	0.00154511843254333\\
0.522	0.00168856069683266\\
0.54	0.00174371316563686\\
0.558	0.00184332359379293\\
0.576	0.00186084745261475\\
0.594	0.00189728886844038\\
0.612	0.00190620853854875\\
0.63	0.00194876462453377\\
0.648	0.00211448310087969\\
0.666	0.00210138249253489\\
0.684	0.00211135656775147\\
0.702	0.00212358852039442\\
0.72	0.00222927389985594\\
0.738	0.00214309477194115\\
0.756	0.00216641743991932\\
0.774	0.00216401501456084\\
0.792	0.00209012849669106\\
0.81	0.00208349529665628\\
0.828	0.00199663296751526\\
0.846	0.00196421648319109\\
0.864	0.00191491805440833\\
0.882	0.00185672614049456\\
0.9	0.00183131879814795\\
0.918	0.00169730829804649\\
0.936	0.00160835053514388\\
0.954	0.00154375817018252\\
0.972	0.00138167369021324\\
0.99	0.00136324643951267\\
1.008	0.00128964771873101\\
1.026	0.00114902132109303\\
1.044	0.00108002279720396\\
1.062	0.000944124280000308\\
1.08	0.000884689517334836\\
1.098	0.00081967913136264\\
1.116	0.000835459078750622\\
1.134	0.000691046773005446\\
1.152	0.000659226939248433\\
1.17	0.000586978592138873\\
1.188	0.000537615263670446\\
1.206	0.000487984789578852\\
1.224	0.000419705212514741\\
1.242	0.000371770601183\\
1.26	0.00038871067973772\\
1.278	0.000293563724587062\\
1.296	0.000254723988420334\\
1.314	0.000227610542617831\\
1.332	0.000206987661560627\\
1.35	0.000186245553094038\\
1.368	0.000165702995516923\\
1.386	0.000153169220850349\\
1.404	0.000146157511292319\\
1.422	0.000113951236982607\\
1.44	0.000115797181622791\\
1.458	9.47861873686096e-05\\
1.476	7.0798323712737e-05\\
1.494	6.80628501989703e-05\\
1.512	5.6684823748299e-05\\
1.53	6.06412348453941e-05\\
1.548	5.86342699509779e-05\\
1.566	4.01521803981882e-05\\
1.584	3.27914290235969e-05\\
1.602	3.11667874734115e-05\\
1.62	3.02235500783975e-05\\
1.638	2.07497755003837e-05\\
1.656	1.68978080556158e-05\\
1.674	1.49275321294347e-05\\
1.692	1.06177454865941e-05\\
1.71	9.93972380604752e-06\\
1.728	9.25748273354705e-06\\
1.746	1.03075159003468e-05\\
1.764	7.36795230431797e-06\\
1.782	7.29469617425207e-06\\
1.8	4.99958335305306e-06\\
};
\addlegendentry{$\displaystyle \sigma^2_{\displaystyle \hat K^\epsilon_N}/\epsilon$}

\addplot [color=black, dashed, line width=0.5pt]
  table[row sep=crcr]{%
0.018	0.0154128230337303\\
0.036	0.011565224777333\\
0.054	0.00239671392412724\\
0.072	0.00307774375539524\\
0.09	0.00281408932019519\\
0.108	0.00374849235742984\\
0.126	0.00347748932977725\\
0.144	0.00356765311764312\\
0.162	0.00287256070584672\\
0.18	0.00323873309344266\\
0.198	0.00299123699729842\\
0.216	0.00317327761188248\\
0.234	0.00317066636832216\\
0.252	0.00286260175151018\\
0.27	0.00317039792789449\\
0.288	0.00318283581723917\\
0.306	0.00301727759613195\\
0.324	0.00305228458722601\\
0.342	0.00312670383981114\\
0.36	0.00317262482932463\\
0.378	0.00293486165138719\\
0.396	0.0031295383421677\\
0.414	0.00295709947774374\\
0.432	0.00304469663027569\\
0.45	0.00312367823535215\\
0.468	0.00298585330192011\\
0.486	0.00326142948618831\\
0.504	0.00306571117568121\\
0.522	0.00323479060695911\\
0.54	0.00322909845488307\\
0.558	0.00330344730070418\\
0.576	0.00323063793856728\\
0.594	0.00319408900410838\\
0.612	0.00311471983422998\\
0.63	0.00309327718179963\\
0.648	0.00326309120506124\\
0.666	0.0031552289677701\\
0.684	0.00308677860782379\\
0.702	0.00302505487235673\\
0.72	0.00309621374979992\\
0.738	0.00290392245520481\\
0.756	0.00286563153428481\\
0.774	0.00279588503173235\\
0.792	0.00263905113218568\\
0.81	0.00257221641562503\\
0.828	0.00241139247284452\\
0.846	0.00232176889266086\\
0.864	0.00221634034075038\\
0.882	0.00210513167856526\\
0.9	0.00203479866460883\\
0.918	0.0018489197146476\\
0.936	0.00171832322130756\\
0.954	0.00161819514694184\\
0.972	0.00142147498993131\\
0.99	0.00137701660556836\\
1.008	0.001279412419376\\
1.026	0.0011199038217281\\
1.044	0.00103450459502295\\
1.062	0.000889005913371288\\
1.08	0.000819156960495219\\
1.098	0.000746520156067978\\
1.116	0.000748619246192314\\
1.134	0.000609388688717324\\
1.152	0.000572245606986487\\
1.17	0.000501691104392199\\
1.188	0.000452538100732699\\
1.206	0.000404630837130059\\
1.224	0.00034289641545322\\
1.242	0.000299332207071659\\
1.26	0.000308500539474381\\
1.278	0.000229705574794258\\
1.296	0.000196546287361369\\
1.314	0.000173219591033357\\
1.332	0.000155396142312783\\
1.35	0.000137959668958547\\
1.368	0.000121127920699505\\
1.386	0.00011051170335523\\
1.404	0.000104100791518746\\
1.422	8.01344845166009e-05\\
1.44	8.04147094602718e-05\\
1.458	6.5011102447606e-05\\
1.476	4.79663439788191e-05\\
1.494	4.55574633192573e-05\\
1.512	3.7489962796494e-05\\
1.53	3.96347940165975e-05\\
1.548	3.78774353688488e-05\\
1.566	2.56399619400946e-05\\
1.584	2.07016597371193e-05\\
1.602	1.94549235164866e-05\\
1.62	1.86565123940725e-05\\
1.638	1.26677506107348e-05\\
1.656	1.02039903717487e-05\\
1.674	8.91728323144247e-06\\
1.692	6.27526328994922e-06\\
1.71	5.81270398014475e-06\\
1.728	5.35733954487676e-06\\
1.746	5.90350280661326e-06\\
1.764	4.17684370993082e-06\\
1.782	4.09354442999555e-06\\
1.8	2.7775463072517e-06\\
};
\addlegendentry{$\displaystyle \sigma^2_{\displaystyle \hat K^\epsilon_N}/\epsilon^2$}

\end{axis}

\node[] at (2.85,0.4) {(c')};

\end{tikzpicture}%
        \captionlistentry{}
        \label{fig:a1_bounds_PSD2}
    \end{subfigure}%
    
    \caption{Monte Carlo estimation of $P^\epsilon_1$ (for $a=1$ and $X_f=2$) using $\hat I_N^\eps$ (standard MC), $\hat J_N^\eps$ (simple control variate) and $K_N^\eps$ (practical optimal control variate). Results have been obtained with the same methodology as in Fig \ref{fig:P1_PSD1_PSD2}. The left column (a)-(b)-(c) corresponds to PSD1, and the right column (a')-(b')-(c') corresponds to PSD2.}
    \label{fig:a1_PSD1_PSD2}
\end{figure}

\begin{figure}[h!]
	\centering
	\begin{subfigure}{.5\textwidth}
		\centering
%
%
\definecolor{mycolor1}{rgb}{0.00000,0.44700,0.74100}%
\definecolor{mycolor2}{rgb}{0.85000,0.32500,0.09800}%
\begin{tikzpicture}

\begin{axis}[%
width=0.8\linewidth,
height=0.64\linewidth,
scale only axis,
xmin=0,
xmax=1,
xtick distance = 0.25,
xlabel = $a$,
ylabel = $P_1^\epsilon$,
yminorticks=true,
axis background/.style={fill=white},
legend style={legend cell align=left, align=left, draw=white!15!black}
]
\addplot[only marks, mark=*, mark options={}, mark size=3pt, color=mycolor1, fill=mycolor1] table[row sep=crcr]{%
a	P\\
0 0.003285\\
0.25 0.002294000000000\\
0.5 0.00155\\
0.75 0.001157000000000\\
1 0.000859\\
};
\addlegendentry{PSD1}

\addplot[only marks, mark=square*, mark options={}, mark size=3pt, color=mycolor2, fill=mycolor2] table[row sep=crcr]{%
a	P\\
0 0.005282\\
0.25 0.004041000000000\\
0.5 0.00291\\
0.75 0.002373000000000\\
1 0.001752\\
};
\addlegendentry{PSD2}
\end{axis}

\end{tikzpicture}%
		\caption{}
	\end{subfigure}%
\caption{Comparison of $P_1^\epsilon$ (with $\epsilon = 0.5, \:X_f=2$) estimated with our hybrid approach for elasto-plastic stiffness ratios $a=0$, $a=0.25$, $a=0.5$, $a=0.75$, and $a=1$; and driving coloured noise PSD1 (circles) and PSD2 (squares).} 
				
			\label{fig:compare_P1_a}
\end{figure}

\section{Discussion on method's advantages, computational efficiency, and limitations}
Here we propose a qualitative discussion on our proposed approach.
We consider the Monte Carlo estimator with control variate $\hat{K}_N^\epsilon$ as defined in Equation (15). For simplicity, we assume that $\hat{\lambda}_N^\epsilon$ is replaced by $\lambda^\epsilon$. The same reasoning applies when using $\hat{\lambda}_N^\epsilon$, particularly in the asymptotic regime as $N$ becomes large. To assess the performance of the method, we compare it with the standard Monte Carlo estimator $\hat{I}_N^\epsilon$ given in Equation (16). We distinguish two cases:

\begin{itemize}
    \item[(i)] When the expectation of the control variate, $\mathbb{E}[F(U^0)]$, admits an explicit and tractable formula. In this case, the control variate method can significantly reduce the variance of the estimator at a negligible additional computational cost.
    
    \item[(ii)] When $\mathbb{E}[F(U^0)]$ is not known explicitly. In this scenario, a numerical approximation is required. One viable approach is to employ a basic finite difference method applied to the newly derived PDE associated with $\mathbb{E} [F(U^0)]$. While this introduces an additional computational burden, the variance reduction achieved may still justify the cost, depending on the accuracy and efficiency of the numerical solver.
\end{itemize}

To achieve an approximation error $\delta >0$ with the standard MC estimator, i.e.
$$
\mathcal{E} := \sqrt{\mathbb{E} 
\left [ \left ( \hat{I}_N^\epsilon - I^\epsilon \right )^2 \right ]} \leq \delta,
$$
the minimal number of samples is $N_{\hat I} =  {\rm var}(F(U^\epsilon)) \delta^{-2}$.
This comes from the fact that 
$
\mathcal{E} = \sqrt{{\rm var}(F(U^\epsilon))/N}$.
On the other hand, 
to achieve the same accuracy with the control variate estimator the same reasoning implies that 
the minimal number of samples is $N_{\hat K} =  {\rm var}(F(U^\epsilon)-\lambda^\epsilon F(U^0)) \delta^{-2}$. Thus $N_{\hat K} = (1-\rho^2){\rm var}(F(U^\epsilon)) \delta^{-2}$.
From this analysis, we see that it is the factor $1-\rho^2$ that drives the reduction in the number of samples while preserving the approximation error.

Assume that the computational cost of generating a single sample is $T_{\widehat{I}}$ for the standard Monte Carlo estimator $\widehat{I}_N^\epsilon$, and $T_{\widehat{K}}$ for the control variate estimator $\widehat{K}_N^\epsilon$. From a computational efficiency standpoint, the control variate method is preferable if the total computational time required to achieve a given accuracy is lower, that is,
\[
T_{\hat{K}} N_{\hat{K}} < T_{\hat{I}} N_{\hat{I}}.
\]
Using known variance reduction properties of control variates, this condition translates to
\[
T_{\hat{K}} (1 - \rho^2) < T_{\hat{I}},
\]
where $\rho$ denotes the correlation coefficient between $F(U^\epsilon)$ and $F(U^0)$. Since $T_{\hat{K}}$ and $T_{\hat{I}}$ are typically of the same order of magnitude, even a moderate or strong correlation is sufficient to ensure that the control variate estimator is computationally advantageous.

When the expectation $\mathbb{E}[F(U^0)]$ is not available in closed form, an additional computational cost arises, denoted by $T_{\mathrm{PDE}}^{\delta/2}$ (with an accuracy $\delta/2$), corresponding to the numerical solution of the newly derived PDE. In our implementation, we employed a basic finite difference scheme to demonstrate that the numerical resolution behaves well, thereby illustrating the feasibility of the proposed hybrid approach. Naturally, this cost can be  reduced by adopting more advanced discretization techniques.

In this setting, the hybrid method becomes computationally advantageous when
\[
T_{\mathrm{PDE}}^{\delta/2} + T_{\widehat{K}} (1 - \rho^2) \, \mathrm{Var}(F(U^\epsilon)) \, (\delta/2)^{-2} < T_{\widehat{I}} \, \mathrm{Var}(F(U^\epsilon)) \, \delta^{-2},
\]
where $\delta$ denotes the target accuracy.

We have illustrated this regime with an example involving the estimation of a small probability, where the standard Monte Carlo method is prohibitively slow to achieve the desired accuracy. In contrast, our hybrid method attains the same level of accuracy with significantly reduced computational time for a range of $\epsilon$ values that we have identified.

\section{Conclusion}

We have developed an efficient method for computing probabilities with transient modulated,  coloured noise excitation and with target quantities of type \eqref{eq:P1} and \eqref{eq:P2}.
These quantities of interest are representative of an important class of problems in earthquake engineering. Our method has been shown to have good accuracy and is more computationally efficient than standard Monte Carlo methods.
It also does not make any  approximation other than the discretisation of the domain for solving the PDEs, unlike stochastic averaging or other common methods which approximate the process. Previous work was handled on the white noise-driven EPPO \cite{MWS18}, which was computationally less challenging because it can be reduced to a 2-dimensional problem.
That study also used stationary excitation and only considered \eqref{eq:P1} type quantities. Our current method is non-trivial  generalization of this approach. We can take into account the \bepo, non-stationary coloured noise excitation and \eqref{eq:P1},\eqref{eq:P2} -type quantities. As mentioned above, these features are essential for engineering fields like earthquake engineering. With this paper, we show that the concept developed in \cite{MR4373175,MWS18} can indeed be used in engineering in more realistic situations and with commonly used models like \bepo. 

\section*{Acknowledgement}
The authors would like to thank Cyril Feau for useful discussions.
Laurent Mertz is thankful for support through NSFC Grant No. 12271364 and GRF Grant No. 11302823.
The authors  thank the reviewers for their helpful feedback.
\appendix

\section{Time discretization of the coloured noise and the stochastic variational inequalities}
\label{sec:time_discretization}
Fix the final time $T>0$, the number of time steps $N_T \in \mathbb{N}$, and the number of Monte Carlo samples $N$. Define $\delta t$ such that $T = N_T \delta t$.
Furthermore, take a sequence of i.i.d. $d'$-dimensional Gaussian variables
$
\{ \Delta W_n^k \sim \mathcal{N} \left (0, I_{d'} \right ), \: 0 \leq n \leq N_T-1, \: 1 \leq k \leq N\},
$
where $\mathcal{N}(0,\sigma^2)$ denotes a normal distribution with mean $0$ and variance $\sigma^2$.
For the $k$-th sample, $1 \leq k \leq N$, the discretization of the OU process $\eta^{\eps,k}$ is given by the following:

If $A \in \mathbb{R}^{d \times d}, K \in \mathbb{R}^{d \times d'}$ in \eqref{eq:noise} are arbitrary, then 
\begin{equation}
\label{eq:OU_EM_BEPO}
\begin{cases}
\hat \eta_0^{\eps,k} & \sim \mathcal{N} \left (0, C(A,K) \right ),\\
\hat \eta_{n+1}^{\eps,k} & = (I - \frac{\delta t}{\eps^2} A ) \hat \eta_n^{\eps,k} + (\frac{\sqrt{\delta t}}{\eps} K) \Delta W_n^k, \: \: 0 \leq n \leq N_T - 1,
\end{cases}
\end{equation} 
where $C(A,K)$ is the covariance matrix \eqref{eq:cov_matrix}. In particular, we have

\begin{itemize}
\item
If the PSD of $\eta^{\eps,k}$ takes the form \eqref{eq:PSD1} with $q=1$, then $d=d'=1$ and 
$$
\begin{cases}
\hat \eta_0^{\eps,k} & \sim \mathcal{N} \left (0, \lambda_1/2 \right ),\\
\hat \eta_{n+1}^{\eps,k} & = (1 - \frac{\delta t}{\eps^2} \lambda_1 ) \hat \eta_n^{\eps,k} + (\frac{\sqrt{\delta t}}{\eps} \lambda_1 ) \Delta W_n^k, \: \: 0 \leq n \leq N_T - 1.
\end{cases}
$$

\item
If the PSD of $\eta^\eps$ takes the form \eqref{eq:PSD2} with $q=1$, then $d=d'=2$ and 
$$
\begin{cases}
(\hat \eta_0^{\eps,k})_1 & \sim \mathcal{N} \left (0, \lambda_1/2 \right ),\\
(\hat \eta_0^{\eps,k})_2 & \sim \mathcal{N} \left (0, \lambda_1/2 \right ),\\
(\hat \eta_{n+1}^{\eps,k})_1 & = (1 - \frac{\delta t}{\eps^2} \lambda_1) (\hat \eta_n^{\eps,k})_1 + (\frac{\delta t}{\eps^2} \omega_1) (\hat \eta_n^{\eps,k})_2 + (\frac{\sqrt{\delta t}}{\eps} \lambda_1) (\Delta W_n^k)_1, \: \: 0 \leq n \leq N_T - 1, \\
(\hat \eta_{n+1}^{\eps,k})_2 & = (1 - \frac{\delta t}{\eps^2} \lambda_1) (\hat \eta_n^{\eps,k})_2 - (\frac{\delta t}{\eps^2} \omega_1) (\hat \eta_n^{\eps,k})_1 + (\frac{\sqrt{\delta t}}{\eps} \lambda_1) (\Delta W_n^k)_2, \: \: 0 \leq n \leq N_T - 1.
\end{cases}
$$
where $(\cdot)_i$ denotes the $i$-th component.
\end{itemize}

For the time discretization of the coupling between the two stochastic variational inequalities, we proceed as follows :
\begin{itemize}
\item
$\hat X_0^{\eps,k} = x, \:  \hat Y_0^{\eps,k} = y, \:  \hat Z_0^{\eps,k} = z$ and for $0 \leq n \leq N_T - 1$
\begin{equation}
\label{eq:colorednoise_EM_BEPO}
\begin{cases}
& \hat X_{n+1}^{\eps,k} = \hat X_n^{\eps,k} + \delta t \hat Y_n^{\eps,k},\\
& \hat Y_{n+1}^{\eps,k} = \hat Y_n^{\eps,k} + \delta t \mathbf{B}(\hat X_n^{\eps,k},\hat Y_n^{\eps,k},\hat Z_{n}^{\eps,k}) 
+ \sigma(n \delta t) \delta t \: r \cdot \hat \eta_n^{\eps,k}/\eps, \\ 
& \hat Z_{n+1}^{\eps,k} = \textup{proj}_{[-z_{\textup{max}},z_{\textup{max}}]} \left ( \hat Z_n^{\eps,k} + \delta t \hat Y_n^{\eps,k} \right ).
\end{cases}
\end{equation} 
\item
$\hat X_0^{0,k} = x, \:  \hat Y_0^{0,k} = y, \:  \hat Z_0^{0,k} = z$ and for $0 \leq n \leq N_T - 1$
\begin{equation}
\label{eq:whitenoise_EM_BEPO}
\begin{cases}
\hat X_{n+1}^{0,k} = \hat X_n^{0,k} + \delta t \hat Y_n^{0,k},\\
\hat Y_{n+1}^{0,k} = \hat Y_n^{0,k} + \delta t \mathbf{B}(\hat X_n^{0,k},\hat Y_n^{0,k},\hat Z_{n}^{0,k}) 
+ \sigma(n \delta t) \sqrt{\delta t} \: r \cdot A^{-1} K \Delta W_n^k,\\ 
\hat Z_{n+1}^{0,k} = \textup{proj}_{[-z_{\textup{max}},z_{\textup{max}}]} \left ( \hat Z_n^{0,k} + \delta t \hat Y_n^{0,k} \right ).
\end{cases}
\end{equation} 

\end{itemize}
See \cite[p~305-337]{Kloeden2010-wa} for a detailed account of the time-discretization of SDEs.

\section{Finite difference implementation details}
\label{sec:fd_details}
Here, we present the finite difference method discretization of the PDE associated with $P_1^0$. The corresponding discretization for the PDE associated with $P_2^0$ follows similarly, with appropriate modifications.

To numerically approximate the solutions of the equations \eqref{eq:truncatedpde1} and \eqref{eq:truncatedpde2}, we employ a finite difference scheme. Let  $\tilde I,\tilde J,\tilde K$ be integers and define $I = 2 \tilde I+1$, $J = 2 \tilde J+1$, $K = 2 \tilde K+1$. We build our approximation on the following three-dimensional finite difference grid,
$$ 
\mathcal{G} \triangleq \{ (x_i,y_j,z_k) = 
(-\xmax+(i-1) \delta x,
-\ymax+(j-1) \delta y,
-\zmax+(k-1) \delta z), \: 1 \leq i \leq I, 1 \leq j \leq J, 1 \leq k \leq K \} 
$$
\[
\mathcal{G} \triangleq \left\{ (x_i, y_j, z_k) \,:\,
x_i = -\xmax + (i-1)\delta x,\,
y_j = -\ymax + (j-1)\delta y,\,
z_k = -\zmax + (k-1)\delta z
\right\}
\]
\[
\text{for } 1 \leq i \leq I,\quad 1 \leq j \leq J,\quad 1 \leq k \leq K
\]
where $\delta x \triangleq \xmax/\tilde I$, $\delta y \triangleq \ymax/\tilde J$ $\delta z \triangleq \zmax/\tilde K$, and hence there are $P \triangleq IJK$ points in the grid.  We then, denote $t_n \triangleq n \delta t$ with $\delta t = T/N_T$, where $N_T$ is the number of time steps. We use the notation $v_{i,j,k}^n$ for the approximation of $v(x_i,y_j,z_k,t_n)$, $\mathbf{v}^n$ for the vector of $P$ unknown when $n$ is fixed, and $f_{i,j,k}$, $\sigma_n$, $B_{i,j,k}$ for $f(x_i,y_j,z_k)$, $\sigma(t_n)$, $B(x_i,y_j,z_k)$, respectively.
The computation of $\mathbf{v}^n$ is obtained via a finite difference scheme in space. 
We use the following short-hand notation for centered finite differences of orders 2 and 1,
$$\m{\partial_x^2 w}_{i,j,k} \triangleq \frac{w_{i+1,j,k}-2w_{i,j,k}+w_{i-1,j,k}}{\delta x^2}, 
\: 
\mbox{ and }
\:
\m{\partial_x w}_{i,j,k} \triangleq \frac{w_{i+1,j,k}-w_{i-1,j,k}}{2\delta x}, $$
and 
forward and backward finite differences of order 1
$$
\m{\partial_x^f w}_{i,j,k} \triangleq \frac{w_{i+1,j,k}-w_{i,j,k}}{\delta x}
\: 
\mbox{ and }
\:
\m{\partial_x^b w}_{i,j,k} \triangleq \frac{w_{i,j,k}-w_{i-1,j,k}}{\delta x}.
$$
Moreover, we also denote for any function of $y$, $c(y)$,
$$
\m{c(y) \hat \partial_x w}_{i,j,k} 
\triangleq
\min(0,c(y_j)) \m{\partial_x^b w}_{i,j,k} 
+
\max(0,c(y_j)) \m{\partial_x^f w}_{i,j,k}.
$$
The centered, forward, and backward finite differences w.r.t. $y$ and $z$ are defined similarly.
At each time step $n \in \{0,...,N_T-1\}$ (decreasing in $n$), $\mathbf{v}^{n+1}$ is known and we solve for $\mathbf{v}^{n}$. This leads to the following set of $P$ linear equations: 
\begin{itemize}
\item
when $1 \leq i \leq I, 2 \leq j \leq J-1, 1 \leq k \leq K, $
$$ \frac{v_{i,j,k}^{n+1} -  v_{i,j,k}^n}{\delta t} + \frac{1}{2} \left ( \m{L^{n+1} v^{n+1}}_{i,j,k} + \m{L^n v^n}_{i,j,k} \right ) = 0,$$
\item
when $1 \leq i \leq I , j = \{1,J\}, 1 \leq k \leq K, $
$$ \m{\partial_y^f w^n}_{i,1,k} = 0 
\: \mbox{ and } \:
\m{\partial_y^b w^n}_{i,J,k} = 0.
$$
\end{itemize}
Here when $2 \leq i \leq I-1, 
\: 2 \leq j \leq J-1, 
\: 2 \leq k \leq K-1, $
$$ \m{L^{n} w^n}_{i,j,k} \triangleq \frac{\sigma_n^2}{2} \m{\partial_y^2 w^n}_{i,j,k} + B_{i,j,k} \m{\partial_y w^n}_{i,j,k} + \m{y \hat \partial_x w^n}_{i,j,k} 
+ \m{y \hat \partial_z w^n}_{i,j,k}.
$$
Otherwise, we modify $(L^n w^n)_{i,j,k}$ at the boundary. The last two terms describe the transport of the underlying process in the directions $x$ and $z$.
They contain four terms 
$$
\min(0,y_j) \m{\partial_x^b w^n}_{i,j,k} 
+
\max(0,y_j) \m{\partial_x^f w^n}_{i,j,k}
+
\min(0,y_j) \m{\partial_z^b w^n}_{i,j,k} 
+
\max(0,y_j) \m{\partial_z^f w^n}_{i,j,k}.
$$
To be consistent with Equations \eqref{eq:truncatedpde1} and \eqref{eq:truncatedpde2}, the outward flux terms are removed from $(L^n w^n)_{i,j,k}$ at the boundary.
We explicitly modify the operator as follows: in all cases $2 \leq j \leq J-1$, whenever $i=1$ or $k=1$, the terms to be removed from $(L^n w^n)_{i,j,k}$ are 
$$
\min(0,y_j) \m{\partial_x^b w^n}_{i,j,k} \: \mbox{ or }\:
\min(0,y_j) \m{\partial_z^b w^n}_{i,j,k}, \: \textup{respectively}.
$$
Similarly, whenever $i=I$ or $k=K$, the terms to be removed from $(L^n w^n)_{i,j,k}$ are 
$$
\max(0,y_j) \m{\partial_x^f w^n}_{i,j,k} \: \mbox{ or }\:
\max(0,y_j) \m{\partial_z^f w^n}_{i,j,k}, \: \textup{respectively}.
$$
Note that at most two terms can be removed.

\bibliography{references}

@article{housner1963behavior,
  title={The behavior of inverted pendulum structures during earthquakes},
  author={Housner, G. W.},
  journal={Bulletin of the seismological society of America},
  volume={53},
  number={2},
  pages={403--417},
  year={1963},
  publisher={The Seismological Society of America}
}

@article{Karnopp1966PlasticDI,
  title={Plastic Deformation in Random Vibration},
  author={Karnopp, D. C.and Scharton, T. D.},
  journal={Journal of the Acoustical Society of America},
  year={1966},
  volume={39},
  pages={1154-1161}
}

@article{alamilla2001evolutionary,
  title={Evolutionary properties of stochastic models of earthquake accelerograms: Their dependence on magnitude and distance},
  author={Alamilla, J. and Esteva, L. and Garcia-Perez, J. and Diaz-Lopez, O.},
  journal={Journal of Seismology},
  volume={5},
  number={1},
  pages={1--21},
  year={2001},
  publisher={Springer}
}

@article{milligan1966bauschinger,
  title={The Bauschinger effect in a high-strength steel},
  author={Milligan, R. V. and Koo, W. H. and Davidson, T. E.},
  journal={Journal of Basic Engineering},
  volume={88},
  number={2},
  pages={480--488},
  year={1966},
  publisher={American Society of Mechanical Engineers}
}

@article{sowerby1979review,
  title={A review of certain aspects of the Bauschinger effect in metals},
  author={Sowerby, R. and Uko, D. K. and Tomita, Y.},
  journal={Materials Science and Engineering},
  volume={41},
  number={1},
  pages={43--58},
  year={1979},
  publisher={Elsevier}
}

@article{patinet2020origin,
  title={Origin of the Bauschinger effect in amorphous solids},
  author={Patinet, S. and Barbot, A. and Lerbinger, M. and Vandembroucq, D. and Lema{\^\i}tre, A.},
  journal={Physical Review Letters},
  volume={124},
  number={20},
  pages={205503},
  year={2020},
  publisher={APS}
}

@article{qiang2019seismic,
  title={Seismic responses of postyield hardening single--degree-of-freedom systems incorporating high-strength elastic material},
  author={Qiang, H. and Feng, P. and Qu, Z.},
  journal={Earthquake Engineering \& Structural Dynamics},
  volume={48},
  number={6},
  pages={611--633},
  year={2019},
  publisher={Wiley Online Library}
}

@article{li2020isotropic,
  title={An isotropic-kinematic hardening model for cyclic shakedown and ratcheting of sand},
  author={Li, Z. and Liu, H.},
  journal={Soil Dynamics and Earthquake Engineering},
  volume={138},
  pages={106329},
  year={2020},
  publisher={Elsevier}
}

@article{rossi2015importance,
  title={Importance of isotropic hardening in the modeling of buckling restrained braces},
  author={Rossi, P. P.},
  journal={Journal of Structural Engineering},
  volume={141},
  number={4},
  pages={04014124},
  year={2015},
  publisher={American Society of Civil Engineers}
}

@article{chaboche1979modelization,
  title={Modelization of the strain memory effect on the cyclic hardening of 316 stainless steel},
  author={Chaboche, J. L. and Van, K. D. and Cordier, G.},
  year={1979},
  publisher={IASMiRT}
}

@article{lipinski1989elastoplasticity,
  title={Elastoplasticity of micro-inhomogeneous metals at large strains},
  author={Lipinski, P. and Berveiller, M.},
  journal={International Journal of Plasticity},
  volume={5},
  number={2},
  pages={149--172},
  year={1989},
  publisher={Elsevier}
}

@article{dirrenberger2012elastoplasticity,
  title={Elastoplasticity of auxetic materials},
  author={Dirrenberger, J. and Forest, S. and Jeulin, D.},
  journal={Computational Materials Science},
  volume={64},
  pages={57--61},
  year={2012},
  publisher={Elsevier}
}

@article{feau2015experimental,
  title={Experimental and numerical investigation of the earthquake response of crane bridges},
  author={Feau, C. and Politopoulos, I. and Kamaris, G. S. and Mathey, C. and Chaudat, T. and Nahas, G.},
  journal={Engineering Structures},
  volume={84},
  pages={89--101},
  year={2015},
  publisher={Elsevier}
}

@article{campbell1983inelastic,
  title={Inelastic response of piping systems subjected to in-structure seismic excitation},
  author={Campbell, R. D. and Kennedy, R. P. and Trasher, R. D.},
  journal={American Society of Mechanical Engineers. Paper},
  pages={10},
  year={1983}
}

@article{kasinos2015performance,
  title={Performance-based seismic analysis of light SDoF secondary substructures},
  author={Kasinos, S. and Palmeri, A. and Lombardo, M.},
  year={2015},
  publisher={University of British Columbia/{\copyright} The Authors}
}

@article{touboul1999seismic,
  title={Seismic behaviour of piping systems with and without defects: experimental and numerical evaluations},
  author={Touboul, F. and Sollogoub, P. and Blay, N.},
  journal={Nuclear engineering and design},
  volume={192},
  number={2-3},
  pages={243--260},
  year={1999},
  publisher={Elsevier}
}

@article{MWS18,
title = "A Feynman–Kac formula approach for computing expectations and threshold crossing probabilities of non-smooth stochastic dynamical systems",
journal = "Physica D: Nonlinear Phenomena",
volume = "397",
pages = "25 - 38",
year = "2019",
issn = "0167-2789",
doi = "https://doi.org/10.1016/j.physd.2019.05.003",
url = "http://www.sciencedirect.com/science/article/pii/S0167278918306183",
author = "Mertz, L. and Stadler, G. and Wylie, J."
}

@article{BT08,
  title={Degenerate Dirichlet problems related to the invariant measure of elasto-plastic oscillators},
  author={Bensoussan, A. and Turi, J.},
  journal={Applied Mathematics and Optimization},
  volume={58},
  number={1},
  pages={1--27},
  year={2008},
  publisher={Springer}
}

@book{eurocode8,
      title         = "{Eurocode 8: Design provisions for earthquake resistance
                       of structures : part. 1-3 : General rules, specific rules
                       for various materials and element}",
      publisher     = "BSI",
      address       = "London",
      year          = "1996",
      url           = "https://cds.cern.ch/record/848046",
      note          = "Merged with DD-ENV-1998-1-1 and DD-ENV-1998-1-2 into
                       prEN-1998-1",
}

@inproceedings{aci2008building,
  title={Building code requirements for structural concrete (ACI 318-08) and commentary},
  author={ACI Committee},
  year={2008},
  organization={American Concrete Institute}
}

@article{bazzurro2004nonlinear,
  title={Nonlinear soil-site effects in probabilistic seismic-hazard analysis},
  author={Bazzurro, P. and Cornell, C. A.},
  journal={Bulletin of the seismological society of America},
  volume={94},
  number={6},
  pages={2110--2123},
  year={2004},
  publisher={Seismological Society of America}
}

@article{liu1969spectral,
  title={Spectral simulation and earthquake site properties},
  author={Liu, S.-C. and Jhaveri, D. P.},
  journal={Journal of the Engineering Mechanics Division},
  volume={95},
  number={5},
  pages={1145--1168},
  year={1969},
  publisher={American Society of Civil Engineers}
}

@article{bourinet2011assessing,
  title={Assessing small failure probabilities by combined subset simulation and support vector machines},
  author={Bourinet, J.-M. and Deheeger, F. and Lemaire, M.},
  journal={Structural Safety},
  volume={33},
  number={6},
  pages={343--353},
  year={2011},
  publisher={Elsevier}
}

@article{katzgraber2006feedback,
  title={Feedback-optimized parallel tempering Monte Carlo},
  author={Katzgraber, H. G. and Trebst, S. and Huse, D. A. and Troyer, M.},
  journal={Journal of Statistical Mechanics: Theory and Experiment},
  volume={2006},
  number={03},
  pages={P03018},
  year={2006},
  publisher={IOP Publishing}
}

@article{au2003subset,
  title={Subset simulation and its application to seismic risk based on dynamic analysis},
  author={Au, S.-K. and Beck, J. L.},
  journal={Journal of engineering mechanics},
  volume={129},
  number={8},
  pages={901--917},
  year={2003},
  publisher={American Society of Civil Engineers}
}

@book {MR3308895,
    AUTHOR = {Pardoux, E. and R\u{a}\c{s}canu, A.},
     TITLE = {Stochastic differential equations, backward {SDE}s, partial
              differential equations},
    SERIES = {Stochastic Modelling and Applied Probability},
    VOLUME = {69},
 PUBLISHER = {Springer, Cham},
      YEAR = {2014},
     PAGES = {xviii+667},
      ISBN = {978-3-319-05713-2; 978-3-319-05714-9},
   MRCLASS = {60H05 (34F05 35D40 35R60 60H10 60J60)},
  MRNUMBER = {3308895},
MRREVIEWER = {Mark\ A.\ McKibben},
       DOI = {10.1007/978-3-319-05714-9},
       URL = {https://doi.org/10.1007/978-3-319-05714-9},
}

@book {MR0521262,
    AUTHOR = {Duvaut, G. and Lions, J.-L.},
     TITLE = {Inequalities in mechanics and physics},
    SERIES = {Grundlehren der Mathematischen Wissenschaften},
    VOLUME = {219},
      NOTE = {Translated from the French by C. W. John},
 PUBLISHER = {Springer-Verlag, Berlin-New York},
      YEAR = {1976},
     PAGES = {xvi+397},
      ISBN = {3-540-07327-2},
   MRCLASS = {69.00},
  MRNUMBER = {521262},
}

@article {MR4373175,
    AUTHOR = {Garnier, J. and Mertz, L.},
     TITLE = {A control variate method driven by diffusion approximation},
   JOURNAL = {Comm. Pure Appl. Math.},
  FJOURNAL = {Communications on Pure and Applied Mathematics},
    VOLUME = {75},
      YEAR = {2022},
    NUMBER = {3},
     PAGES = {455--492},
      ISSN = {0010-3640,1097-0312},
   MRCLASS = {60H10 (60F05 65C99)},
  MRNUMBER = {4373175},
       DOI = {10.1002/cpa.21976},
       URL = {https://doi.org/10.1002/cpa.21976},
}

@BOOK{Kloeden2010-wa,
  title     = "Numerical solution of stochastic differential equations",
  author    = "Kloeden, P. E. and Platen, E.",
  publisher = "Springer, Berlin",
  series    = "Stochastic Modelling and Applied Probability",
  year      =  1992,
  doi       = "https://doi.org/10.1007/978-3-662-12616-5"
}

@article{Giles_2015,
title={Multilevel Monte Carlo methods},
volume={24},
DOI={10.1017/S096249291500001X},
journal={Acta Numerica},
author={Giles, M. B.},
year={2015},
pages={259–328}}

@article{GOODMAN20097127,
title = {Coupling control variates for Markov chain Monte Carlo},
journal = {Journal of Computational Physics},
volume = {228},
number = {19},
pages = {7127-7136},
year = {2009},
issn = {0021-9991},
doi = {https://doi.org/10.1016/j.jcp.2009.03.043},
url = {https://www.sciencedirect.com/science/article/pii/S0021999109001685},
author = {Goodman, J. B. and Lin, K. K. }
}

@article{DerKiureghian1989,
title = {An evolutionary model for earthquake ground motion},
journal = {Structural Safety},
volume = {6},
number = {2},
pages = {235-246},
year = {1989},
issn = {0167-4730},
doi = {https://doi.org/10.1016/0167-4730(89)90024-6},
url = {https://www.sciencedirect.com/science/article/pii/0167473089900246},
author = {{Der Kiureghian}, A. and Crempien, J.}
}

@article{Rodolfo1973,
author = {Rodolfo Saragoni, G. and Hart, G. C.},
title = {Simulation of artificial earthquakes},
journal = {Earthquake Engineering \& Structural Dynamics},
volume = {2},
number = {3},
pages = {249-267},
doi = {https://doi.org/10.1002/eqe.4290020305},
url = {https://onlinelibrary.wiley.com/doi/abs/10.1002/eqe.4290020305},
eprint = {https://onlinelibrary.wiley.com/doi/pdf/10.1002/eqe.4290020305},
year = {1973}
}

@article{MATHEY2018,
title = {Experimental and numerical analyses of variability in the responses of imperfect slender free rigid blocks under random dynamic excitations},
journal = {Engineering Structures},
volume = {172},
pages = {891-906},
year = {2018},
issn = {0141-0296},
doi = {https://doi.org/10.1016/j.engstruct.2018.06.064},
url = {https://www.sciencedirect.com/science/article/pii/S0141029617319338},
author = {Mathey, C. and Feau, C. and Clair, D. and Baillet, L. and Fogli, M.}
}

@incollection{Bormann2013,
  title={Seismic Signals and Noise},
  author={Bormann, P. and Wielandt, E.},
  year={2013},
  booktitle={New Manual of Seismological Observatory Practice 2 (NMSOP2)},
  publisher={Potsdam : Deutsches GeoForschungsZentrum GFZ},
  doi={https://doi.org/10.2312/GFZ.NMSOP-2_ch4}
}

@article{marano2017,
  title={Fitting Earthquake Spectra: Colored Noise and Incomplete Data},
  author={Maran{\`o}, S. and Edwards, B. and Ferrari, G. and F{\"a}h, D.},
  journal={Bulletin of the Seismological Society of America},
  volume={107},
  number={1},
  pages={276-291},
  year={2017}
}

@article{chen2023,
    author = {Chen, C. and Lin, X. and Li, W. and Cheng, L. and Wang, H. and Zhang, Q. and Wang, Z.},
    title = {Adaptive coloured noise multirate Kalman filter and its application in coseismic deformations},
    journal = {Geophysical Journal International},
    volume = {234},
    number = {2},
    pages = {1236-1253},
    year = {2023},
    month = {03},
    doi = {10.1093/gji/ggad117}
}

@article{hillers2015,
author = {Hillers, G. and Husen, S. and Obermann, A. and Planès, T. and Larose, E. and Campillo, M.},
title = {Noise-based monitoring and imaging of aseismic transient deformation induced by the 2006 Basel reservoir stimulation},
journal = {Geophysics},
volume = {80},
number = {4},
pages = {KS51-KS68},
year = {2015},
doi = {10.1190/geo2014-0455.1}
}

@article{er2019,
author = {Er, G.-K. and Iu, V. P. and Du, H. E.},
title = {Probabilistic solutions of a stretched beam discretized with finite difference scheme and excited by Kanai-Tajimi ground motion},
journal = {Archives of Mechanics},
volume = {71},
number = {4-5},
pages = {433-457},
year = {2019},
doi = {http://doi.org/10.24423/aom.3145}
}

@article{er2018,
author = {Er, G.-K. and Wang, K. and Iu, V. P.},
title = {Probabilistic Solutions of the In-Plane Nonlinear Random Vibrations of Shallow Cables Under Filtered Gaussian White Noise},
journal = {International Journal of Structural Stability and Dynamics},
volume = {18},
number = {04},
pages = {1850062},
year = {2018},
doi = {10.1142/S0219455418500621}
}

@article{er2014,
title = {Probabilistic solutions of some multi-degree-of-freedom nonlinear stochastic dynamical systems excited by filtered Gaussian white noise},
journal = {Computer Physics Communications},
volume = {185},
number = {4},
pages = {1217-1222},
year = {2014},
issn = {0010-4655},
doi = {https://doi.org/10.1016/j.cpc.2013.12.019},
url = {https://www.sciencedirect.com/science/article/pii/S0010465513004311},
author = {Er G. K.}
}

@article{er2011,
author = {Er, G.-K.},
title = {Methodology for the solutions of some reduced Fokker-Planck equations in high dimensions},
journal = {Annalen der Physik},
volume = {523},
number = {3},
pages = {247-258},
doi = {https://doi.org/10.1002/andp.201010465},
year = {2011}
}

@article{er2012,
  title = {State-space-split method for some generalized Fokker-Planck-Kolmogorov equations in high dimensions},
  author = {Er, G.-K. and Iu, V. P.},
  journal = {Phys. Rev. E},
  volume = {85},
  issue = {6},
  pages = {067701},
  numpages = {4},
  year = {2012},
  month = {Jun},
  publisher = {American Physical Society},
  doi = {10.1103/PhysRevE.85.067701},
  url = {https://link.aps.org/doi/10.1103/PhysRevE.85.067701}
}

@article{er2017,
author = {Er, G.-K. and Iu, V. P.},
title = {Probabilistic Solutions of a Nonlinear Plate Excited by Gaussian White Noise Fully Correlated in Space},
journal = {International Journal of Structural Stability and Dynamics},
volume = {17},
number = {09},
pages = {1750097},
year = {2017},
doi = {10.1142/S0219455417500973},
}

@article{er2025,
title = {Evolutionary stochastic characteristics of nonlinear oscillator with one side barrier due to multiple modulated Gaussian white noise},
journal = {Communications in Nonlinear Science and Numerical Simulation},
volume = {145},
pages = {108696},
year = {2025},
issn = {1007-5704},
doi = {https://doi.org/10.1016/j.cnsns.2025.108696},
url = {https://www.sciencedirect.com/science/article/pii/S1007570425001078},
author = {Er, G.-K. and Luo, J. and Iu, V. P.},
}

@article{luo2024,
title = {Transient probability density of nonlinear oscillator under parametric harmonic and external modulated stochastic excitations},
journal = {Communications in Nonlinear Science and Numerical Simulation},
volume = {130},
pages = {107754},
year = {2024},
issn = {1007-5704},
doi = {https://doi.org/10.1016/j.cnsns.2023.107754},
url = {https://www.sciencedirect.com/science/article/pii/S1007570423006755},
author = {Luo, J. and Er, G.-K. and Iu, V. P.},
}
\bibliographystyle{unsrt}

\end{document}